\documentclass[aps,pra,reprint,onecolumn,nofootinbib,showpacs,notitlepage]{revtex4-1}
\usepackage{amsfonts}
\usepackage{amsmath,amssymb}
\usepackage{physics}
\usepackage{tikz}
\usepackage{qcircuit}
\usepackage{graphicx}
\usepackage{blkarray}
\usepackage{float}
\usepackage{subfigure}

\usepackage{bm}
\usepackage{xcolor}
\usepackage{hyperref}
\hypersetup{
     colorlinks   = true,
     citecolor    = blue
}

\def\be{\begin{equation}}
\def\ee{\end{equation}}
\def\bea{\begin{eqnarray}}
\def\eea{\end{eqnarray}}
\begin{document}
	\title{Quantum Computation of Phase Transition in the Massive Schwinger Model}
	\author{Shane Thompson}
	\email{sthomp78@tennessee.edu}
	\author{George Siopsis}
	\email{siopsis@tennessee.edu}
	\affiliation{Department of Physics and Astronomy, The University of Tennessee, Knoxville, TN 37996-1200, USA}

	\date{\today}
	\begin{abstract}
As pointed out by Coleman, physical quantities in the Schwinger model depend on a parameter $\theta$ that determines the background electric field. There is a phase transition for $\theta = \pi$ only. We develop a momentum space formalism on a lattice and use it to perform a quantum computation of the critical point of this phase transition on the NISQ device IMB Q Lima. After error mitigation, our results give strong indication of the existence of a critical point at $m/e\simeq 0.32$, where $m$ is the bare fermion mass and $e$ is the coupling strength, in good agreement with the classical numerical result $m/e \simeq 0.3335$.
	\end{abstract}
	\maketitle

\section{Introduction}
Gauge theories form a class of quantum field theories that play an essential role in the description of fundamental particle physics. Thus, devising methods that provide an efficient calculation of physical quantities within these theories, especially in a non-perturbative setting, is of primary importance. Non-perturbative calculations rely primarily on lattice formulations of gauge theories requiring a prohibitive amount of computational resources due to Fock spaces that grow exponentially with the size of the system. Quantum computers promise to address or bypass many of the complications arising in classical computing methods, such as the aforementioned exponentially growing Fock spaces as well as the infamous sign problem occurring in Monte Carlo simulations.

In addition to some theoretical schematics of such quantum simulations (see, for example, \cite{Byrnes2006,Banuls_2020,Shaw2020,Davoudi2020}), the recent emergence of real quantum hardware has ushered in intense focus on how to translate these techniques towards implementation on existing devices \cite{Martinez_2016,Muschik_2017,ORNL,Klco2020,Yang_2020}. The limited efficacy of these devices, termed noisy intermediate scale quantum (NISQ) hardware, is one major hurdle to overcome. Gauge theories present the additional challenge of requiring that quantum computers process only quantum states which live within the physical, or gauge invariant, subspace of the full Fock space. 

The Schwinger model has been studied extensively serving as a toy model for Quantum Chromodynamics (QCD) \cite{Schwinger1962,COLEMAN1975267,COLEMAN1976}. Gauge invariance is usually implemented with constraints derived from Gauss's law \cite{Kogut1975,Indrakshi2020}. In \cite{Martinez_2016}, the gauge degrees of freedom were completely removed by solving Gauss's law following \cite{Hamer1997}. While this resulted in a pure spin formulation, the new Hamiltonian required long-range couplings between the spins. The qubit connectivity needed to encode such couplings is currently only realizable on trapped ion systems.  In addition, accidental ``leakage" of the pure spin system into unphysical states is inevitable on NISQ devices. The authors of Ref.\ \cite{ORNL} evaded the latter issue by encoding only states that were constructed by hand to satisfy Gauss's law, at the cost of leaving in the original gauge degrees of freedom. Ref.\ \cite{ORNL} also reduced the size of the Hilbert space using symmetry arguments, which was necessary for an actual quantum computation.

The Schwinger model depends on an additional parameter $\theta$ that parametrizes the background electric field. It is known that a phase transition occurs only for $\theta=\pi$ \cite{COLEMAN1976}.
Here we study this phase transition using quantum computation. We show that, even though the phase transition occurs in the continuum, we can estimate the critical point with a small number of points in the spatial dimension (as low as four). We implement the quantum algorithm on quantum hardware (IBM Q Lima) by minimizing the number of qubits and circuit depth to avoid quantum error. We achieve this minimization by employing the formalism developed in \cite{bib1}, adapted to accommodate a spatial lattice. This formalism not only reduces the number of gauge degrees of freedom to one in a fashion similar to \cite{Hamer1997}, but it also enforces Gauss's law automatically in the encoded basis states. In addition, it facilitates the restriction of the Hilbert space to states obeying certain symmetries. Thus we are able to adopt the advantages of both approaches (Refs.\ \cite{Martinez_2016} and \cite{ORNL}).

The discussion proceeds as follows. In Section \ref{Section:Continuum}, we review the Schwinger model and its properties. In Section \ref{Section:Lattice}, we discuss the lattice Hamiltonian in which all but one gauge degrees of freedom are eliminated with the help of Gauss's Law. 
In Section \ref{Section: Phase Transition}, we discuss the critical point of the phase transition for $\theta=\pi$ in the continuum, and how it can be approximated to a significant degree of accuracy using small lattice sizes with as low as four spatial sites. In Section \ref{Section: Quantum Computation}, we perform a quantum calculation employing the Variational Quantum Eigensolver algorithm (VQE) applied to very simple Ans\"atze. We discuss two error mitigation techniques and present our results. Our conclusions are summarized in Section \ref{sec:Conclusion}.

\section{The model} \label{Section:Continuum}
In this Section, we review the massive Schwinger model and its pertinent features. We also briefly review the momentum space formalism discussed in Ref.\ \cite{bib1} that arrives at an exact solution in the massless case. We adapt this formalism to lattice discretization in Section  \ref{Section:Lattice}.

The massive Schwinger model describes a massive charged fermion in one spatial dimension. Despite being a $U(1)$ abelian gauge theory, it shares many important features with QCD (an $SU(3)$ gauge theory), making it an effective toy model. The Higg's mechanism, charge screening, ``quark" confinement, and spontaneous chiral symmetry breaking can all be found within the theory \cite{COLEMAN1975267}. 

Most importantly for our discussion, the model features topological ``$\theta$-vacua." \footnote{In QCD the ``strong CP problem" centers around Nature's apparent choice of $\theta\approx 0$ \cite{thooft1976,Graner2016}.} 
The case $\theta = \pi$ exhibits a phase transition when the mass gap vanishes somewhere between the asymptotic limits $e\gg m$ and $e\ll m$ \cite{COLEMAN1976,Shimizu2014,buyens2017,Azcoiti2018}. The latter limit corresponds to a phase where charge conjugation and parity (C/P) symmetry is spontaneously broken. The transition has been shown to belong in the universality class of the Ising model \cite{Byrnes2002,Shimizu2014}.

The Lagrangian density for the model is given by
\be \mathcal{L} = -\frac{1}{4}F_{\mu\nu}F^{\mu\nu}+\overline{\psi}\left(i\partial_\mu -eA_\mu\right)\gamma^\mu\psi - m\overline{\psi}\psi,  \ee
where 
$F_{\mu\nu} = \partial_\mu A_\nu - \partial_\nu A_\mu$, $A_\mu$ is an abelian gauge potential,
$\psi$ is a two-component fermionic field, $\gamma^0$ and $\gamma^1$ are Dirac matrices, and $\overline{\psi}\equiv \psi^\dagger \gamma^0$. There are two parameters in the model, the coupling strength $e$ and the bare fermion mass $m$. In the case $m=0$, the model can be solved exactly and is equivalent to a massive scalar boson of mass $\frac{e}{\sqrt\pi}$ \cite{LOWENSTEIN1971172}.

Gauss's Law is the constraint
\be \partial_1E = ej_0 \ee
where $E= F_{01}$ is the electric field and $j_0 = \psi^\dagger \psi$ is the charge density. The solution to Gauss's Law involves a constant of integration which is the background electric field \cite{COLEMAN1976}
\be F = \frac{e\theta}{2\pi} \ee
written in terms of a third parameter $\theta$ that determines the system, in addition to $e$ and $m$.

The Hamiltonian is
\be \label{eq:continuum H} H = \int_0^{2L} dx \left( \frac{1}{2}E^2 +i\overline{\psi}\gamma^1\left(\partial_1+ieA_1\right)\psi +  m\overline{\psi}\psi\right) \ee
where $2L$ is the length of the spatial dimension. We apply periodic boundary conditions.



The Gauss law constraint generates  gauge transformations. Thus in the quantum theory physical (gauge invariant) states $\ket{\Psi}$ must satisfy
\be \mathcal{G}(x)\ket{\Psi}=0,\qquad \mathcal{G}(x)=\partial_1 E(x)-ej_0(x)\ee
Gauss's Law does not completely eliminate all gauge degrees of freedom. This can be seen by, e.g., applying the Coulomb gauge $\partial_1 A_1 = 0$ \cite{MANTON1985220}, or a Fourier expansion of the gauge field $A_1$ in the temporal gauge $A_0=0$ \cite{bib1}. The residual gauge degree of freedom is the zero mode
\be\label{eq:5} q = \frac{1}{2L}\int_0^{2L} dx A_1 \ee
It determines the exponent of the Wilson loop operator $e^{ie\oint Adl}$ which is a gauge-invariant quantity \cite{Wilson1974,Aharonov1959}.

The scheme employed by Ref.\ \cite{bib1} yields two important features that we will use in the lattice formulation. First, the residual gauge degree of freedom is the only bosonic (zero) mode, and it plays an important role on the lattice. All other (higher) modes are purely fermionic. Second, the massless theory yields an infinite and degenerate set of ``$\theta$-vacua" of the form
\be \label{eq:cont theta vacua} \ket{0}_\theta = \sum_N e^{-iN\theta}\ket{0}_N, \ee
where $N$ characterizes the level of ``displacement" of some  low-energy fermionic state (Dirac sea). This displacement is implemented by large gauge transformations (LGT) which are topologically non-trivial. LGTs and $\theta$-vacua will be key ingredients in the development of a basis to be used on the lattice.

\section{Lattice}\label{Section:Lattice}
In this Section, we introduce the lattice Hamiltonian using the Kogut-Susskind staggering formulation \cite{Kogut1975,LatticeFermions,bib2} which is expressed in the position representation. We then transform it to the momentum space representation in order to facilitate  restricting to the relevant sector of the Hilbert space and employing the two advantageous features of the formalism of \cite{bib1} discussed in the previous Section. By using Gauss's Law to eliminate all but one of the gauge degrees of freedom, the electric part of the Hamiltonian is split into a zero-mode bosonic term and a fermionic term corresponding to the higher Fourier modes. 

We also discuss the parameter $\theta$ in the context of LGTs which we use to identify states of the form \eqref{eq:cont theta vacua}. In Appendix \ref{app:B}, we explain how $\theta$ can be related to a background field in our lattice formulation.

The Kogut-Susskind Hamiltonian for the Schwinger model consists of an electromagnetic (gauge) and a fermionic part,
\be
    H = H_{EM} + H_f \ee 
    where
    \bea H_{EM} &=& \frac{1}{2}\sum_{r=0}^{2L-1}E^2(r) \nonumber\\
    H_f &=& \frac{i}{2} \sum_{r=0}^{2L-1}\Big(\psi^\dagger(r) e^{ieA_1(r)} \psi(r+1) - \text{h.c.}\Big) + m\sum_{r=0}^{2L-1}(-1)^r\psi^\dagger(r)\psi(r)
\eea
We work in units in which the lattice spacing (between fermionic sites) is $a=1$, so $L$ is the number of spatial sites and $2L$ is the number of fermionic sites. 
The fermionic part is simplified by defining
\be \label{eq: res gauge xform} \widetilde{\psi} (r) = e^{ie \sum_{r'=0}^{ r-1} A_1(r')} \psi (r) \ee
This transformation is similar to that used in  \cite{Hamer1997}, in which a purely fermionic formulation is obtained for open boundary conditions. A free parameter exists at the boundary, which Refs.\ \cite{Martinez_2016} and \cite{Hamer1997}  set to zero. In our approach, we use periodic boundary conditions which results in a single gauge degree of freedom (zero mode) \cite{bib1}.

Using \eqref{eq: res gauge xform}, the fermionic part of the Hamiltonian $H_f$ simplifies to
\be H_f = \frac{i}{2} \sum_{r=0}^{2L-1}\left( \widetilde{\psi}^\dagger(r) \widetilde{\psi}(r+1) - \text{h.c.} \right) + m\sum_{r=0}^{2L-1}(-1)^r\widetilde\psi^\dagger(r)\widetilde\psi(r) \ee
It is convenient to express the fields in momentum space. We have
\be\label{eq:11}
\widetilde{\psi} (r) = \frac{1}{\sqrt{2L}} \sum_{l=0}^{2L-1} b_l e^{-ik_l r} \ , \ k_l=\frac{(2l+1)\pi}{2L} - eq
\ee
where $q$ is the zero mode of the gauge field $A_1$ (Eq.\ \eqref{eq:5}), while for the fermion expansion coefficients the usual anti-commutation relations hold:
\be \label{eq: b anticommies} \{b_l,b_{l'}^\dagger\}=\delta_{ll'} \ee
Since $A(r)$ obeys periodic boundary conditions, the expansion \eqref{eq:11} gives anti-periodic boundary conditions for $\psi(r)$.
We obtain 
\be H_f =  \sum_{l=0}^{2L-1} b_l^\dagger b_l \sin k_l  + m\sum_{l=0}^{2L-1} b_{l+L}^\dagger b_l\ee
For the gauge part, we introduce the Fourier expansion of the electric field,
\be E(r) = \frac{p+(-1)^r a_{L}}{2L} - \frac{i}{2L} \sum_{l=1}^{L-1}\left(a_{l} e^{\pi ilr/L} - \text{h.c.} \right) \ee
The modes $p, a_{L}$ are hermitian, and $p$ is the momentum conjugate to the gauge zero mode $q$,
\be [q,p] = i \ee
The Hamiltonian for the gauge field becomes
\be H_{EM} =  \frac{p^2 +a_{L}^2}{4L} + \frac{1}{2L}\sum_{l=1}^{L-1} a_l^\dagger a_l \ee
Gauss's Law reads
\be E(r) - E(r-1) = e j(r) - e \langle j(r) \rangle  \ee
where $j(r) = \psi^\dagger(r)\psi(r)$ is the charge density and we normal-ordered by subtracting its vacuum expectation value.
In terms of modes,
\be a_l = ie \frac{j_l - \langle j_l \rangle}{1-e^{-\pi in/L}} \ , \ \ a_{L} = \frac{e}{2} (j_{L} - \langle j_{L} \rangle ) \ , \ \ j_l = \sum_{s=0}^{2L-1} b_{s+l}^\dagger b_s \ee
where indices are mod$(2L)$ and $j_l = \sum_{r=0}^{2L-1} j(r) e^{-\pi i lr/L}$ are Fourier modes of the charge density $j(r)$. Notice that $j_{2L-l} = j_l^\dagger$. It is also easy to show that $[j_l,j_{l'}]=0$, $\forall (l,l')$.

Moreover, the total charge vanishes,
\be\label{eq:11Q} Q = j_0 - \langle j_0 \rangle = 0 \ee
The gauge part of the Hamiltonian is written entirely in terms of fermionic modes and the bosonic zero mode momentum $p$,
\be H_{EM} =  \frac{p^2}{4L} + \frac{e^2}{16L} \sum_{l=1}^{2L-1} \csc^2\frac{\pi l}{2L} \left( j_l^\dagger - \langle j_l \rangle^\ast \right) \left( j_l - \langle j_l \rangle \right) \ee
We adopt the ground state in which $\langle j(r)\rangle = r\ \text{mod} (2)$. In Ref.\ \cite{bib2} it was shown that this is the ground state in the strong-coupling limit $e\gg1$. Therefore,
\be\label{eq:21} \langle j_l\rangle =  L\left(\delta_{l,0}-\delta_{l,L}\right)\ee 
and the Hamiltonian $H_{EM}$ becomes
\be \label{eq:normal-ordered lattice H} H_{EM}=\frac{p^2}{4L}+\frac{e^2}{8L}\sum_{l=1}^{L-1}\csc^2{\frac{\pi l}{2L}}j_l^\dagger j_l+\frac{e^2}{16L}\left(j_{L}+L \right)^2 \ee
The remaining constraint \eqref{eq:11Q} needs to be imposed on all physical states. From \eqref{eq:21}, we see that the requirement of vanishing charge restricts our Hilbert space to half-filling: $j_0=L$.

By solving Gauss's law, we have restricted the Hilbert space to states that are invariant under small gauge transformations. 
In order to account for large gauge transformations, first consider a general gauge transformation,
\be {\psi} (r) \to e^{ie\omega (r)} {\psi} (r) \ , \ \ A(r)\to A(r) - \omega(r+1) + \omega(r)\ee
From this we have
\be q \to q - \frac{\omega(2L) - \omega(0)}{2L} \ , \ \ \tilde{\psi} (r) \to e^{ie\omega (0)} \tilde{\psi} (r) \ee
Gauge transformations are classified by their winding numbers $w\in \mathbb{Z}$,
\be \omega (2L) - \omega(0) = \frac{2\pi}{e} w \ee
transforming
\be q\to q-\frac{\pi}{eL}w\, \ \ k_l \to k_{l+w} \ , \ \ b_l \to b_{l+w} \ee
with indices taking values mod$(2L)$. Gauss's Law itself only yields \eqref{eq:11Q} as a necessary condition for a physical state. However, if $w=0$ (small gauge transformation) one can show in a manner similar to the continuum case \cite{bib1} that it is a sufficient condition as well. 

LGTs correspond to $w\ne0$. Let $U_1$ implement the LGT with $w=1$,
\be U_1 q U_1^\dagger = q - \frac{\pi}{eL}\ , \ \ U_1 b_l U_1^\dagger = b_{l+1} \ee
Notice that $U_1^{2L} = \mathbb{I}$. Physical quantities must be invariant under $U_1$. For physical states, this implies
\be U_1\ket{\text{phys}} = e^{i\theta} \ket{\text{phys}}
\ee
Owing to the periodicity $U^{2L} = \mathbb{I}$, $\theta$ can take on $2L$ values,
\be\label{eq:30} \theta = 0, \pm\frac{\pi}{L}, \dots , \pm \frac{(L-1)\pi}{L} , \pi \ee 
Each $\theta$ labels a distinct sector of the Hilbert space defining an independent physical system. The parameter $\theta$ is related to the background electric field (see Appendix \ref{app:B} for details). The system considered in \cite{Martinez_2016,ORNL} corresponds to the $\theta = 0$ sector.


The Fock space is a tensor product of fermionic states and a one-dimensional system described by the residual gauge degree of freedom $q$. A basis for fermionic states is provided by the states
\be \ket{\bm{x}} \equiv \ket{x_0 x_1\dots x_{2L-1}} = (b_0^\dagger)^{x_0} (b_1^\dagger)^{x_1} \ldots (b_{2L-1}^\dagger)^{x_{2L-1}}\ket{\Omega} \ee
where $\ket{\Omega}$ is the fermionic vacuum annihilated by all $b_l$, and $x_l = 0,1$ ($l=0,1\dots,2L-1$) are occupation numbers.

For the gauge degree of freedom, a basis is provided by plane waves
\be \langle q |p=ne\rangle = \frac{1}{\sqrt{2\pi}}e^{ineq} \ee
where $n\in \mathbb{Z}$.

Imposing gauge invariance under LGTs, we obtain states in the $\theta$ sector,
\be \label{eq:eigenstate} |p=ne, \bm{x} ; \theta \rangle \propto \sum_{l=0}^{2L-1} e^{il\theta} U_1^l |p=ne\rangle \otimes |\bm{x} \rangle,  \ee
where normalization is neglected for the moment since there may be repeated terms in the sum. 

Under the LGT $U_1$ we have
\be U_1\ket{n\theta}=e^{-i\theta}\ket{n\theta} \ee
for the values \eqref{eq:30} of the parameter $\theta$.

We define the action of $U_1$ on the fermionic vacuum by
\be \label{eq:U on Omega} U_1\ket{\Omega}=(-)^{L}\ket{\Omega} \ee
so that systems in the $\theta$ sector with odd $L$ behave similar to those with even $L$.

It is convenient to introduce the lattice translation operator
\be \label{eq:Tdef} \mathcal{T}\equiv (-)^{L}e^{2i\mathcal{P}} \ , \ \ \mathcal{P}\equiv\sum_{l=0}^{2L-1}\frac{(2l+1)\pi}{2L} b_l^\dagger b_l \ee
where $\mathcal{P}$ is the momentum operator (notice the extra factor of 2 owing to the smallest translation on a staggered lattice being by two fermionic sites). The $(-)^{L}$ factor is inserted for a similar reason to \eqref{eq:U on Omega}.
It is straightforward to check that the translation operator $\mathcal{T}$ commutes with the Hamiltonian and is gauge-invariant (under both small and large gauge transformations). 

We also define the lattice parity operator $\Pi$ by \cite{Berruto1998}
\be \Pi b_l \Pi = -b_{L-l-1},\ \Pi q \Pi = -q,\ \Pi p \Pi = -p \ee
where indices are defined mod($2L$). 
We define its action on $\ket{\Omega}$ including a phase that depends on $L$:
\be \Pi\ket{\Omega}=(-)^{L(L+1)/2}\ket{\Omega}\ , \ee
which ensures that the Dirac sea state is even under parity.

It is straightforward to show that $H$ is invariant under $\Pi$. In particular, we have $\Pi j_l \Pi = j_{2L-l}$. It should be noted that $\Pi$ is not gauge-invariant on the entire Hilbert space. Indeed, it is straightforward to show that
\be \Pi\ket{p=ne,\bm{x};\theta} = \ket{p=-ne,\bm{x}';-\theta} \ee
with $\ket{\bm{x}'} = \Pi\ket{\bm{x}}$, i.e., $\Pi$ maps states in the $\theta$ sector to states in the $-\theta$ sector.
Thus $\Pi$ is only gauge-invariant for $\theta=0,\pi$, which are the sectors we concentrate on here.

To reduce the size of the Fock space for the calculation of the critical point, we invoke the symmetries of the lattice Schwinger model under $\mathcal{T}$ and $\Pi$, enabling us to focus on the sectors that contain the ground and first-excited states. We take advantage of the gauge invariance of the states.

Details of the calculation of matrix elements of the Hamiltonian can be found in Appendix \ref{app:A}. 
\begin{figure}[H]
    \centering
    \subfigure[]{\includegraphics[scale=0.4]{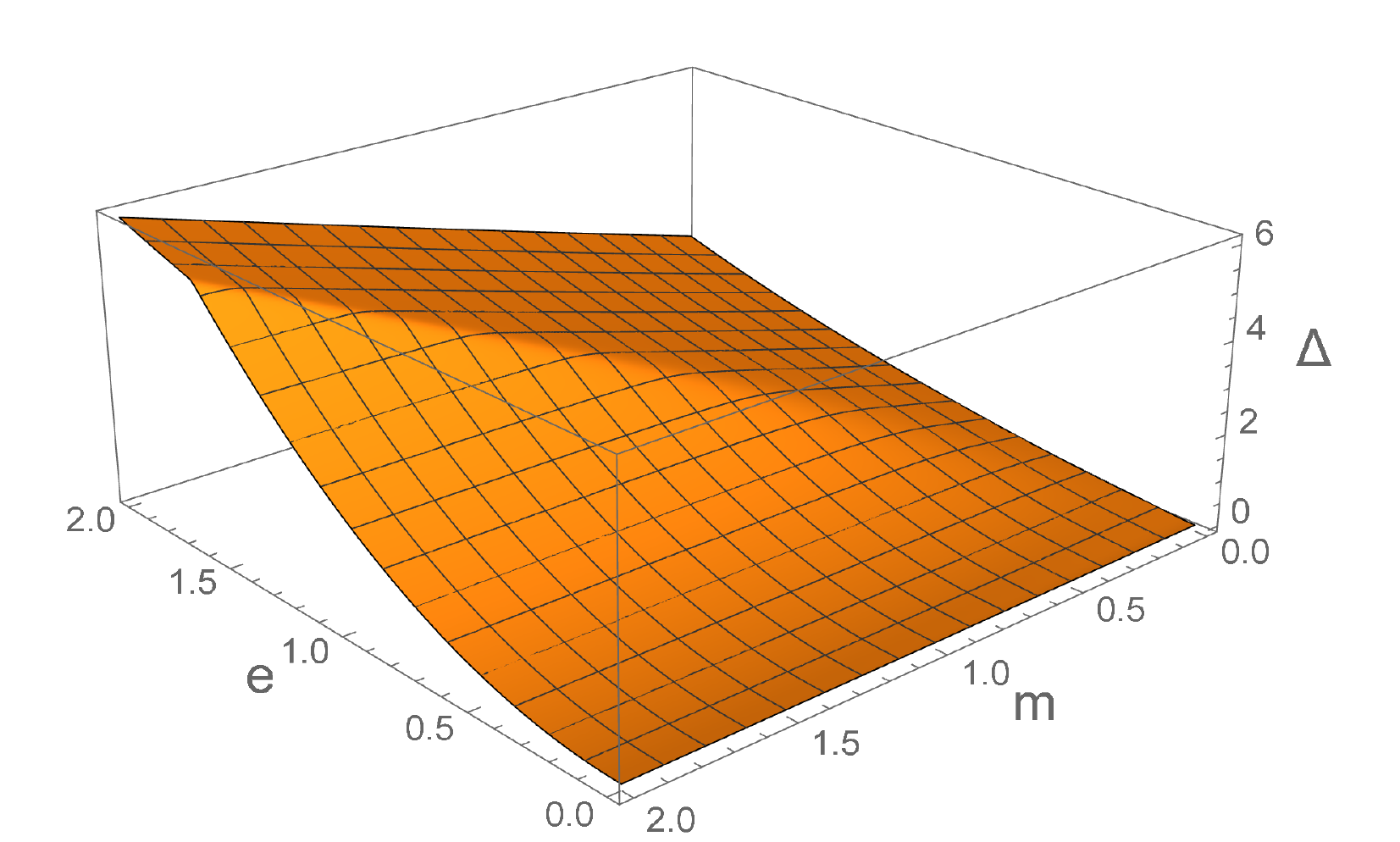}}
    \subfigure[]{\includegraphics[scale=0.4]{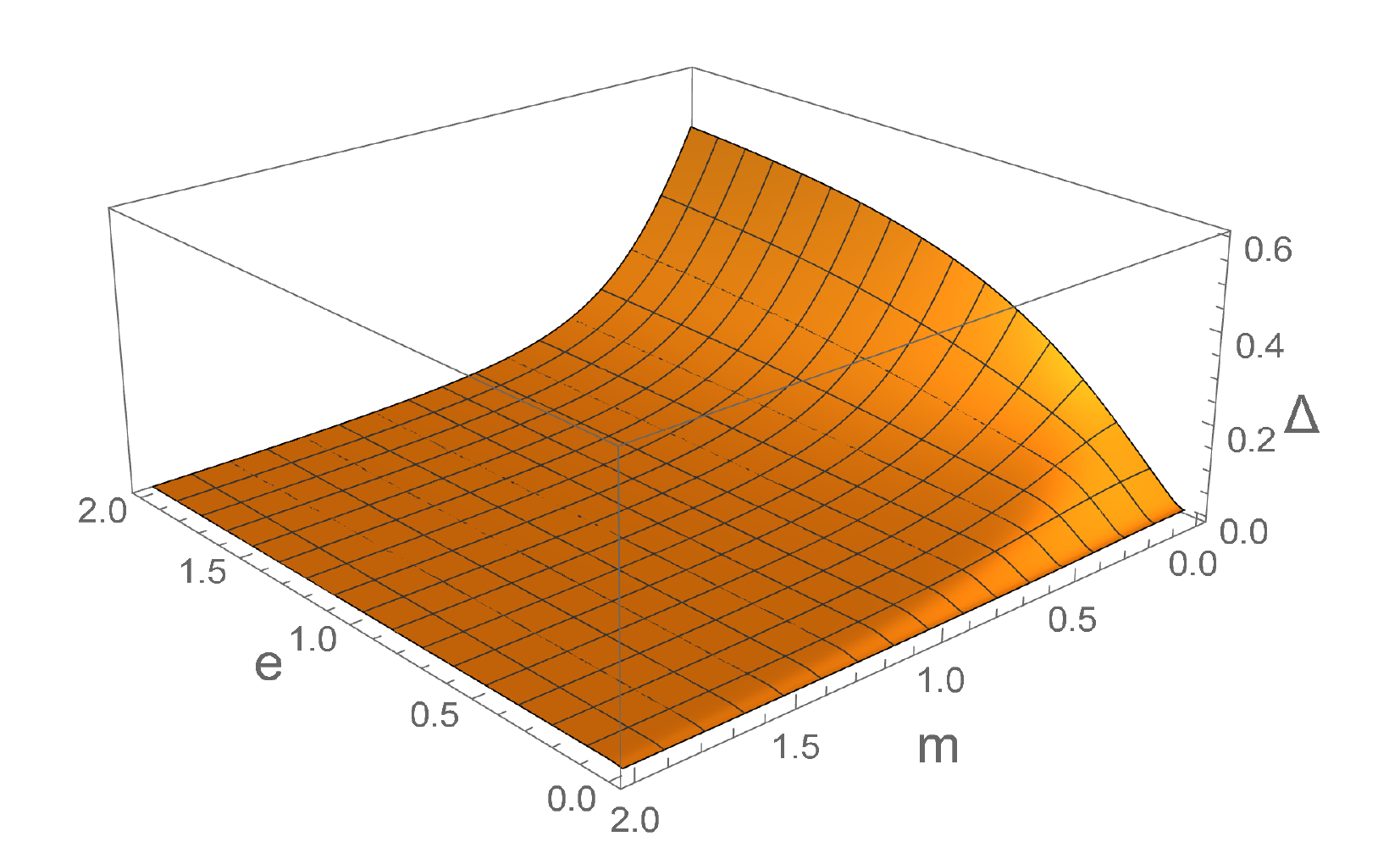}}
    \caption{Mass gap $\Delta$ for $L=2$ as a function of coupling $e$ and bare mass $m$, for (a) $\theta=0$ and (b) $\theta=\pi$. The Hilbert space is truncated to $n_{\text{max}}=20$.}
    \label{fig:dim=8}
\end{figure}

To calculate eigenvalues numerically using the matrix elements of the Hamiltonian, it is necessary to truncate the infinite dimensional Hilbert space to a cutoff, $n\le n_{\text{max}}$. For $e\gg 1$ and/or $m\gg 1$, we obtain accurate values of the mass gap (difference between the energies of the first-excited state and ground state) with modest values of the cutoff. 

In Figure \ref{fig:dim=8}, the mass gap is plotted as a function of $e$ and $m$, for two spatial points ($L=2$), with the Hilbert space truncated to $n_{\text{max}}=20$. For $\theta=\pi$ (right panel), the gap approaches zero asymptotically for all values of $e$, as $m$ varies from $m\ll e$ (where the gap is noticeably finite) to $m\gg e$ indicating the presence of a phase transition. For $\theta=0$ (left panel), as we increase the mass $m$, the gap initially increases and then levels out when the spectrum of the mass term in the Hamiltonian becomes dominant, indicating the absence of a phase transition.  

\begin{figure}[H]
    \centering
    \subfigure[]{\includegraphics[scale=0.4]{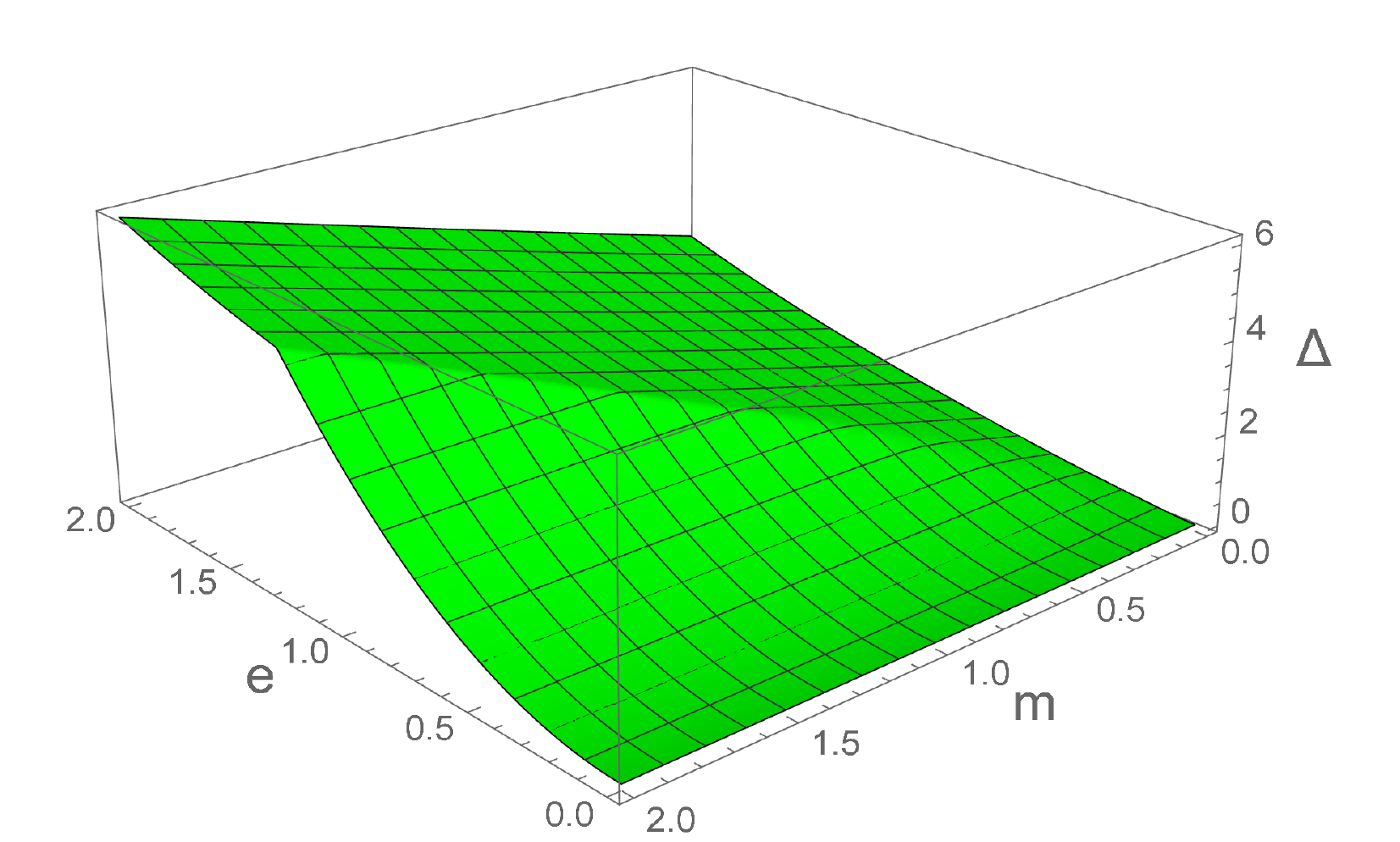}}
    \subfigure[]{\includegraphics[scale=0.4]{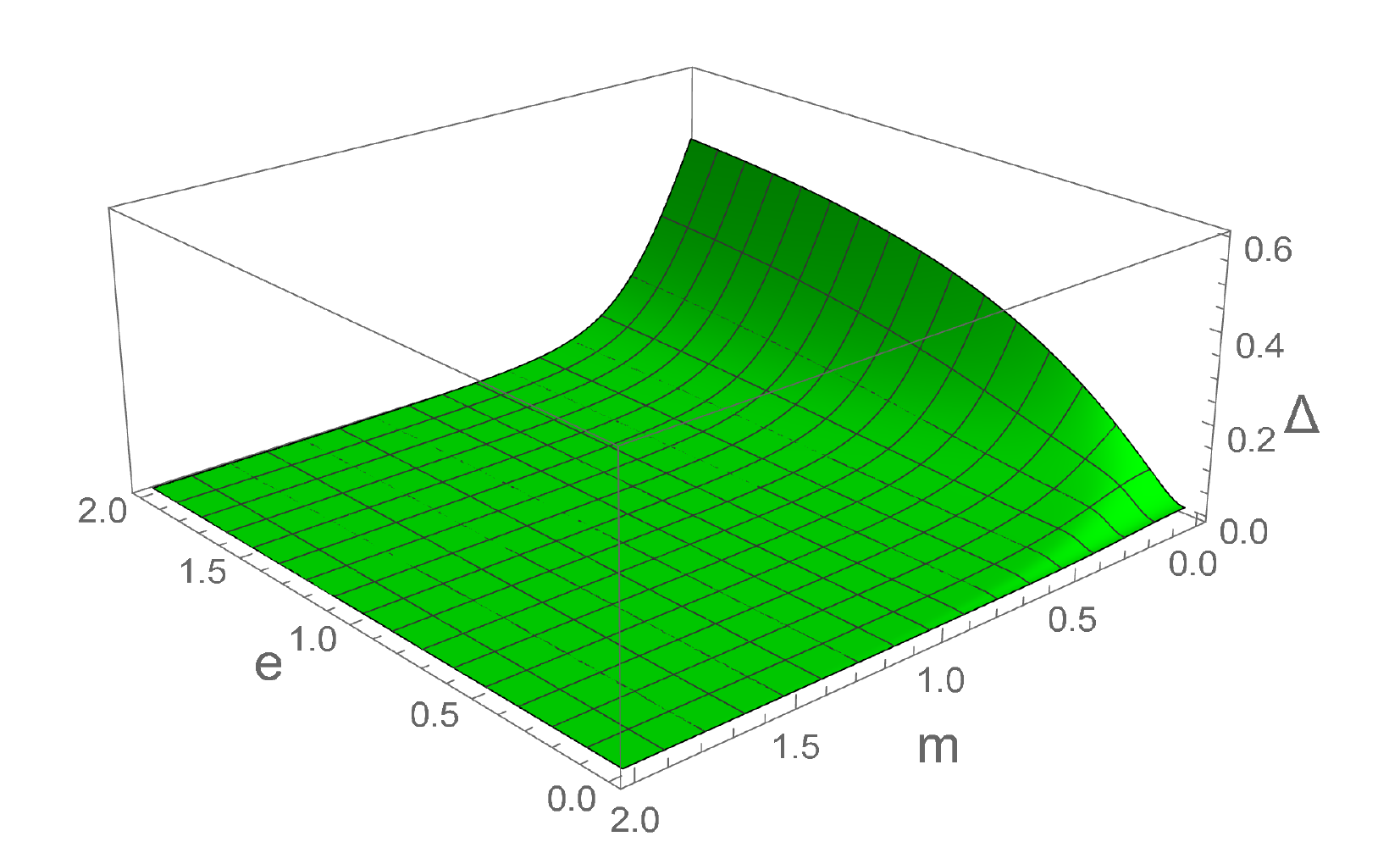}}
    \caption{Mass gap $\Delta$ for $L=3$ as a function of coupling $e$ and bare mass $m$, for (a) $\theta=0$ and (b) $\theta=\pi$. The Hilbert space is truncated to $n_{\text{max}}=20$.}
    \label{fig:n=47}
\end{figure}

In Figure \ref{fig:n=47}, the mass gap is plotted as a function of $e$ and $m$, for three spatial points ($L=3$), with the Hilbert space truncated to $n_{\text{max}}=20$. As expected, these plots exhibit similar behavior to those for two spatial points ($L=2$) (Figure \ref{fig:dim=8}). Notice that for $\theta = \pi$, the drop of the gap to zero asymptotically is sharper, in accord with the expectation of a phase transition in the continuum limit, where the gap vanishes at a finite value of $m$ (critical point) for each $e$.

In Appendix \ref{Section:Truncation}, it is shown that a large cutoff in the Fock space is not needed for a wide range of parameters. Accurate results can be obtained for $n_{\text{max}} \sim 2L$. This is advantageous for quantum computation on NISQ devices.

\section{Phase Transition} \label{Section: Phase Transition}
In this Section, we  compute the critical value of $\frac{m}{e}$ for the phase transition that occurs in the $\theta=\pi$ sector. To this end, we apply finite-size scaling theory to small lattice sizes, going up to $L=5$ (five spatial sites). For each lattice size and coupling $e$ the scaling theory yields key values of $\frac{m}{e}$ known as ``pseudo-critical points." Using weighted linear regression we can extrapolate to $e=0$, which corresponds to zero lattice spacing (since the dimensionless quantity $ea\to 0$, where $a$ is the lattice spacing). With this we find the critical point to considerable accuracy, showing that these small lattice sizes are close enough to the continuum limit. We perform exact diagonalization classically. A quantum calculation using $L=3,4$ is implemented on quantum hardware (IBM Q) in the next Section.

\begin{figure}[H]
    \centering
    \subfigure[\ $e=0.1$]{\includegraphics[scale=0.55]{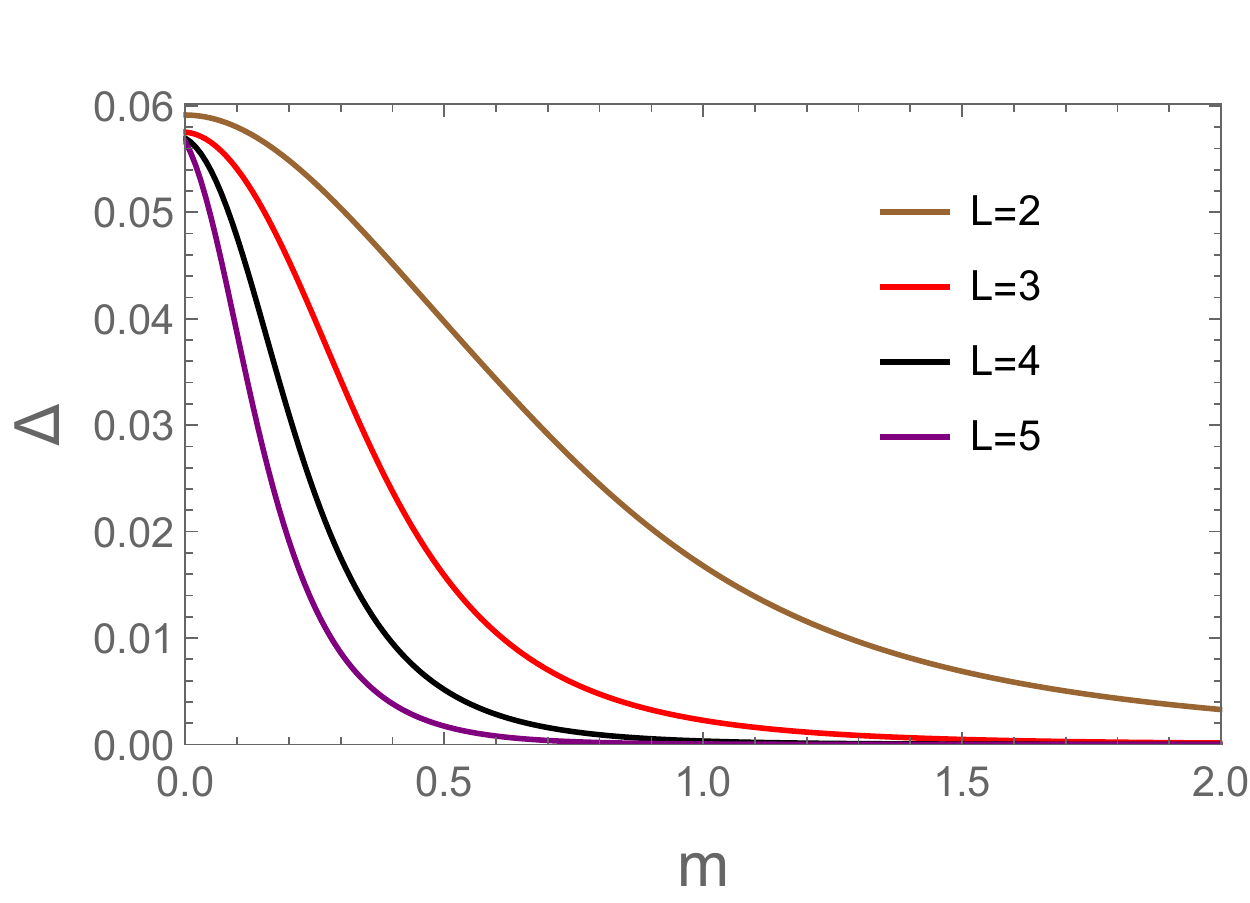}}
    \subfigure[\ $e=1$]{\includegraphics[scale=0.55]{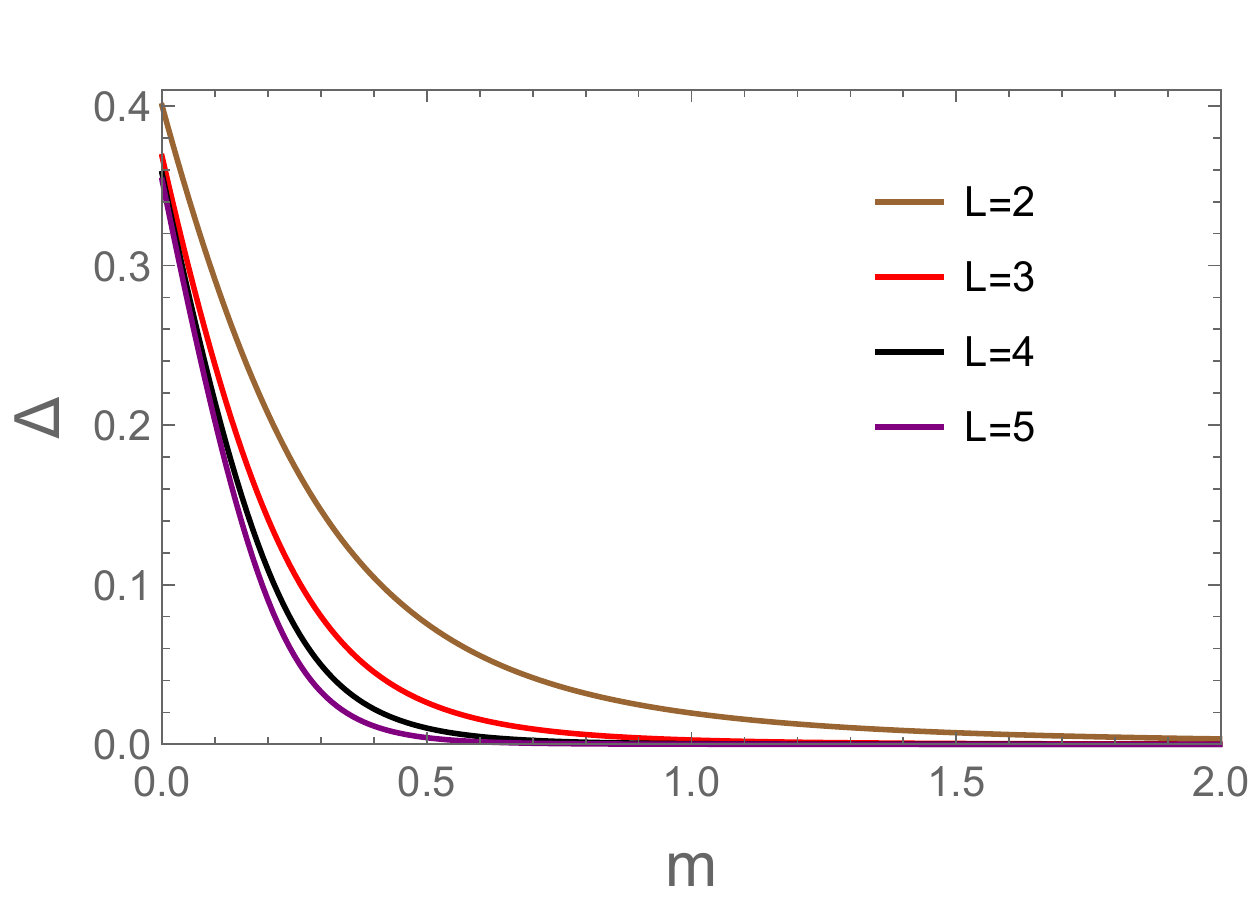}}
    \caption{Mass gap $\Delta$ \emph{vs.}\ bare mass $m$ in the $\theta=\pi$ sector for $L=2,3,4,5$. The coupling is set at \emph{(a)} $e=0.1$ and \emph{(b)} $e=1$. Notice that the $L\to\infty$ limit is approached faster for larger $e$.}
    \label{fig:gap_vs_m_e=01_N=2,3,4,5,6}
\end{figure}

As shown in Figure \ref{fig:gap_vs_m_e=01_N=2,3,4,5,6}, the mass gap $\Delta$ approaches zero for $m>0$ for $\theta= \pi$, as expected. For other values of $\theta$, this behavior is not observed, which is consistent with the absence of a phase transition. For $\theta = 0$, it vanishes for a negative value of $m$, because the system is equivalent to $\theta = \pi$ if we transform $m\to -m$ \cite{CREUTZ1995,Zache2019,Funcke2020}. This is shown explicitly in Appendix \ref{Section:other theta}.

\begin{figure}[H] 
    \centering
    \subfigure[]{\includegraphics[scale=0.65]{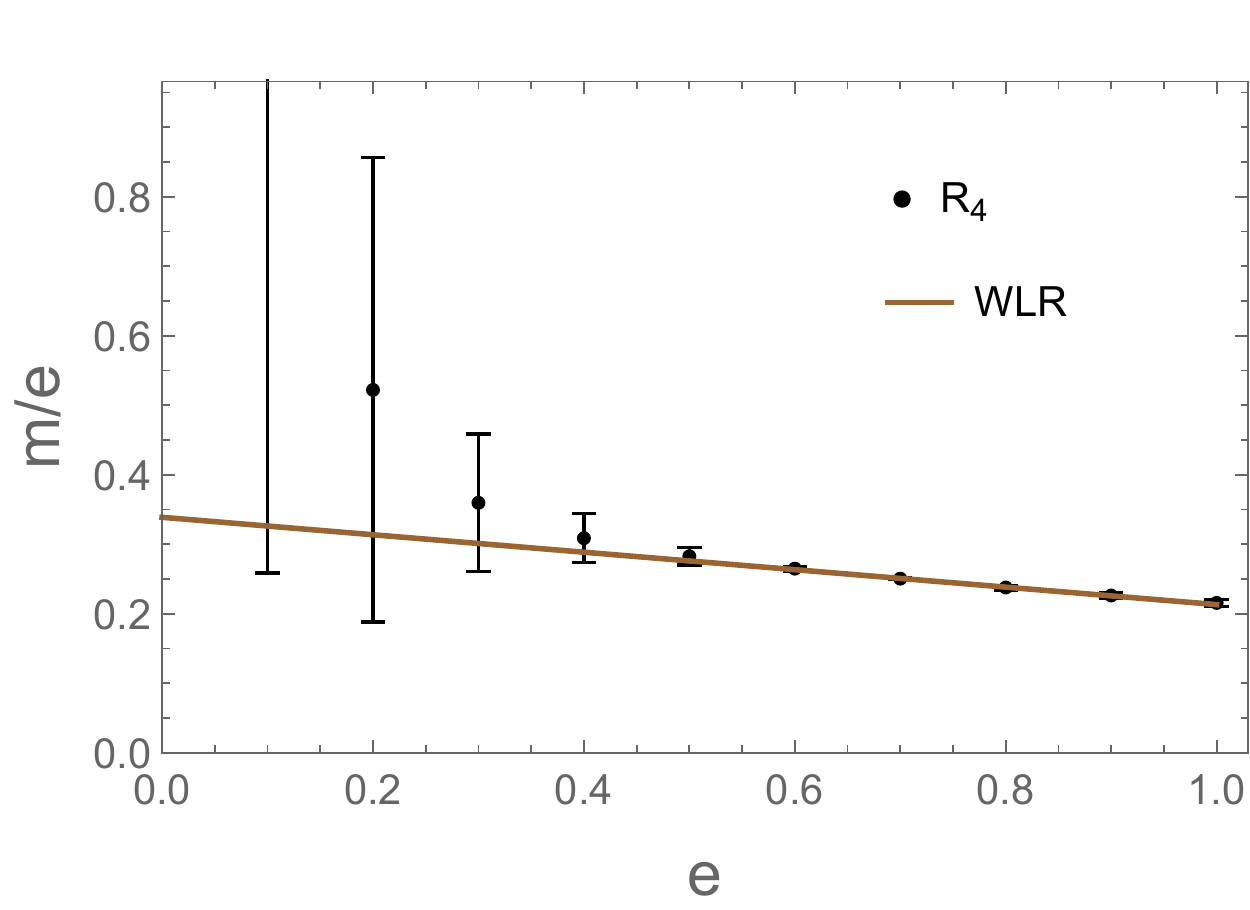}}
    \subfigure[]{\includegraphics[scale=0.65]{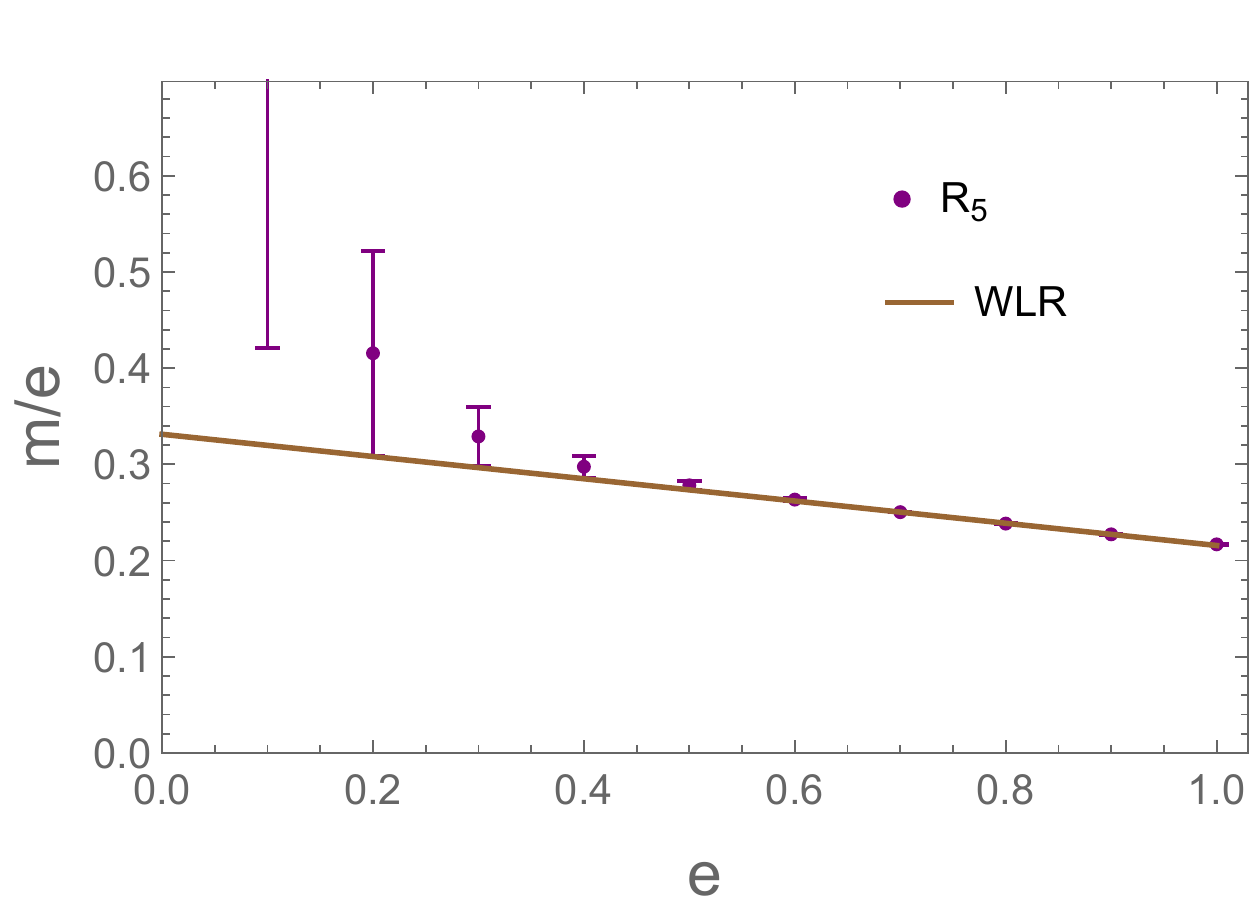}}
    \caption{Pseudo-critical points \emph{vs.}\ coupling constant $e$ for \emph{(a)} $L=4$ and \emph{(b)} $L=5$, along with a weighted linear regression. The error bars are estimated as the deviation of the $L=3$ and $L=4$ points, respectively. While errors increase as $e\to 0$, the weighted linear regression yields the values $0.339$ for $L=4$ and $0.331$ for $L=5$ at $e=0$, in good agreement with the continuum result $0.3335\pm 0.0002$ \cite{Byrnes2002}.}
    \label{fig:pseudo_vs_coupling}
\end{figure}

Following finite-size scaling theory \cite{Fisher1972,Hamer_1980}, we estimate the critical point by finding values of the bare mass $m$ and coupling constant $e$ such that the scaled energy-gap ratio $R=1$, where for a given lattice size $L$, the ratio is defined by
\be \label{eq:finite size} R_{L}(m/e)\equiv \frac{L\Delta_{L}(m/e)}{(L-1)\Delta_{L-1}(m/e)} \ee
and $\Delta_L$ is the mass gap. This form was developed by Hamer, \emph{et al.}, \cite{Hamer_1980} for Hamiltonian field theory from the statistical mechanics scaling theory of Fisher, \emph{et al.}, \cite{Fisher1972}, and was applied to the Schwinger model in \cite{HAMER1982} and \cite{Byrnes2002}. Solutions to $R_L=1$ are the aforementioned pseudo-critical points approximating the critical point obtained in the continuum limit $L\to\infty$. Pseudo-critical points are plotted for $L=4$ and $e\in[0.1,1]$ in Figure \ref{fig:pseudo_vs_coupling}. A weighted linear regression, using the deviation of the $L=3$ points as an estimate of the error,  yields a value of $m/e=0.339$ for $e=0$ which corresponds to the continuum limit. This result agrees with Ref.\ \cite{HAMER1982} where the critical value was found to be $0.325\pm 0.02$. It is also a good approximation to $0.3335\pm 0.0002$, obtained in Ref.\ \cite{Byrnes2002} through the use of very large lattice sizes.

In order to obtain Figures \ref{fig:gap_vs_m_e=01_N=2,3,4,5,6} and \ref{fig:pseudo_vs_coupling}, we diagonalized finite matrices by imposing a sufficiently high cutoff on the value of $p=ne,\ n\in\mathbb{Z}$ in the Fock space. In our quantum computation we had limited resources to work with, which necessitated truncating the space at a lower cutoff. Fortunately, this truncation did not have a significant effect on the mass gap in the $\theta = \pi$ sector where the phase transition is observed (see Appendix \ref{Section:Truncation} for details).

\section{Quantum Computation} \label{Section: Quantum Computation}
In this Section, we locate the critical point in the $\theta=\pi$ sector using IBM's quantum hardware. The algorithm used is the Variational Quantum Eigensolver (VQE) \cite{Peruzzo_2014}, which we find useful in our case since the ground and first excited states appear as ground states in the even/odd parity sectors, respectively (see Appendix \ref{app:A} for details). We develop simple Ans\"atze involving only three qubits for this algorithm, and describe the error mitigation techniques employed for improved accuracy. To reduce machine error from long circuit depth and numerous two-qubit gates, we set the maximum lattice size we put on quantum hardware to $L=4$. In the previous section it was shown that this is sufficient for a good estimate of the critical point. 

For the quantum computation, we restrict our scope to the points carrying small error bars in Figure \ref{fig:pseudo_vs_coupling}, which eliminates the need for a weighted linear regression involving the calculation of two ratios, $R_3$ and $R_4$. Thus we may restrict our attention to $L=3$ and $L=4$ and only calculate $R_4$. In this range of the coupling constant $e$, we can write simple Ans\"atze for the ground and first excited states that can be implemented with only three qubits. The Ans\"atze introduced below are effective in the range $e\in[0.5,1.0]$.\footnote{Classical calculations reveal that the pseudo-critical points for $R_3$ and $R_4$ are nearly equal in this range. After extrapolating these to $e=0$, however, it is $R_4$ which delivers a critical value right at the upper bound of Hamer's result \cite{HAMER1982} (with improvement if we include weights), while that from $R_3$ lies well above it. It is also important to note that the error bars for $R_3$ require results from the simple $L=1$ (one spatial site) case, and these error bars turn out to be significant even for $e\ge 0.8$.}

According to the discussion in Appendix \ref{section:larger lattices}, we expect to require a maximum of $2\text{log}_2 L+1$ qubits for $L\le 5$. Thus one might expect to need five qubits for our purposes ($L\le 4$). However the results of Appendix \ref{Section:Truncation} indicate that three qubits suffice for this range of $e$.

\subsection{Trial Functions}
To form our Ans\"atze we select appropriate basis states $\ket{p=ne,\bm{x};\theta}$ which contribute the most to the full eigenstates. For $L=3\ (L=4)$ we take the six (eight) most significant basis states. This yields a three-qubit problem. From perturbation theory it can be deduced that for large $e$ we should populate our trial function with basis states with small $n$. On the other hand if $e$ is decreased, we expect higher $n$ to contribute, but it is the ``Dirac-sea" fermionic configurations which give the largest contributions (e.g., $\ket{00001111}$ for $L=4$). We find from exact diagonalization that the former choice of basis states will be more useful in forming our trial functions. As demonstrated classically in Appendix \ref{Section:Truncation}, this truncation does not affect the value of the gap significantly.

To form simple Ans\"atze it is helpful to permute the basis so that the states are ordered from highest to lowest in terms of contribution to the exact eigenstates (although one additional swap is useful in order to construct the $L=3$ ground state). Instead of including the permutation unitary within the ansatz, which will increase circuit depth, we instead transform the Hamiltonian into the newly-ordered basis.

For $L=4$, we adopt trial functions with a simple structure, composed of pairs of basis states that contribute nearly equally,
\be \ket{\Psi_{\text{trial}}}_{L=4} = \left(a_0\ket{00}+a_1\ket{01}+a_2\ket{10}+a_3\ket{11}\right)\otimes\frac{1}{\sqrt2}\left(\ket{0}+\ket{1}\right),\quad a_i\in \mathbb{R},\ \sum_i a_i^2=1 \ee
where we used IBM Qiskit's convention where the last qubit is ``qubit 0". Thus the following circuits suffice for the ground and first excited states, respectively:
\[\Qcircuit @C=0.5em @R=0.5em {
    \lstick{\ket{0}_0} & \qw & \gate{R_Y(\pi/2)} & \qw &\qw & \qw & \qw & \qw & \qw & \qw & \qw \\
    \lstick{\ket{0}_1} & \qw & \gate{R_Y(\theta_0)} & \qw& \gate{R_Y(\theta_2)} & \qw & \targ & \qw & \gate{R_Y(-\theta_2)} & \qw & \qw \\
    \lstick{\ket{0}_2} & \qw & \gate{R_Y(\theta_1)} & \qw& \qw & \qw & \ctrl{-1} & \qw & \qw & \qw & \qw
}
\hspace{2 cm}
\Qcircuit @C=0.5em @R=0.5em {
    \lstick{\ket{0}_0} & \qw & \gate{R_Y(\pi/2)} & \qw & \gate{R_Y(\theta_2)} & \targ & \gate{R_Y(-\theta_2)} & \qw & \qw & \qw & \qw & \qw \\
    \lstick{\ket{0}_1} & \qw & \gate{R_Y(\theta_0)} & \qw & \qw & \ctrl{-1} & \qw & \qw & \qw & \qw &\qw &\qw \\
    \lstick{\ket{0}_2} & \qw & \gate{R_Y(\theta_1)} & \qw & \qw & \qw & \qw & \qw & \qw & \qw & \qw & \qw
}\]
For $L=3$ we use the same circuit for both the ground and first excited states:
\[\Qcircuit @C=0.5em @R=0.5em {
    \lstick{\ket{0}_0} & \qw & \gate{R_Y(\pi/2)} & \qw & \gate{R_Y(\theta_2)} & \qw & \qw & \qw & \qw & \qw & \targ & \qw & \gate{R_Y(-\theta_2)} & \qw \\
    \lstick{\ket{0}_1} & \qw & \gate{R_Y(\theta_0)}  & \qw & \gate{X} & \qw & \ctrl{1} & \qw & \gate{X} & \qw & \qw & \qw & \qw & \qw \\
    \lstick{\ket{0}_2} & \qw & \gate{R_Y(\theta_1)} & \qw & \qw & \qw & \targ & \qw & \gate{R_Y(-\theta_1)} & \qw & \ctrl{-2} & \qw & \qw & \qw 
}\]
This circuit constructs a trial function in a six dimensional subspace:
\be \label{eq:six_dim} \ket{\Psi_{\text{trial}}}_{L=3} = \left(a_0\ket{00}+a_1\ket{01}\right)\otimes\frac{1}{\sqrt2}\left(\ket{0}+\ket{1}\right)+\ket{10}\otimes\left(a_2\ket{0}+a_3\ket{1}\right),\quad a_i\in \mathbb{R},\ \sum_i a_i^2=1 \ee

\subsection{Error Mitigation}
We employ two forms of error mitigation to improve the accuracy of our results. The first source of error to address is readout (RO) error, also known as measurement error, in which unintended classical bit-flips occur upon measurement. The probabilities of such bit-flips occurring can be obtained by preparing a single basis state on the quantum device and counting the populations of all basis states that are measured upon readout. From this we may obtain a ``calibration matrix" which can be used to modify raw results obtained on hardware. IBM's \textit{qiskit-ignis} contains the tools to do this, namely employing a least-squares approach to best approximate the original probabilities before readout. A re-calibration is performed every $30$ minutes during a run.  

\begin{figure}[ht]
    \centering
    \subfigure[\ L=3 Ground State]{\includegraphics[scale=0.5]{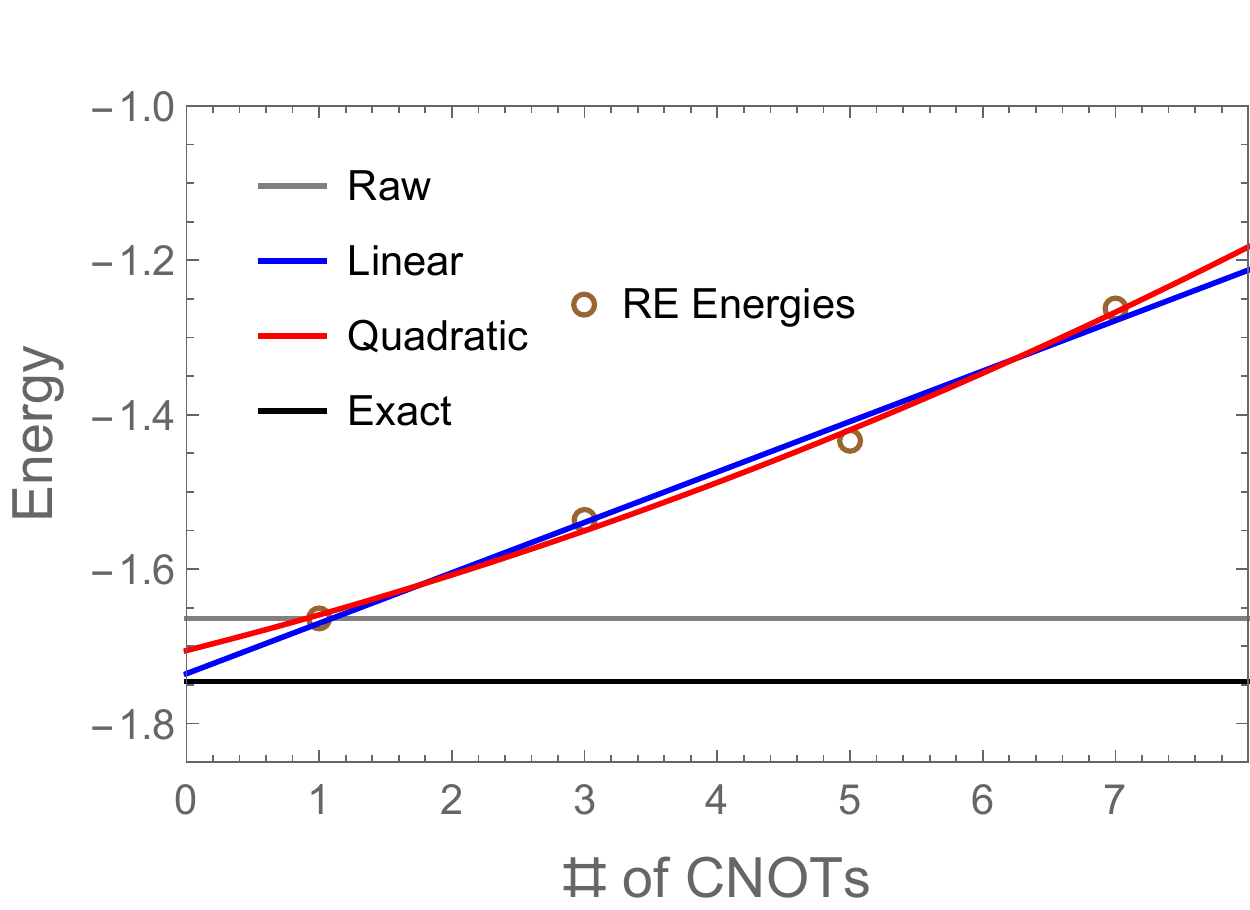}}
    \subfigure[\ L=3 First Excited State]{\includegraphics[scale=0.5]{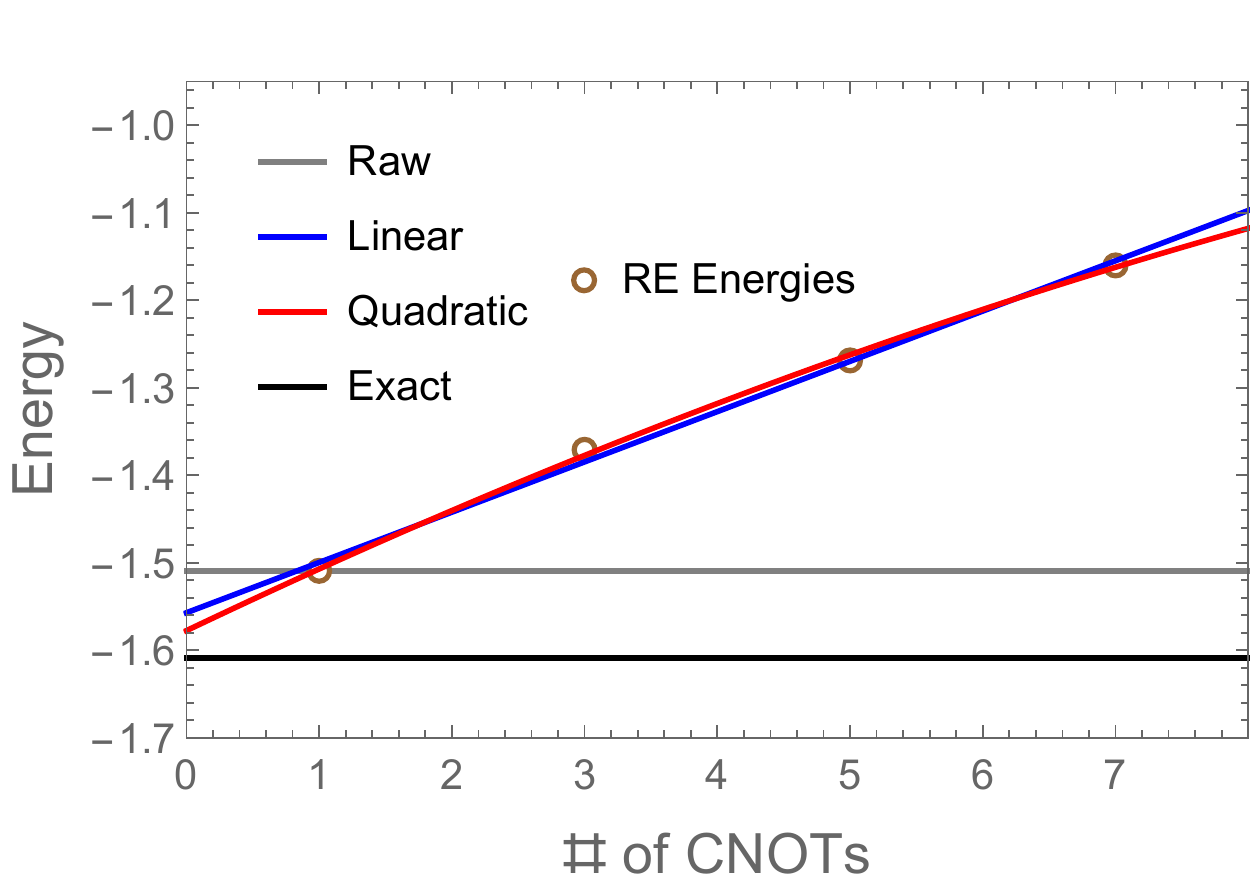}}
    \subfigure[\ L=4 Ground State]{\includegraphics[scale=0.5]{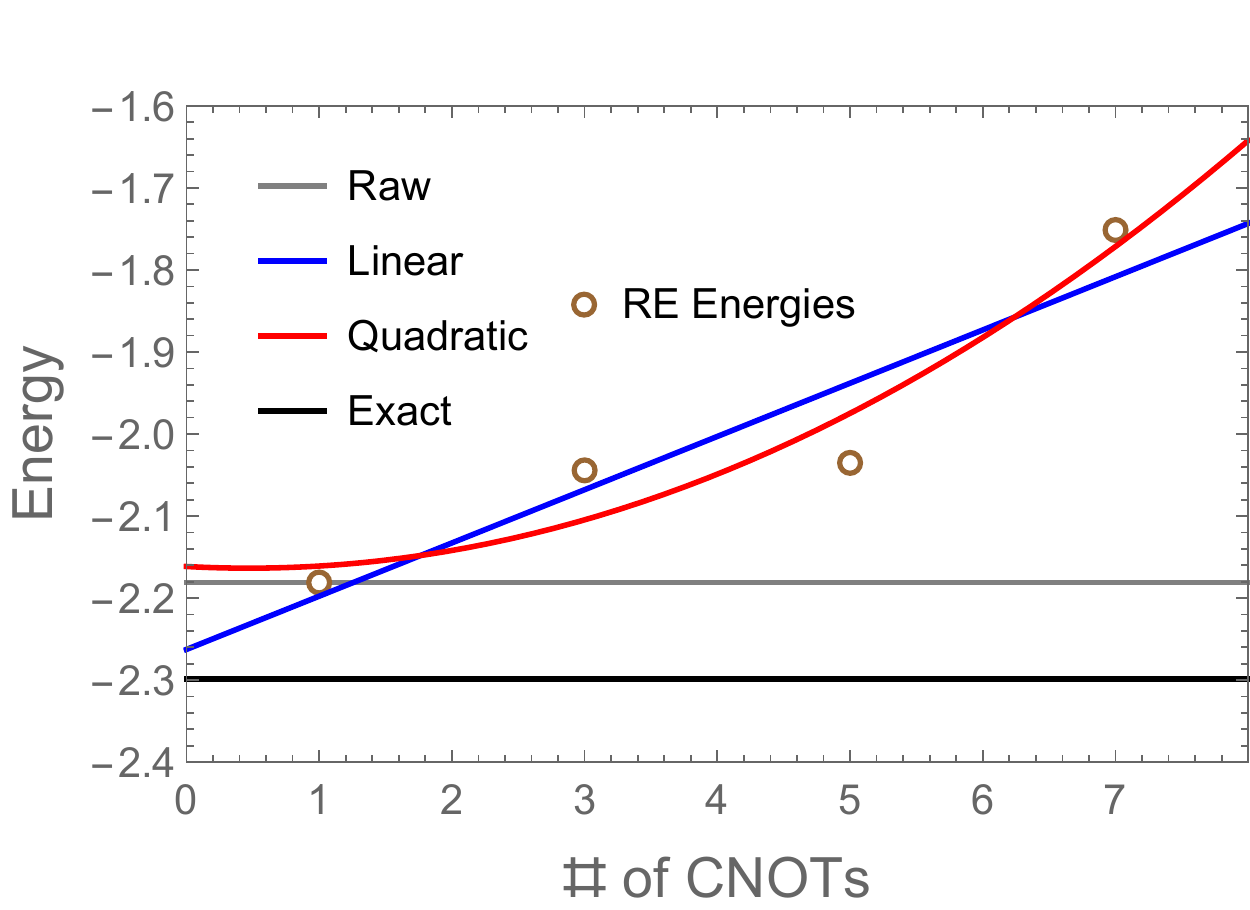}}
    \subfigure[\ L=4 First Excited State]{\includegraphics[scale=0.5]{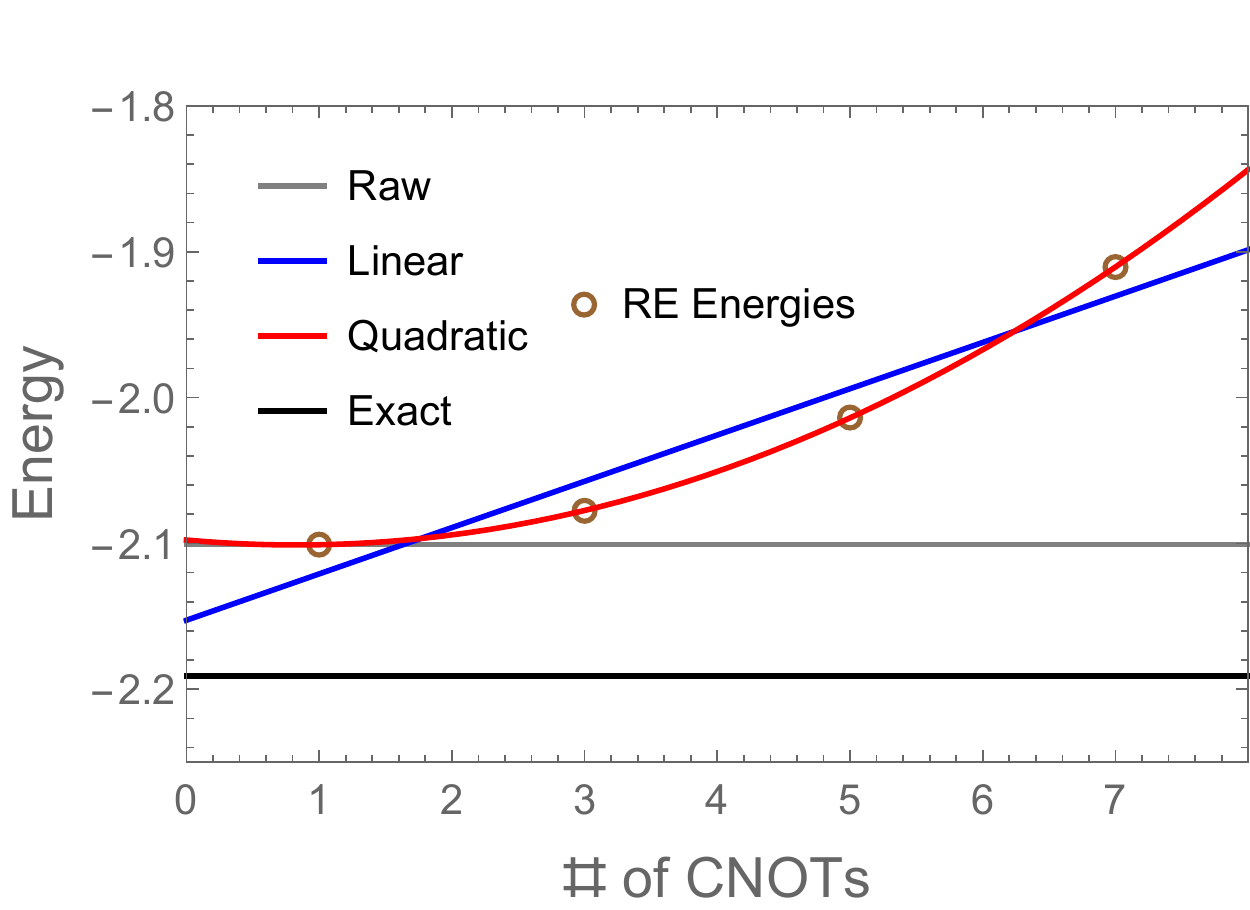}}
    \caption{Example of the Richardson extrapolation procedure, using both linear and quadratic extrapolation. Here we calculate the ground and first excited state energies for $L=3$ and $L=4$, with the parameters $e=0.5$ and $m=0.12$. For reference we added horizontal lines indicating the energy obtained by diagonalizing the Hamiltonian ``exactly" (i.e., using a very large cutoff for the gauge field) and that obtained from a raw quantum calculation (without extrapolation). Results are obtained from IBM Q Lima.}
    \label{fig:extrapolation_exact_energies}
\end{figure}

To deal with two-qubit gate errors we use Richardson extrapolation \cite{Li2017}. Here we add pairs of CNOT gates for each CNOT gate appearing in our original Ans\"atze, and do four such runs so that, for each original CNOT in the Ans\"atze, we have $1,3,5,\text{ and }7$ CNOTs in its place (as is done in \cite{Dumitresu2018,Kubra2020} as well). We then perform both a linear and quadratic extrapolation to the 0-CNOT gate case, thereby achieving two levels of zero-noise extrapolation. An example is shown in Figure\ \ref{fig:extrapolation_exact_energies} for the parameters $e=0.5$ and $m=0.12$.

\subsection{Results}
Due to long queue and run times on IBM's quantum hardware, it is not practical to locate the pseudo-critical points solely on hardware. Not only are there many points in the two-dimensional parameter space ($e\mbox{-}m$ plane) we need to sample, but with run times scaling with the number of shots, results with error bars small enough to distinguish each point in the space are inaccessible. Instead, we proceed by first locating the pseudo-critical points using noiseless simulations, and then checking that the results obtained on hardware for the scaled gap ratio, evaluated at these points, contain the value $R_4=1$ within error bars. 

We vary $e$ from $0.5$ to $1$ in steps of $0.1$. Then using noiseless simulations, for each of these values of $e$, we sample different values of $m$ in steps of $0.01$ and determine the value of $m$ that gives a gap ratio closest to $1$. This way only six separate points in the $e\mbox{-}m$ plane need be implemented on actual quantum hardware. Also, in order to reduce the number of optimization iterations performed on hardware, we feed the optimizer as an initial point the optimal VQE parameters obtained from noiseless simulations.

\begin{figure}[ht!]
    \centering
    \includegraphics[scale=0.65]{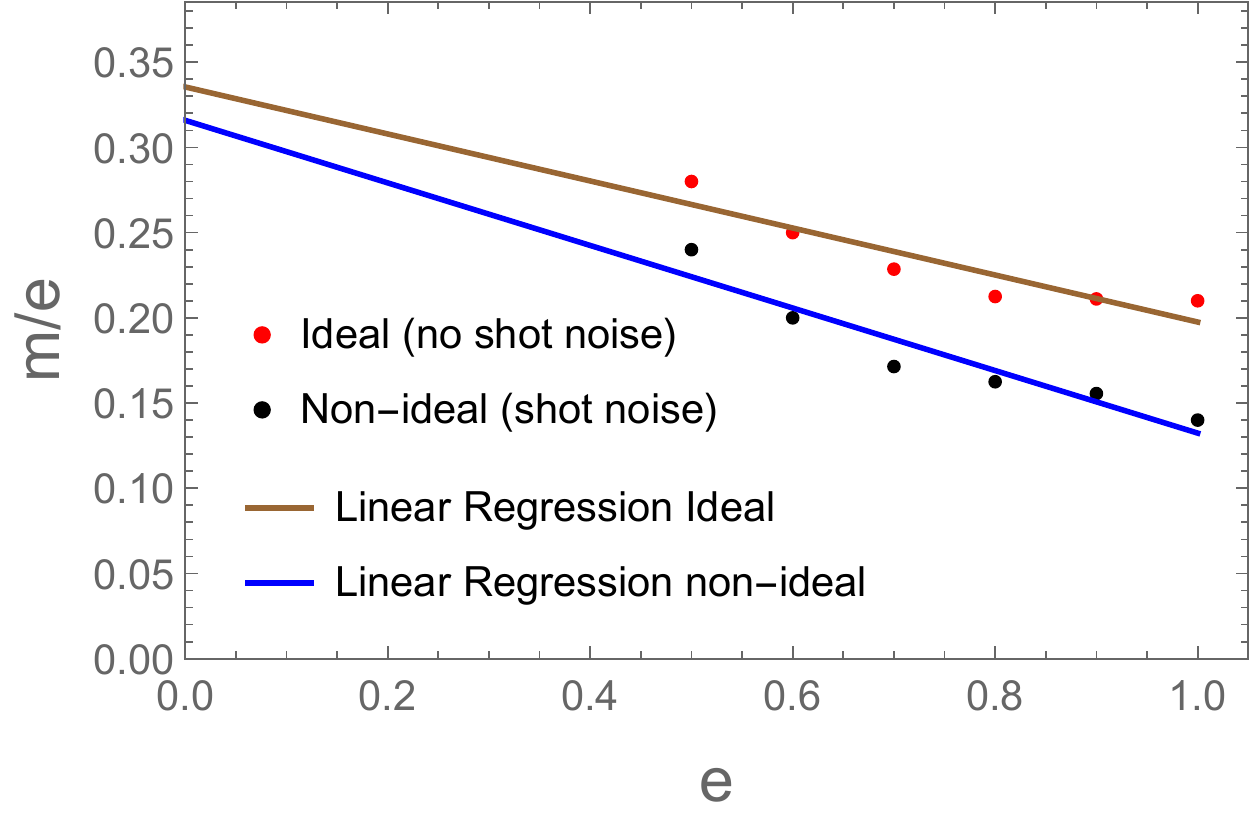}
    \caption{Pseudo-critical points vs.\ coupling $e$ for $R_4$, along with linear regressions to extrapolate to the continuum critical point, found using: $(1)$ ideal applications of the variational method, using the \textit{statevector simulator} from \textit{qiskit-aer}, and $(2)$ VQE with shot noise, using the \textit{qasm simulator} from \textit{qiskit-aer}. }
    \label{fig:statevector_vs_qasm}
\end{figure}
The scaled gap ratio is very sensitive to variations in the calculated energies. This means that shot noise can have a significant effect on results. To demonstrate this, we plot the pseudo-critical points determined from simulations both with and without shot noise in Figure\ \ref{fig:statevector_vs_qasm}. The pseudo-critical points seem to differ significantly but we still end up with final critical points which are relatively close to the true value: $m/e=0.316$ and $m/e=0.335$, with and without shot noise, respectively. To best mimic usage of an actual device we use the pseudo-critical points with shot noise for our hardware runs.



\begin{figure}[ht!]
    \centering
    \subfigure[\ $L=3$ ground and first excited state energies]{\includegraphics[scale=0.65]{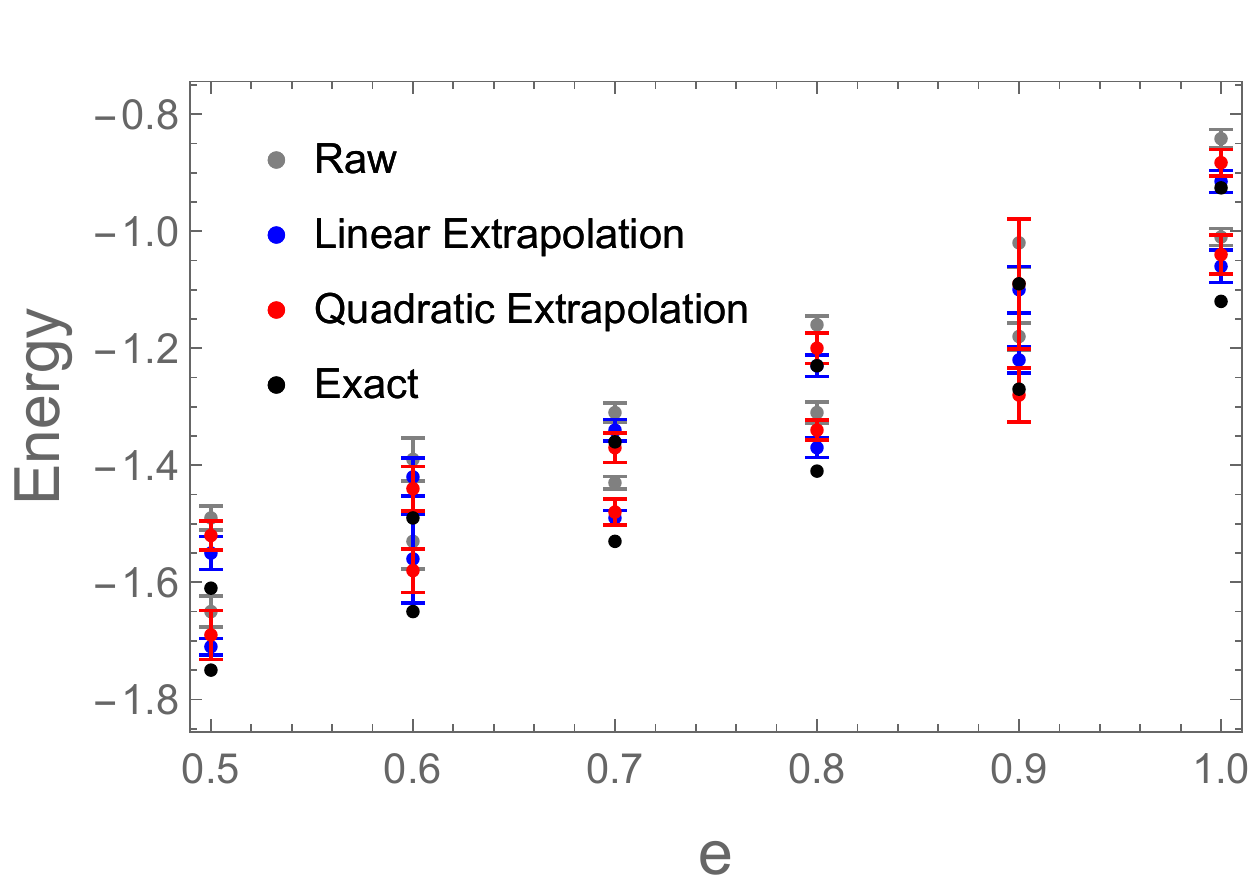}}
    \subfigure[\ $L=4$ ground and first excited state energies]{\includegraphics[scale=0.65]{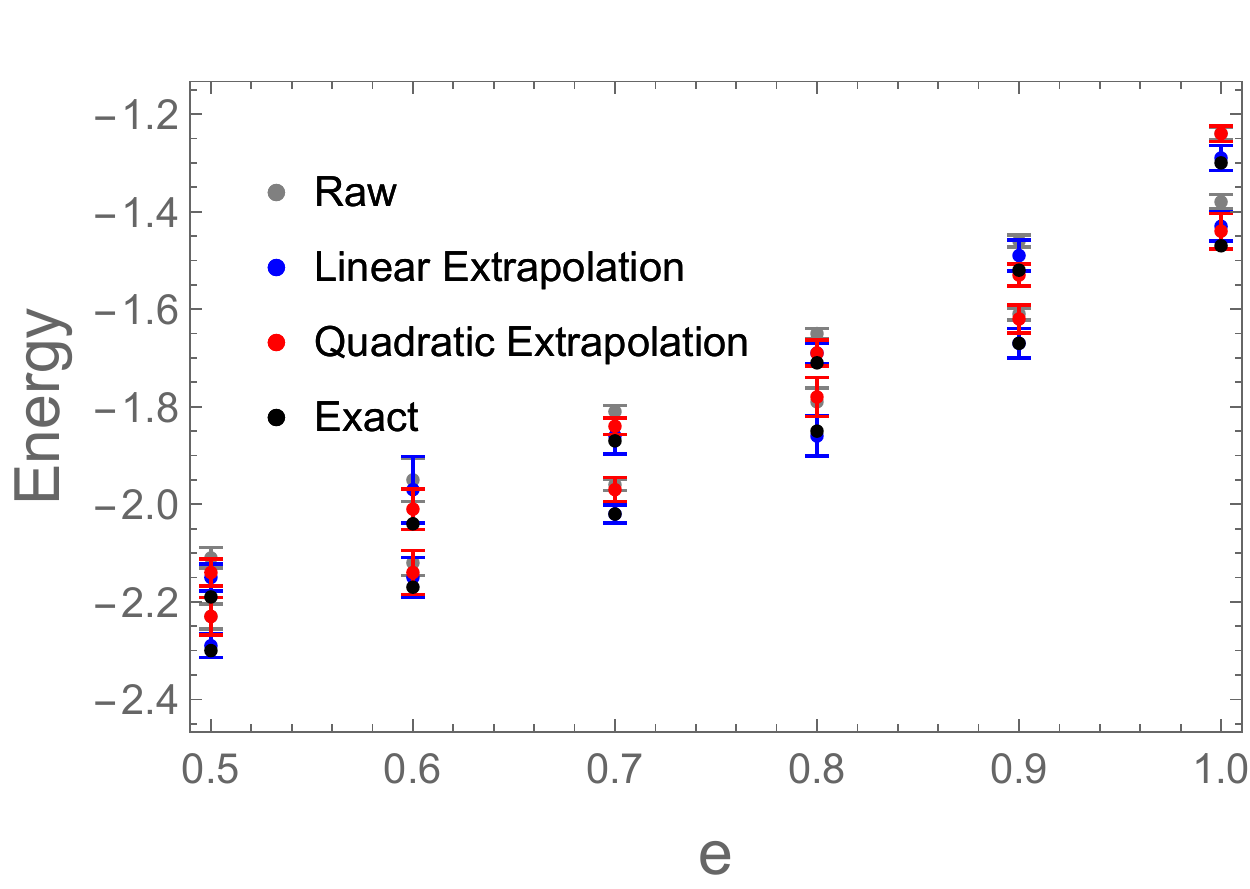}}
    \caption{Hardware results for the $L=3$ and $L=4$ ground and first excited state energies, vs.\ coupling $e$. For each $e$, $m$ is chosen so that the points $(e,m)$ in the parameter space are pseudo-critical points as identified by the black points in Figure \ref{fig:statevector_vs_qasm}. The means and error bars describe the data from 10 runs. Results without Richardson extrapolation and with linear and quadratic extrapolation are compared with the eigenenergies found by diagonalizing the Hamiltonian ``exactly" (exact in the sense that we take a very large cutoff for the gauge field). Results are obtained from IBM Q Lima.}
    \label{fig:energy_results_lima_concat_exact_energies}
\end{figure}
\begin{figure}[H]
    \centering
    \subfigure[]{\includegraphics[scale=0.45]{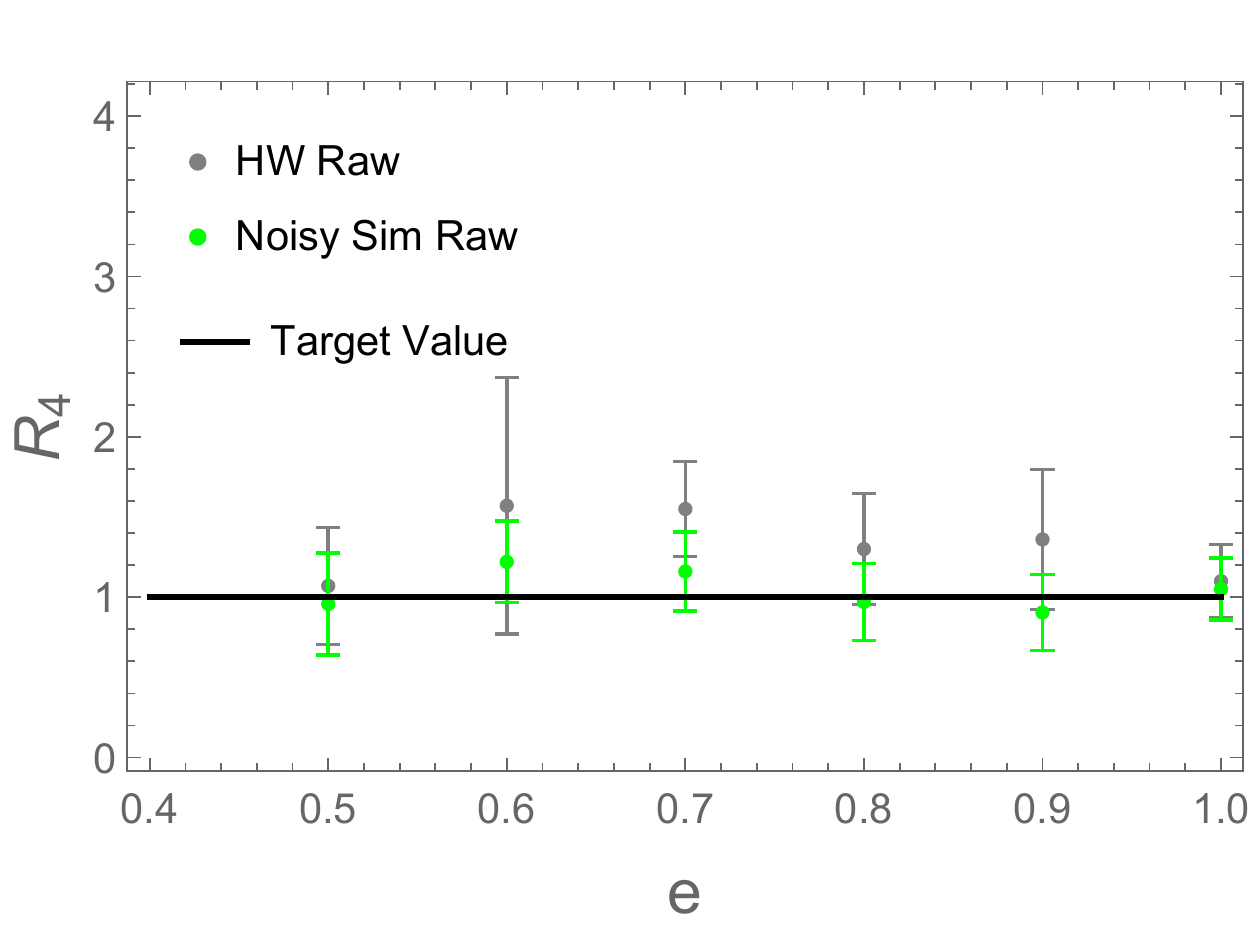}}
    \subfigure[]{\includegraphics[scale=0.45]{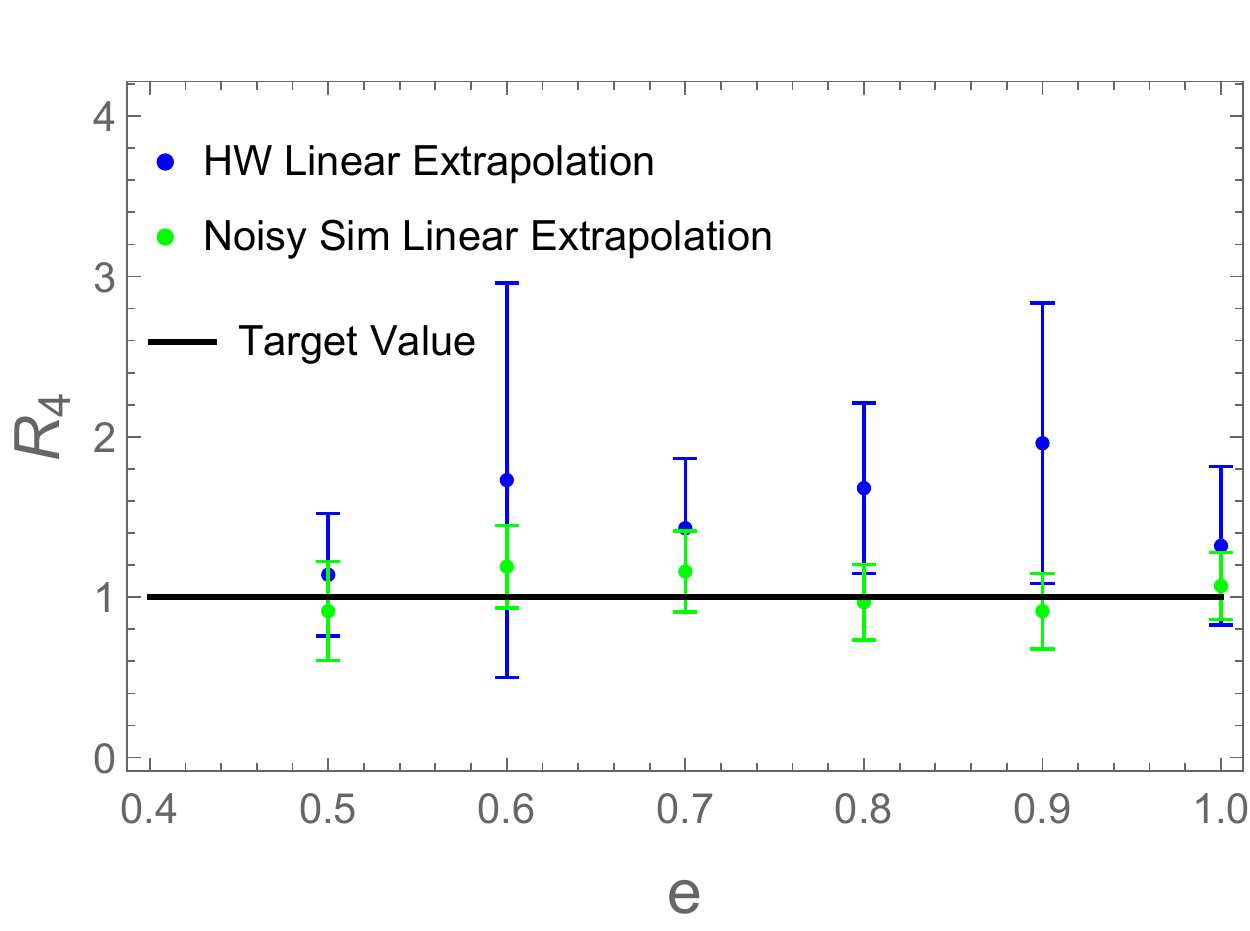}}
    \subfigure[]{\includegraphics[scale=0.45]{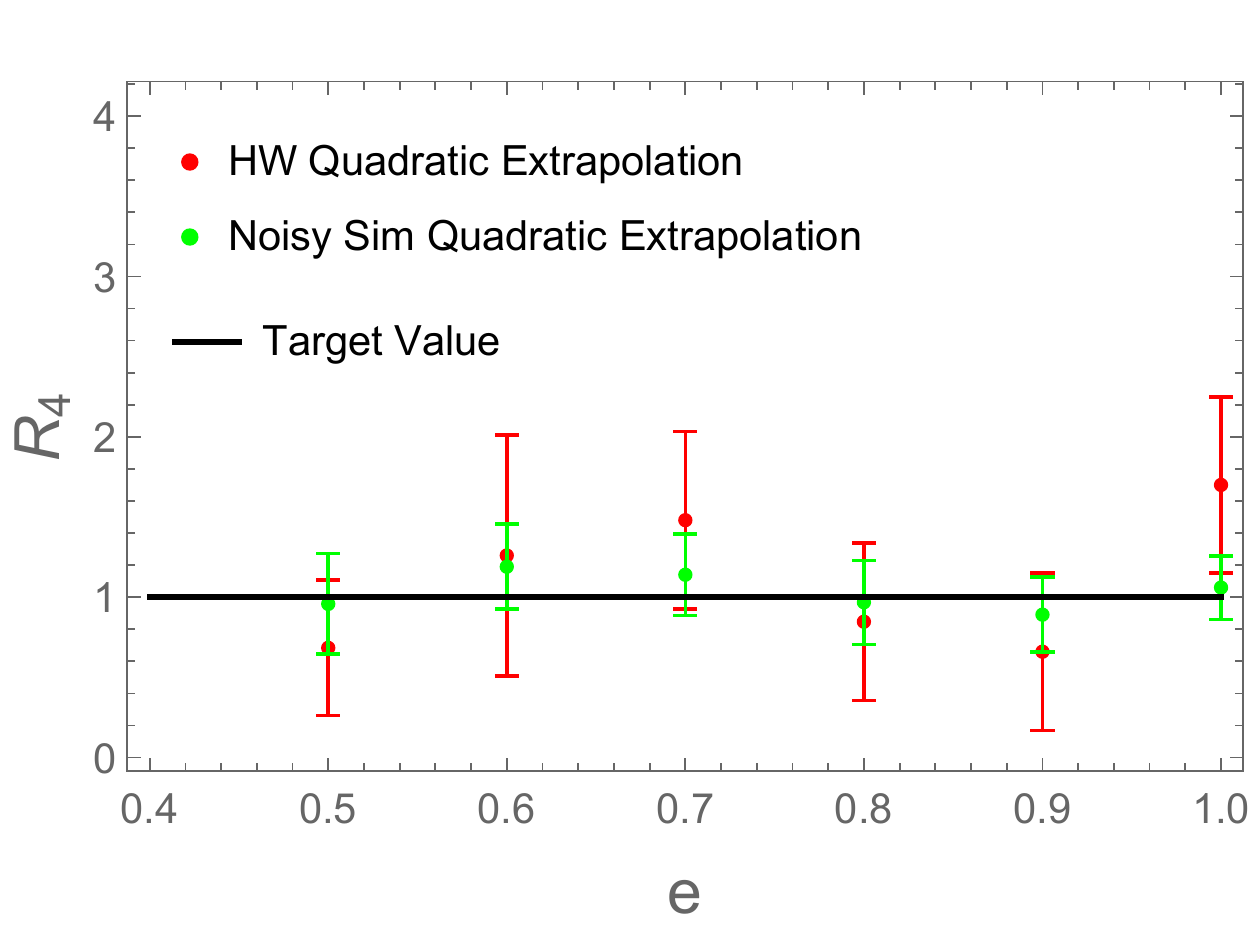}}
    \caption{Hardware (HW) results for the scaled energy-gap ratio $R_4$ with (a) no Richardson extrapolation, (b) linear extrapolation, and (c) quadratic extrapolation. The energies used to compute the ratio are those from Figure \ref{fig:energy_results_lima_concat_exact_energies}. The parameter values $(e,m)$ we evaluate this ratio at should correspond to pseudo-critical points, and so we expect $R_4\approx1$. Also given are results from noisy simulations, with ten runs for each point.}
    \label{fig:gap_ratio_results_lima_concat}
\end{figure}

Next, we present our final results with and without Richardson extrapolation. The device used was IBM Q Lima. To obtain these results we performed two groups of five runs on hardware. While the two groups were run at separate times, the jobs within each group (for a given circuit and point) were submitted in parallel. In our first set of results, given by Figures\ \ref{fig:energy_results_lima_concat_exact_energies} and \ref{fig:gap_ratio_results_lima_concat}, the means and statistical error bars correspond to the full ten runs.
In Figure \ref{fig:energy_results_lima_concat_exact_energies} we plot the ground and first excited state energies for $L=3,4$. It can be seen that accuracy improves in general when extrapolation is performed. In Figure \ref{fig:gap_ratio_results_lima_concat} we plot the ratio $R_4$ obtained from these energies. We see that the ratio is very sensitive to errors and variations in the energies. It can also be seen that extrapolation is a somewhat unstable technique, sometimes producing significantly larger error bars for certain points when compared with no extrapolation. Also given are results from noisy simulations, with ten runs for each point. They overlap with hardware results as expected.

If we are to accept each point in Figure\ \ref{fig:gap_ratio_results_lima_concat} as being close enough to $1$, then we can confirm the pseudo-critical points depicted in Figure \ref{fig:statevector_vs_qasm} (black points) as the true pseudo-critical points, which after extrapolation lead us to a continuum critical value of $m/e\simeq 0.32$.

\begin{figure}[ht!]
    \centering
    \subfigure[\ $L=3$ ground and first excited state energies]{\includegraphics[scale=0.65]{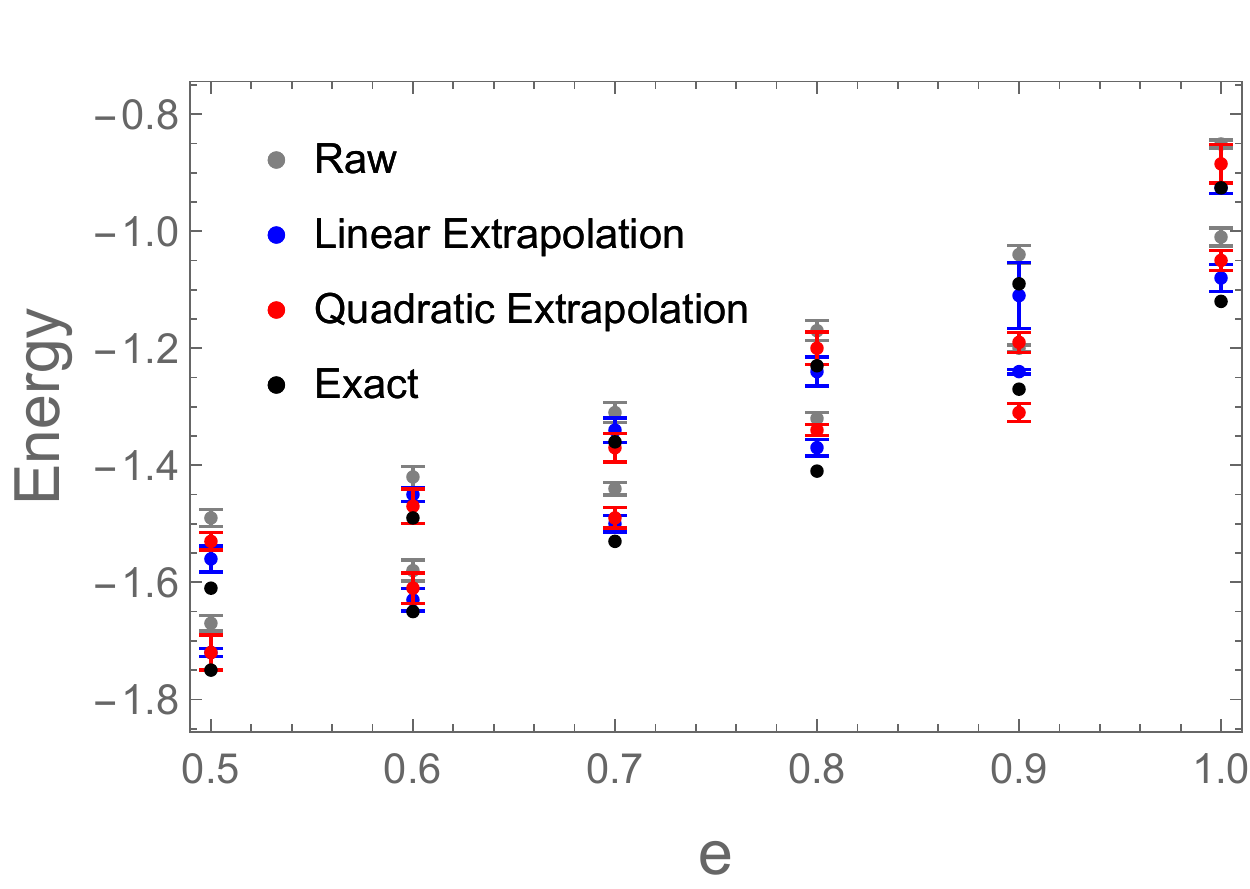}}
    \subfigure[\ $L=4$ ground and first excited state energies]{\includegraphics[scale=0.65]{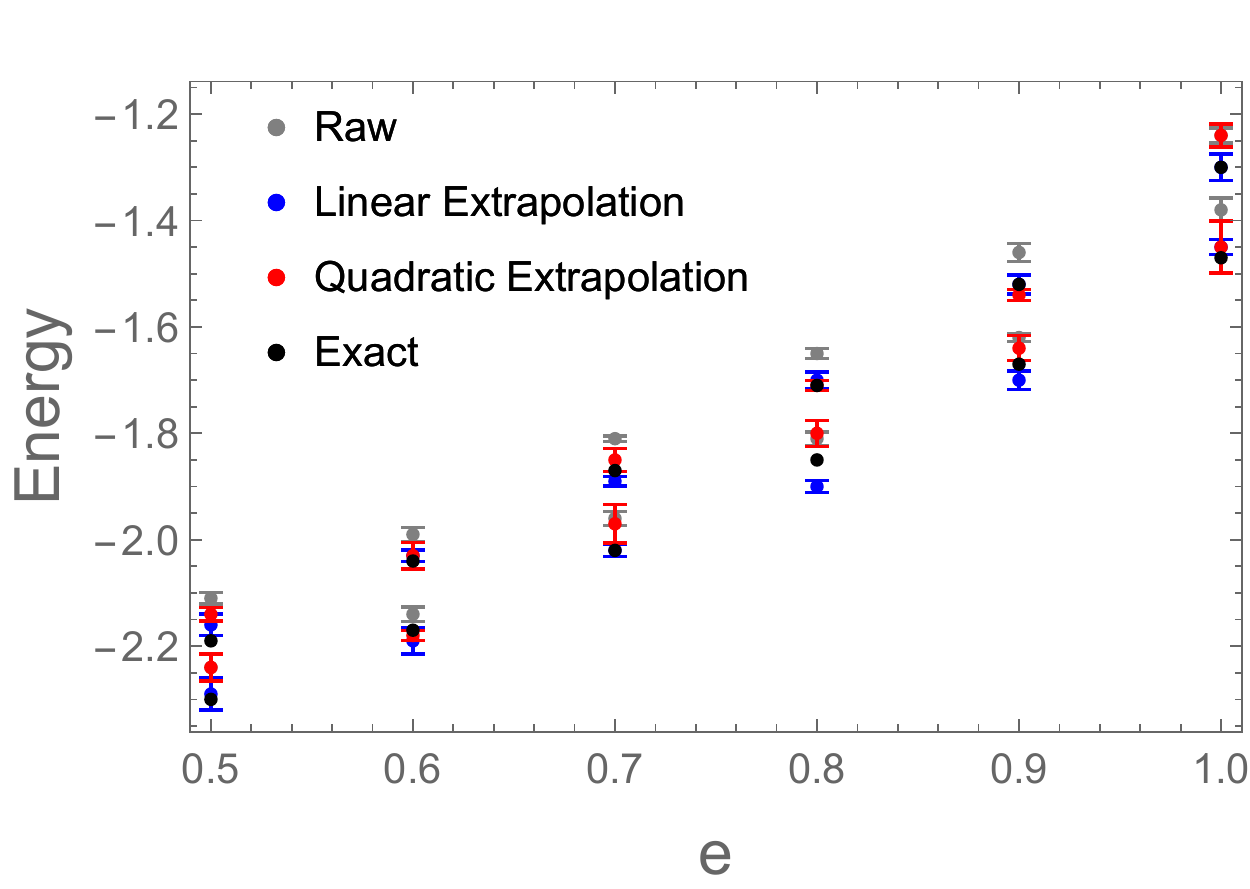}}
    \caption{Hardware results for the $L=3$ and $L=4$ ground and first excited state energies, vs coupling $e$. For each $e$, $m$ is chosen so that the points $(e,m)$ in the parameter space are pseudo-critical points as identified by the black points in Figure \ref{fig:statevector_vs_qasm}. Here the energies are found by obtaining results from two groups of five runs each, and then finding which group gives the lowest mean energy, in accordance with the variational method. Statistical error bars correspond only to the group whose mean is displayed, and are not the error bars for all 10 runs. Results without Richardson extrapolation and with linear and quadratic extrapolation are compared with the eigenenergies found by diagonalizing the full Hamiltonian ``exactly" (exact in the sense that we take a very large cutoff for the gauge field). Results are obtained from IBM Q Lima.}
    \label{fig:energy_results_lima_lowest_energies_exact_energies}
\end{figure}

\begin{figure}[ht]
    \centering
    \subfigure[]{\includegraphics[scale=0.45]{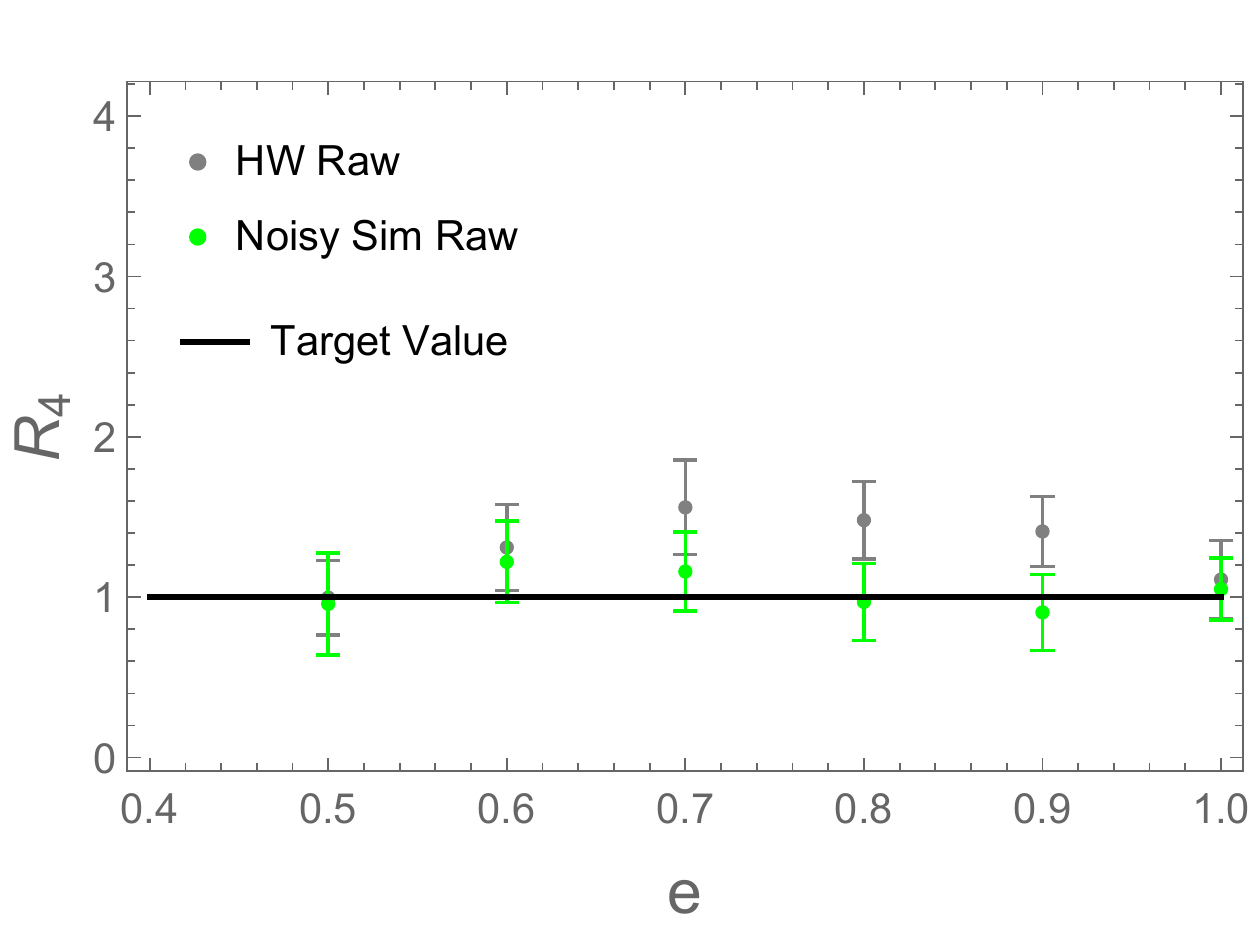}}
    \subfigure[]{\includegraphics[scale=0.45]{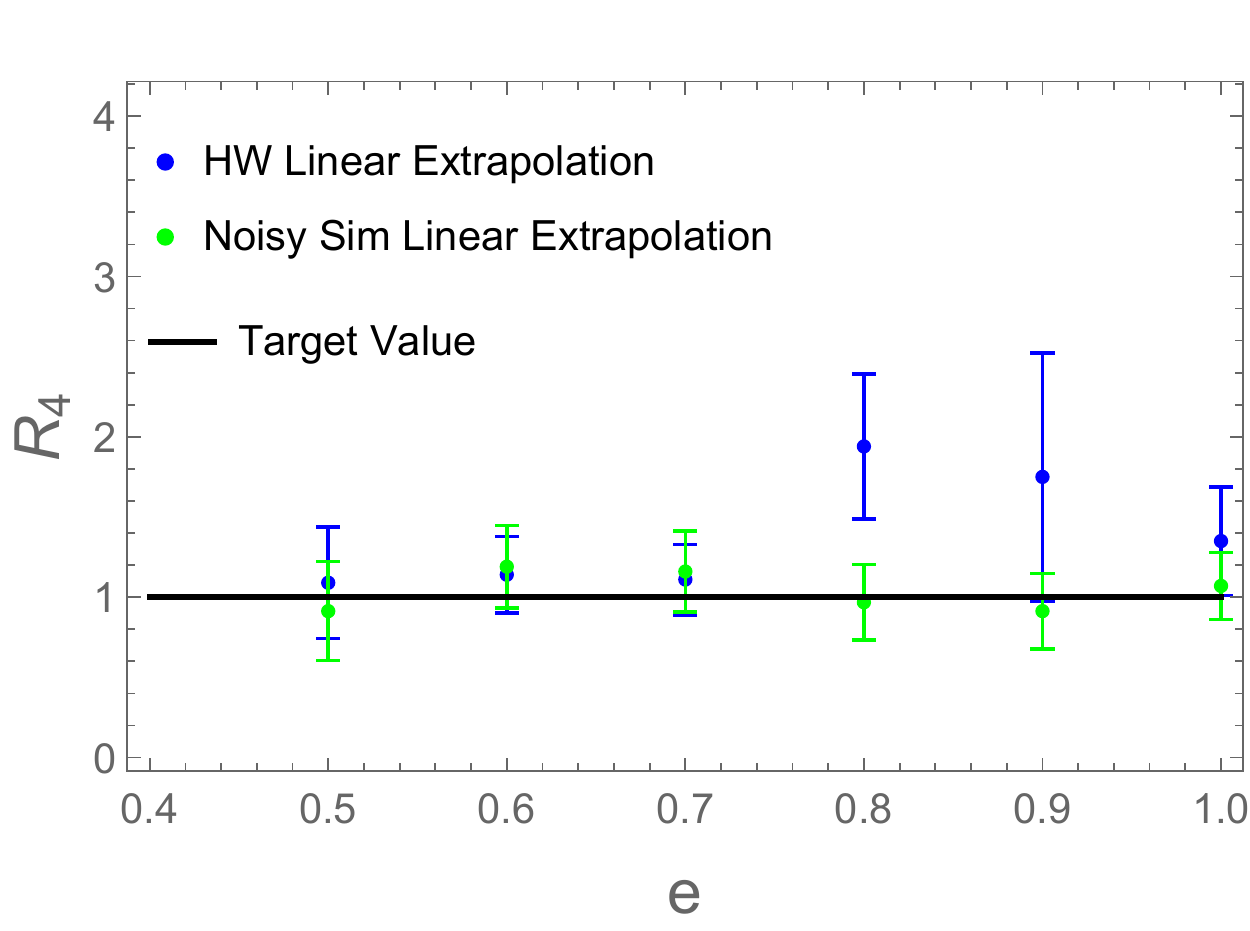}}
    \subfigure[]{\includegraphics[scale=0.45]{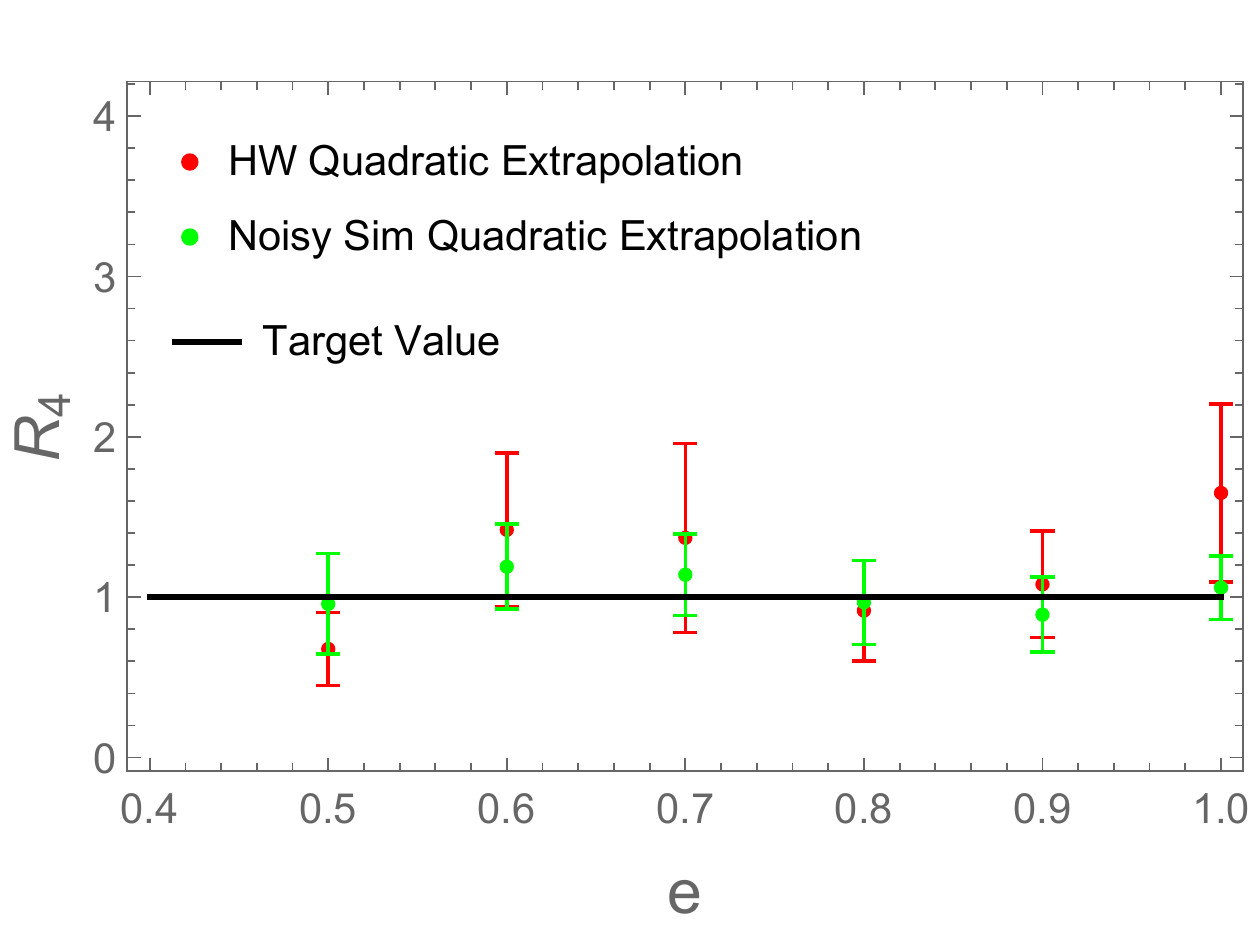}}
    \caption{Hardware (HW) results for $R_4$ with (a) no Richardson extrapolation, (b) linear extrapolation, and (c) quadratic extrapolation. This time the energies used to compute the ratio are those from Figure \ref{fig:energy_results_lima_lowest_energies_exact_energies}. The parameter values $(e,m)$ we evaluate this ratio at should correspond to pseudo-critical points, and so we expect $R_4\approx1$. We still use 10 runs for our noisy simulations.}
    \label{fig:gap_ratio_results_lima_lowest_energies}
\end{figure}

Another option is to instead pick out the lowest mean energies from the two groups of five runs each. All we are doing here is recalling that we are employing the variational method, and so the lowest energies should be the most accurate. This method is displayed in Figures\ \ref{fig:energy_results_lima_lowest_energies_exact_energies} and  \ref{fig:gap_ratio_results_lima_lowest_energies}.
As before, accepting the points in Figure\ \ref{fig:gap_ratio_results_lima_lowest_energies} to be close enough to $1$ implies a continuum critical point of $m/e\simeq 0.32$. This approach proves effective as well because even though the results from a single group may be consistent from a machine error perspective (since the jobs are submitted in parallel), two separate groups can give very different results. For example the $L=3$ ground state energy for $e=0.6$ (without extrapolation) was found to be $-1.58\pm0.18$ in the first group of five runs and $-1.49\pm0.12$ in the second group. This indicates how quickly the machine error can change. 

While it is clear that several of the mean energies did not improve in accuracy (after all, we only have two sets of runs), the error bars did shorten for all points since we are only choosing the ``best" set of results. Notice however that extrapolation can occasionally give us energies below the exact values (see Figure\ \ref{fig:energy_results_lima_lowest_energies_exact_energies}), and so the lowest energy is not always the ``best." This is because this should not occur for noiseless simulations, which we use to locate the pseudo-critical points for our quantum computation.


\subsection{Discussion}
One immediate observation from Figures\ \ref{fig:gap_ratio_results_lima_concat} and \ref{fig:gap_ratio_results_lima_lowest_energies} is that there does not seem to be a major improvement in the calculation of $R_4$ when extrapolation is performed. This can be attributed to the fact that we are calculating the difference between the ground and first excited state energies, whose corresponding quantum circuits are very similar for $L=4$ and identical (even up to qubit layout on hardware) for $L=3$ . This would suggest that in calculating the gap, most of the machine error from each circuit should cancel. Once extrapolation is introduced, the unstable effects of adding additional CNOTs (as evidenced by the larger error bars) can disrupt this cancellation, even if the energies themselves are generally more accurate post-extrapolation. 

\begin{figure}[ht]
    \centering
    \subfigure[\ Figure\ \ref{fig:energy_results_lima_concat_exact_energies} energies ]{\includegraphics[scale=0.65]{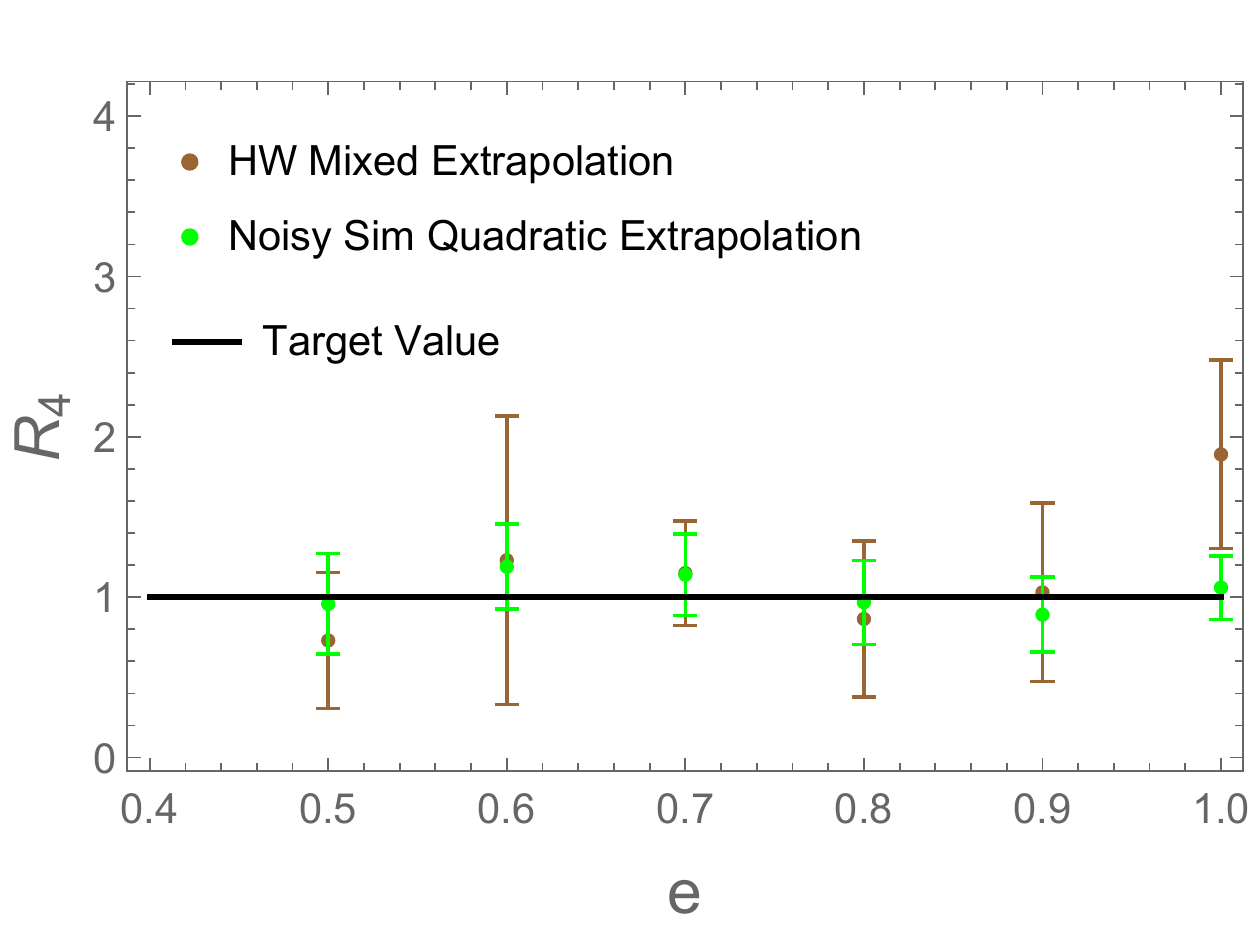}}
    \subfigure[\ Figure\ \ref{fig:energy_results_lima_lowest_energies_exact_energies} energies]{\includegraphics[scale=0.65]{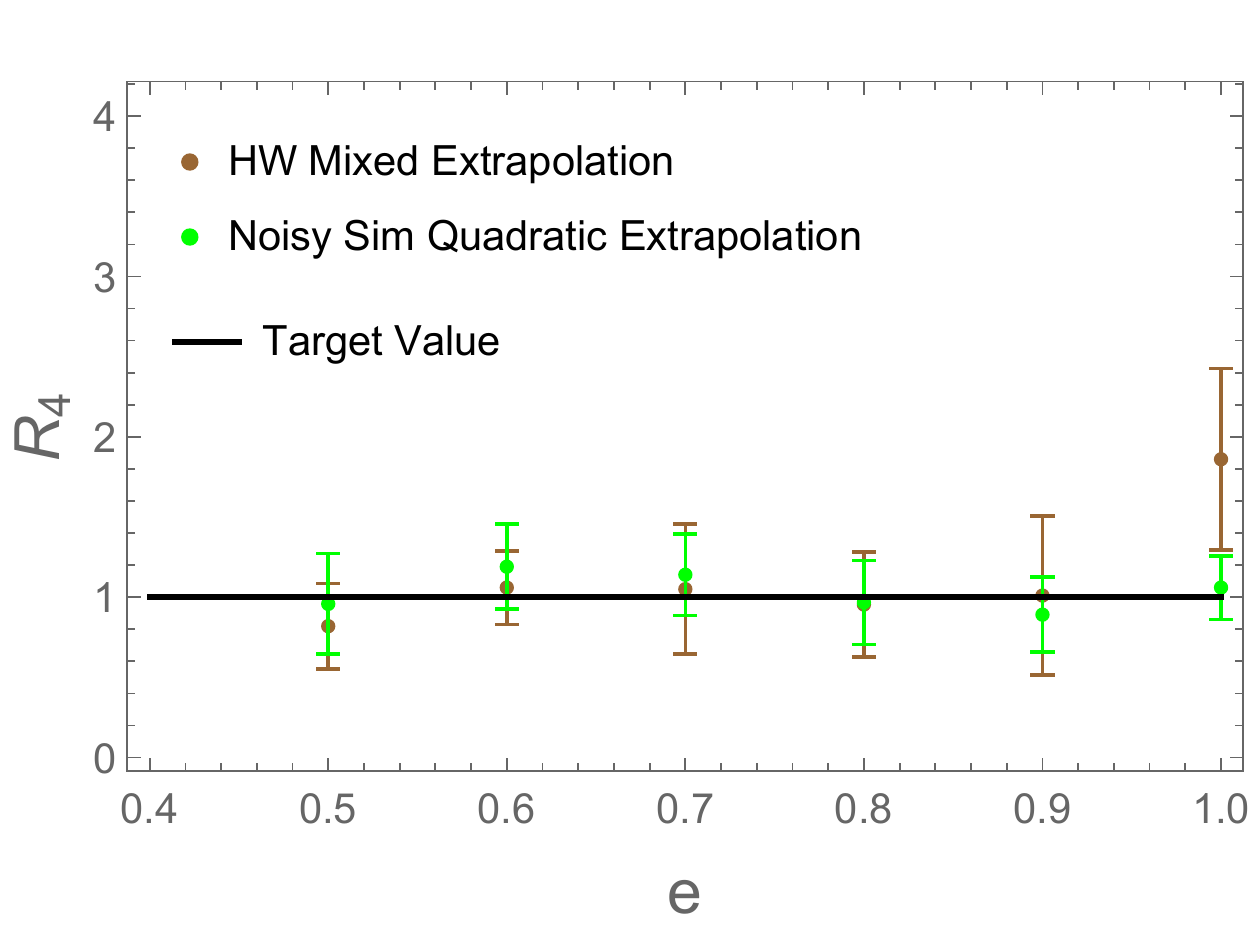}}
    \caption{Hardware (HW) results for $R_4$ with linear extrapolation employed for $L=3$ and quadratic for $L=4$. The energies used for (a) come from Figure\ \ref{fig:energy_results_lima_concat_exact_energies}, while the energies used for (b) come from Figure\ \ref{fig:energy_results_lima_lowest_energies_exact_energies}. The results show improvement over Figures\ \ref{fig:gap_ratio_results_lima_concat} and \ref{fig:gap_ratio_results_lima_lowest_energies}, except for the point $e=1$.}
    \label{fig:gap_ratio_results_mixed_lima_concat}
\end{figure}

Upon further inspection of these figures, linear extrapolation appears to be plagued by a systematic error which resulted in an overestimate of $R_4$, despite it generally performing better in terms of getting more accurate energies (see Figures\ \ref{fig:extrapolation_exact_energies}, \ref{fig:energy_results_lima_concat_exact_energies}, and \ref{fig:energy_results_lima_lowest_energies_exact_energies}). On the other hand, quadratic extrapolation appears more consistent with results from noisy simulations. Thus one could instead attempt a mixed extrapolation approach as shown in Figure\ \ref{fig:gap_ratio_results_mixed_lima_concat}, where we use linear extrapolation for $L=3$ and quadratic for $L=4$.



With this we see an improvement over Figures\ \ref{fig:gap_ratio_results_lima_concat} and \ref{fig:gap_ratio_results_lima_lowest_energies}, except for the point $e=1$. This may be seen as a mitigation of the systematically enlarged gaps corresponding to $L=4$ for linear extrapolation (see Figures\ \ref{fig:energy_results_lima_concat_exact_energies} (b) and \ref{fig:energy_results_lima_lowest_energies_exact_energies} (b)).

We conclude this section by noting that it is possible to reduce the size of the error bars by increasing the number of shots. However the above results use the maximum offered on hardware, $8192$. To get around this, one may use circuit bundling in which the same circuits are run several times within each job so that the counts (probabilities) for each computational basis state correspond more accurately to actual probability amplitudes in the quantum states. Unfortunately, in order to see a significant decrease in the size of the error bars, so many circuits would need to be ran that the device run time would increase beyond a practical duration (this has been tested with five runs per job; while there was a small improvement in the size of error bars, the increase in run time was far more significant).

\section{Conclusion}\label{sec:Conclusion}

In summary, we have developed a momentum space formalism for the massive Schwinger model which, unlike other lattice approaches, leaves us with one gauge degree of freedom. This was done in the spirit of \cite{bib1}. This gauge degree of freedom provides us with an infinite dimensional Hilbert space that we truncate for our quantum computation. 
By using plane waves for the gauge field wavefunctions and invoking symmetry considerations, we identified a basis in the truncated Hilbert space that minimized the number of qubits needed for a quantum computation.

We analyzed the phase transition that occurs when $\theta=\pi$. We found via exact diagonalization (of matrices written in our carefully selected basis) that we did not need large lattice sizes in order to obtain a good approximation of the critical point for the ratio $m/e$. This allowed us to calculate the critical point on a quantum computer (IBM Q) using only three qubits and low-depth circuits. We identified simple Ans\"atze for the Variational Quantum Eigensolver (VQE) algorithm with which we could get good approximations to the ground and first excited state energies for lattices with $L=3$ and $L=4$ spatial sites. We utilized readout error mitigation and Richardson extrapolation in our quantum computation, and compared the effects of linear vs.\ quadratic extrapolation. We found that the calculation of the critical point was highly sensitive to the effects of shot and machine noise, leading to large error bars. We saw that if we separated our results from ten runs into two sets of five runs each and picked the lowest energies from each set, we could reduce the size of the error bars. The results from our quantum computation on IBM Q were in good agreement with results obtained by classical means.

It would be interesting to apply our calculation to an ion trap quantum device (IBM's devices utilize superconducting qubits), with which we would have access to better connectivity between qubits. This would allow us, for example, to obtain the ground and first excited state energies via the Quantum Imaginary Time Evolution (QITE) algorithm \cite{Motta_2019}. This would lead to Ans\"atze that better approximate the desired states. It would also allow us to more freely increase the number of qubits, providing the opportunity to obtain accurate results for larger lattices ($L>4$).

It would also be interesting to better understand the behavior of the model for $e\lesssim 1$ and $m\lesssim 1$, which is a region relevant to the continuum regime. If we can further develop our formalism in this regime, going beyond the use of a plane-wave basis for the gauge zero mode, we would be in a position to improve the accuracy of the quantum computation of pseudo-critical points.

\acknowledgments

Research supported by DOE ASCR funding under the Quantum Computing Application Teams Program, FWP No. ERKJ347, ARO grant W911-NF-19-1-0397, and NSF grant OMA-1937008.

\bibliographystyle{unsrt}
\bibliography{main}

\begin{thebibliography}{10}

\bibitem{Byrnes2006}
Tim Byrnes and Yoshihisa Yamamoto.
\newblock
  \href{https://journals.aps.org/pra/abstract/10.1103/PhysRevA.73.022328}{Simulating
  lattice gauge theories on a quantum computer}.
\newblock {\em Phys. Rev. A}, 73:022328, Feb 2006.

\bibitem{Banuls_2020}
Mari~Carmen Ba{\~{n}}uls and Krzysztof Cichy.
\newblock
  \href{https://iopscience.iop.org/article/10.1088/1361-6633/ab6311}{Review on
  novel methods for lattice gauge theories}.
\newblock {\em Reports on Progress in Physics}, 83(2):024401, Jan 2020.

\bibitem{Shaw2020}
Alexander~F. Shaw, Pavel Lougovski, Jesse~R. Stryker, and Nathan Wiebe.
\newblock \href{https://quantum-journal.org/papers/q-2020-08-10-306/}{Quantum
  {A}lgorithms for {S}imulating the {L}attice {S}chwinger {M}odel}.
\newblock {\em {Quantum}}, 4:306, Aug 2020.

\bibitem{Davoudi2020}
Zohreh Davoudi, Mohammad Hafezi, Christopher Monroe, Guido Pagano, Alireza
  Seif, and Andrew Shaw.
\newblock
  \href{https://journals.aps.org/prresearch/abstract/10.1103/PhysRevResearch.2.023015}{Towards
  analog quantum simulations of lattice gauge theories with trapped ions}.
\newblock {\em Phys. Rev. Research}, 2:023015, Apr 2020.

\bibitem{Martinez_2016}
Esteban~A. Martinez, Christine~A. Muschik, Philipp Schindler, Daniel Nigg,
  Alexander Erhard, Markus Heyl, Philipp Hauke, Marcello Dalmonte, Thomas Monz,
  Peter Zoller, and et~al.
\newblock \href{https://www.nature.com/articles/nature18318}{Real-time dynamics
  of lattice gauge theories with a few-qubit quantum computer}.
\newblock {\em Nature}, 534(7608):516–519, Jun 2016.

\bibitem{Muschik_2017}
Christine Muschik, Markus Heyl, Esteban Martinez, Thomas Monz, Philipp
  Schindler, Berit Vogell, Marcello Dalmonte, Philipp Hauke, Rainer Blatt, and
  Peter Zoller.
\newblock
  \href{https://iopscience.iop.org/article/10.1088/1367-2630/aa89ab}{U(1)
  Wilson lattice gauge theories in digital quantum simulators}.
\newblock {\em New Journal of Physics}, 19(10):103020, Oct 2017.

\bibitem{ORNL}
N.~Klco, E.~F. Dumitrescu, A.~J. McCaskey, T.~D. Morris, R.~C. Pooser, M.~Sanz,
  E.~Solano, P.~Lougovski, and M.~J. Savage.
\newblock
  \href{https://journals.aps.org/pra/abstract/10.1103/PhysRevA.98.032331}{Quantum-classical
  computation of Schwinger model dynamics using quantum computers}.
\newblock {\em Physical Review A}, 98(3), Sep 2018.

\bibitem{Klco2020}
Natalie Klco, Martin~J. Savage, and Jesse~R. Stryker.
\newblock
  \href{https://journals.aps.org/prd/abstract/10.1103/PhysRevD.101.074512}{SU(2)
  non-Abelian gauge field theory in one dimension on digital quantum
  computers}.
\newblock {\em Phys. Rev. D}, 101:074512, Apr 2020.

\bibitem{Yang_2020}
Bing Yang, Hui Sun, Robert Ott, Han-Yi Wang, Torsten~V. Zache, Jad~C. Halimeh,
  Zhen-Sheng Yuan, Philipp Hauke, and Jian-Wei Pan.
\newblock \href{https://www.nature.com/articles/s41586-020-2910-8}{Observation
  of gauge invariance in a 71-site $\text{Bose}\mbox{-}\text{Hubbard}$ quantum
  simulator}.
\newblock {\em Nature}, 587(7834):392–396, Nov 2020.

\bibitem{Schwinger1962}
Julian Schwinger.
\newblock
  \href{https://journals.aps.org/pr/abstract/10.1103/PhysRev.128.2425}{Gauge
  Invariance and Mass. II}.
\newblock {\em Phys. Rev.}, 128:2425--2429, Dec 1962.

\bibitem{COLEMAN1975267}
Sidney Coleman, R~Jackiw, and Leonard Susskind.
\newblock
  \href{https://www.sciencedirect.com/science/article/pii/0003491675902122}{Charge
  shielding and quark confinement in the massive schwinger model}.
\newblock {\em Annals of Physics}, 93(1):267--275, 1975.

\bibitem{COLEMAN1976}
Sidney Coleman.
\newblock
  \href{https://www.sciencedirect.com/science/article/pii/0003491676902803}{More
  about the massive Schwinger model}.
\newblock {\em Annals of Physics}, 101(1):239 -- 267, 1976.

\bibitem{Kogut1975}
John Kogut and Leonard Susskind.
\newblock
  \href{https://journals.aps.org/prd/abstract/10.1103/PhysRevD.11.395}{Hamiltonian
  formulation of Wilson's lattice gauge theories}.
\newblock {\em Phys. Rev. D}, 11:395--408, Jan 1975.

\bibitem{Indrakshi2020}
Indrakshi Raychowdhury and Jesse~R. Stryker.
\newblock
  \href{https://journals.aps.org/prd/abstract/10.1103/PhysRevD.101.114502}{Loop,
  string, and hadron dynamics in SU(2) Hamiltonian lattice gauge theories}.
\newblock {\em Phys. Rev. D}, 101:114502, Jun 2020.

\bibitem{Hamer1997}
C.~J. Hamer, Zheng Weihong, and J.~Oitmaa.
\newblock
  \href{https://journals.aps.org/prd/abstract/10.1103/PhysRevD.56.55}{Series
  expansions for the massive Schwinger model in Hamiltonian lattice theory}.
\newblock {\em Phys. Rev. D}, 56:55--67, Jul 1997.

\bibitem{bib1}
Satoshi Iso and Hitoshi Murayama.
\newblock
  \href{https://academic.oup.com/ptp/article/84/1/142/1893520}{Hamiltonian
  Formulation of the Schwinger Model: Non-Confinement and Screening of the
  Charge}.
\newblock {\em Progress of Theoretical Physics}, 84(1):142--163, Jul 1990.

\bibitem{thooft1976}
G.~'t~Hooft.
\newblock
  \href{https://journals.aps.org/prl/abstract/10.1103/PhysRevLett.37.8}{Symmetry
  Breaking through Bell-Jackiw Anomalies}.
\newblock {\em Phys. Rev. Lett.}, 37:8--11, Jul 1976.

\bibitem{Graner2016}
B.~Graner, Y.~Chen, E.~G. Lindahl, and B.~R. Heckel.
\newblock
  \href{https://journals.aps.org/prl/abstract/10.1103/PhysRevLett.116.161601}{Reduced
  Limit on the Permanent Electric Dipole Moment of $^{199}\mathrm{Hg}$}.
\newblock {\em Phys. Rev. Lett.}, 116:161601, Apr 2016.

\bibitem{Shimizu2014}
Yuya Shimizu and Yoshinobu Kuramashi.
\newblock
  \href{https://journals.aps.org/prd/abstract/10.1103/PhysRevD.90.074503}{Critical
  behavior of the lattice Schwinger model with a topological term at
  $\ensuremath{\theta}=\ensuremath{\pi}$ using the Grassmann tensor
  renormalization group}.
\newblock {\em Phys. Rev. D}, 90:074503, Oct 2014.

\bibitem{buyens2017}
Boye Buyens, Simone Montangero, Jutho Haegeman, Frank Verstraete, and Karel
  Van~Acoleyen.
\newblock
  \href{https://journals.aps.org/prd/abstract/10.1103/PhysRevD.95.094509}{Finite-representation
  approximation of lattice gauge theories at the continuum limit with tensor
  networks}.
\newblock {\em Phys. Rev. D}, 95:094509, May 2017.

\bibitem{Azcoiti2018}
Vicente Azcoiti, Eduardo Follana, Eduardo Royo-Amondarain, Giuseppe Di~Carlo,
  and Alejandro Vaquero Avil\'es-Casco.
\newblock
  \href{https://journals.aps.org/prd/abstract/10.1103/PhysRevD.97.014507}{Massive
  Schwinger model at finite $\ensuremath{\theta}$}.
\newblock {\em Phys. Rev. D}, 97:014507, Jan 2018.

\bibitem{Byrnes2002}
T.~M.~R. Byrnes, P.~Sriganesh, R.~J. Bursill, and C.~J. Hamer.
\newblock
  \href{https://www.sciencedirect.com/science/article/pii/S0920563202014160}{Density
  matrix renormalization group approach to the massive Schwinger model}.
\newblock {\em Phys. Rev. D}, 66:013002, Jul 2002.

\bibitem{LOWENSTEIN1971172}
J.H Lowenstein and J.A Swieca.
\newblock
  \href{https://www.sciencedirect.com/science/article/pii/0003491671902466}{Quantum
  electrodynamics in two dimensions}.
\newblock {\em Annals of Physics}, 68(1):172--195, 1971.

\bibitem{MANTON1985220}
N.S Manton.
\newblock
  \href{https://www.sciencedirect.com/science/article/pii/000349168590199X}{The
  Schwinger model and its axial anomaly}.
\newblock {\em Annals of Physics}, 159(1):220--251, 1985.

\bibitem{Wilson1974}
Kenneth~G. Wilson.
\newblock
  \href{https://journals.aps.org/prd/abstract/10.1103/PhysRevD.10.2445}{Confinement
  of quarks}.
\newblock {\em Phys. Rev. D}, 10:2445--2459, Oct 1974.

\bibitem{Aharonov1959}
Y.~Aharonov and D.~Bohm.
\newblock
  \href{https://journals.aps.org/pr/abstract/10.1103/PhysRev.115.485}{Significance
  of Electromagnetic Potentials in the Quantum Theory}.
\newblock {\em Phys. Rev.}, 115:485--491, Aug 1959.

\bibitem{LatticeFermions}
Leonard Susskind.
\newblock
  \href{https://journals.aps.org/prd/abstract/10.1103/PhysRevD.16.3031}{Lattice
  fermions}.
\newblock {\em Phys. Rev. D}, 16:3031--3039, Nov 1977.

\bibitem{bib2}
T.~Banks, Leonard Susskind, and John Kogut.
\newblock
  \href{https://journals.aps.org/prd/abstract/10.1103/PhysRevD.13.1043}{Strong-coupling
  calculations of lattice gauge theories: (1 + 1)-dimensional exercises}.
\newblock {\em Phys. Rev. D}, 13:1043--1053, Feb 1976.

\bibitem{Berruto1998}
F.~Berruto, G.~Grignani, G.~W. Semenoff, and P.~Sodano.
\newblock
  \href{https://journals.aps.org/prd/abstract/10.1103/PhysRevD.57.5070}{Chiral
  symmetry breaking on the lattice: A study of the strongly coupled lattice
  Schwinger model}.
\newblock {\em Phys. Rev. D}, 57:5070--5083, Apr 1998.

\bibitem{CREUTZ1995}
Michael Creutz.
\newblock
  \href{https://www.sciencedirect.com/science/article/pii/092056329500187E}{Chiral
  symmetry on the lattice}.
\newblock {\em Nuclear Physics B - Proceedings Supplements}, 42(1):56--66,
  1995.

\bibitem{Zache2019}
T.~V. Zache, N.~Mueller, J.~T. Schneider, F.~Jendrzejewski, J.~Berges, and
  P.~Hauke.
\newblock
  \href{https://journals.aps.org/prl/abstract/10.1103/PhysRevLett.122.050403}{Dynamical
  Topological Transitions in the Massive Schwinger Model with a
  $\ensuremath{\theta}$ Term}.
\newblock {\em Phys. Rev. Lett.}, 122:050403, Feb 2019.

\bibitem{Funcke2020}
Lena Funcke, Karl Jansen, and Stefan K\"uhn.
\newblock
  \href{https://journals.aps.org/prd/abstract/10.1103/PhysRevD.101.054507}{Topological
  vacuum structure of the Schwinger model with matrix product states}.
\newblock {\em Phys. Rev. D}, 101:054507, Mar 2020.

\bibitem{Fisher1972}
Michael~E. Fisher and Michael~N. Barber.
\newblock
  \href{https://journals.aps.org/prl/abstract/10.1103/PhysRevLett.28.1516}{Scaling
  Theory for Finite-Size Effects in the Critical Region}.
\newblock {\em Phys. Rev. Lett.}, 28:1516--1519, Jun 1972.

\bibitem{Hamer_1980}
C~J Hamer and M~N Barber.
\newblock
  \href{https://www.sciencedirect.com/science/article/abs/pii/B9780444871091500212}{Finite-size
  scaling in Hamiltonian field theory}.
\newblock {\em Journal of Physics A: Mathematical and General},
  13(5):L169--L174, May 1980.

\bibitem{HAMER1982}
C.J. Hamer, J.~Kogut, D.P. Crewther, and M.M. Mazzolini.
\newblock
  \href{https://www.sciencedirect.com/science/article/pii/0550321382902292}{The
  massive Schwinger model on a lattice: Background field, chiral symmetry and
  the string tension}.
\newblock {\em Nuclear Physics B}, 208(3):413--438, Dec 1982.

\bibitem{Peruzzo_2014}
Alberto Peruzzo, Jarrod McClean, Peter Shadbolt, Man-Hong Yung, Xiao-Qi Zhou,
  Peter~J. Love, Alán Aspuru-Guzik, and Jeremy~L. O’Brien.
\newblock \href{https://www.nature.com/articles/ncomms5213}{A variational
  eigenvalue solver on a photonic quantum processor}.
\newblock {\em Nature Communications}, 5(1), Jul 2014.

\bibitem{Li2017}
Ying Li and Simon~C. Benjamin.
\newblock
  \href{https://journals.aps.org/prx/abstract/10.1103/PhysRevX.7.021050}{Efficient
  Variational Quantum Simulator Incorporating Active Error Minimization}.
\newblock {\em Phys. Rev. X}, 7:021050, Jun 2017.

\bibitem{Dumitresu2018}
E.~F. Dumitrescu, A.~J. McCaskey, G.~Hagen, G.~R. Jansen, T.~D. Morris,
  T.~Papenbrock, R.~C. Pooser, D.~J. Dean, and P.~Lougovski.
\newblock
  \href{https://journals.aps.org/prl/abstract/10.1103/PhysRevLett.120.210501}{Cloud
  Quantum Computing of an Atomic Nucleus}.
\newblock {\em Phys. Rev. Lett.}, 120:210501, May 2018.

\bibitem{Kubra2020}
K\"ubra Yeter-Aydeniz, Raphael~C. Pooser, and George Siopsis.
\newblock \href{https://www.nature.com/articles/s41534-020-00290-1}{Practical
  quantum computation of chemical and nuclear energy levels using quantum
  imaginary time evolution and Lanczos algorithms}.
\newblock {\em npj Quantum Information}, 6(1), Jul 2020.

\bibitem{Motta_2019}
Mario Motta, Chong Sun, Adrian T.~K. Tan, Matthew~J. O'Rourke, Erika Ye,
  Austin~J. Minnich, Fernando G. S.~L. Brandão, and Garnet Kin-Lic Chan.
\newblock \href{https://www.nature.com/articles/s41567-019-0704-4}{Determining
  eigenstates and thermal states on a quantum computer using quantum imaginary
  time evolution}.
\newblock {\em Nature Physics}, 16(2):205–210, Nov 2019.

\end{thebibliography}

\appendix
\section{Hamiltonian Matrix Elements}\label{app:A}
In this appendix we provide details on the structure of the matrices for the Hamiltonian and discuss its spectrum. 
\subsection{Two spatial sites} \label{Section:Nequals2}

Here, we provide details on the case of two spatial sites ($L=2$). 
Since we are only interested in the ground and first excited states, we may restrict attention to the states obeying the constraint \eqref{eq:21}. They are built out of the set of half-filled fermionic states
\be \{\ket{1100},\ \ket{0110},\ \ket{0011},\ \ket{1001},\ \ket{1010},\ \ket{0101}\} \ee
These states are split into two subsets each of which is gauge-invariant under LGTs, $\{\ket{1100},\ \ket{0110},\ \ket{0011},\ \ket{1001} \}$ and $\{ \ket{1010},\ \ket{0101}\}$. The two subsets correspond to different eigenspaces of the translation operator $\mathcal{T}$ with eigenvalues $\pm 1$, respectively. The lowest energy states have $\mathcal{T} = +1$. Since under a LGT these states transform into each other, we will concentrate on the set of states $\ket{n,0011;\theta}$ without loss of generality. From \eqref{eq:eigenstate} they are given by
\be \label{eq:eigenstate L=4 new} |p=ne,0011;\theta\rangle = \frac{1}{2}\sum_{l=0}^3 e^{il\theta} U_1^l |p=ne\rangle \otimes \ket{0011 } \ ,  \ \ \theta = 0, \pm\frac{\pi}{2}, \pi \ee 
These are the eigenstates of the Hamiltonian in the strong-coupling regime.

For $\theta=0,\pi$, it is not difficult to symmetrize the states built out of $\ket{p=ne,0011;\theta}$ with respect to the parity operator $\Pi$. The advantage is that the ground and first excited states live in separate parity sectors. This fact manifests for example in the well-known eigenstates of the strong-coupling case at $\theta=0$, in which the first excited state is a ``vector meson" excitation above the vacuum \cite{bib2}. We also know that at zero mass and arbitrary $\theta$, the continuum limit yields ground and first excited states in which the wavefunctions $\braket{q}{n}$ are the ground and first excited state harmonic oscillator wavefunctions \cite{bib1}, which are opposite in parity.

To achieve this symmetrization define the states
\be \label{eq:L=4 symmetrized} \ket{n,0011;\theta;\pm} = \frac{1}{\sqrt2}\left(\ket{p=ne,0011;\theta}\pm\ket{p=-ne,0011;\theta}\right) \ee
which are eigenstates of the parity operator,
\be\label{eq:43} \Pi\ket{n,0011;\theta;\pm} = \pm\ket{n,0011;\theta;\pm} \ee
Next, we calculate matrix elements in the basis given by \eqref{eq:eigenstate L=4 new} and then apply this symmetrization to further reduce the size of the Hilbert space for the ground and first excited state energy calculations.

In the $e,m \gg 1$ regime, the Hamiltonian simplifies to
\be H^{(0)} = \frac{p^2}{8} + \frac{e^2}{32}\left(4j_1^\dagger j_1 +(j_2+2)^2\right)+mj_2 \ee 
with all terms mutually commuting. The states $\ket{n, 0011; \theta}$ are eigenstates of $H^{(0)}$ with corresponding energy levels
\be \label{eq: L=4 spectrum} E_{n,\theta}^{(0)} = \frac{n^2e^2}{8} + \frac{e^2}{2} \sin^2 \left( \frac{\theta}{2} - \frac{n\pi}{4} \right) \left[ 1+\cos^2 \left( \frac{\theta}{2} - \frac{n\pi}{4} \right) \right] -2m\cos (\theta - \frac{n\pi}{2}) \ee
For $\theta=0$, the ground state of $H^{(0)}$  (strong-coupling vacuum) is given explicitly by 
\be \braket{q}{0,0011;\theta=0}=\frac{1}{\sqrt{8\pi}}\left(\ket{0011}-\ket{1001}+\ket{1100}+\ket{0110}\right) \ee
where we used $\braket{q}{0}=\frac{1}{\sqrt{2\pi}}$,
and has energy
$ E_{0,0}^{(0)}=-2m $.
For $\theta=\pi$, the ground state is doubly degenerate. These two ground states are $\ket{\pm 2,0011;\theta=\pi}$ with energy $E_{\pm 2,\pi}^{(0)}=\frac{1}{2}e^2-2m$. Looking at Figure \ref{fig:dim=8}, perturbations from the rest of $H$ introduce an energy difference in the degenerate subspace. This gap increases for $e\gg m$ and vanishes asymptotically for $m\gg e$. The states $\ket{\pm 2,0011;\theta=\pi}$ map between one another under parity, and so we expect the ground states to exhibit spontaneous symmetry breaking in the continuum limit, for $m\gg e$.

The remaining part of the Hamiltonian is
\be H^{(1)} = H - H^{(0)} = \sum_{l=0}^3 b_l^\dagger b_l \sin k_l \ , \ \ k_l = \frac{(2l+1)\pi}{4} - eq \ee
Using the $q$-space representation for the gauge degree of freedom, we obtain the matrix elements
\be\label{eq:A5} \langle n',0011;\theta | H^{(1)} | n,0011; \theta\rangle = - \frac{1}{\sqrt{2}} (\delta_{n',n+1} + \delta_{n',n-1} ) \ee
For $\theta=0,\pi$, there are no transitions between states of different parity $\Pi$. We can construct independent Hamiltonian matrices in the even and odd parity sectors using the parity eigenstates given by Eq.\ \eqref{eq:L=4 symmetrized}. Since the energy levels \eqref{eq: L=4 spectrum} are invariant under $(n,\theta) \to (-n,-\theta)$, parity eigenstates are also eigenstates of $H^{(0)}$ with eigenvalues given by \eqref{eq: L=4 spectrum}. The matrix elements of $H^{(1)}$ in the basis of parity eigenstates are easily deduced from \eqref{eq:A5}. 
The ground and first excited states of the system are the ground states of the Hamiltonian in the even and odd parity sectors, respectively.

\subsection{Three spatial sites} \label{Section:Nequals3}

Working as in the $L=2$ case, we construct the ground and first-excited states from the states $|p=ne,000111;\theta\rangle$ and $|p=ne,101010;\theta\rangle$, where
\be \label{eq:eigenstate L=6} |p=ne,\bm{x} ;\theta\rangle \propto \sum_{l=0}^5 e^{il\theta} U_1^l |p=ne\rangle \otimes \ket{\bm{x} } \ ,  \ \ \theta = 0, \pm\frac{\pi}{3}, \pm\frac{2\pi}{3}, \pi \ee 
Thus, we obtain two independent sets of gauge-invariant states, both with $\mathcal{T} = +1$. 
Note that some of these states vanish ($\ket{p=ne,101010;\theta}$ vanishes for $n\ne \frac{3\theta}{\pi}$ mod(3)).

At $e,m\gg 1$ the Hamiltonian is
\be H^{(0)} = \frac{p^2}{12}+\frac{e^2}{48}\left(8j_1^\dagger j_1+\frac{8}{3}j_2^\dagger j_2 + \left(j_3+3\right)^2\right) + m j_3 \ee
Concentrating on $\theta=0,\pi$, the eigenstates of $H^{(0)}$ can easily be found by diagonalizing $j_3$. For $n\ne 0\text{ mod}(3)$, $\ket{p=ne,101010;\theta = 0,\pi}$ vanishes and $\ket{p=ne,000111;\theta=0,\pi}$ is an eigenstate of $j_3$ with eigenvalue $-2\cos \left( \theta -\frac{n\pi}{3} \right)$. For $n= 0\text{ mod}(3)$, the eigenstates of $j_3$ are
\bea \ket{p=ne,\alpha;\theta=0,\pi}&\equiv&\frac{1}{2}\left(\sqrt3\ket{p=ne,000111;\theta=0,\pi}-\ket{p=ne,101010;\theta=0,\pi}\right)\nonumber\\ \ket{p=ne,\beta;\theta=0,\pi}&\equiv&\frac{1}{2}\left(\ket{p=ne,000111;\theta=0,\pi}+\sqrt3\ket{p=ne,101010;\theta=0,\pi}\right) \eea
with eigenvalues $-3\cos \left( \theta -\frac{n\pi}{3} \right)$ and $\cos \left( \theta -\frac{n\pi}{3} \right)$, respectively. 

The Hamiltonian $H^{(0)}$ is block-diagonal with $2\times2$ blocks for $n=0\ \text{mod} (3)$ and $1\times 1$ blocks otherwise. We deduce the spectrum of $H^{(0)}$,
\be E_{n}^{(0)} = \frac{n^2+4}{12} e^2 + \frac{e^2}{48} \left[ 3 - 2 \cos \left( \theta - \frac{n\pi}{3} \right) \right]^2 -2m \cos \left( \theta - \frac{n\pi}{3} \right) \ee 
for $n\ne 0\text{ mod}(3)$, and
\bea E_{n,\alpha}^{(0)} &=& \frac{n^2}{12} e^2 + \frac{3e^2}{4}   \sin^4 \left( \frac{\theta}{2} - \frac{n\pi}{6} \right)  -3m \cos \left( \theta - \frac{n\pi}{3} \right) \nonumber\\
E_{n,\beta}^{(0)} &=& \frac{n^2+8}{12} e^2 + \frac{e^2}{48} \left[ 3 + \cos \left( \theta - \frac{n\pi}{3} \right) \right]^2 +m \cos \left( \theta - \frac{n\pi}{3} \right) \eea
for $n= 0\text{ mod}(3)$.

For $\theta = 0$, the strong-coupling vacuum is the state $\ket{p=0,\alpha;\theta =0}$ with energy $E_{0,\alpha}^{(0)} = -3m$.  As with $L=2$, $\theta=\pi$ contains a doubly degenerate ground state. The states are $\ket{\pm 3,\alpha;\theta=\pi}$ with energy $E_{\pm 3,\alpha}^{(0)}=\frac{3}{4}e^2-3m$. As seen in Figure \ref{fig:n=47}, $H^{(1)}$ once again introduces an energy difference which increases for $e\gg m$ and vanishes asymptotically for $m\gg e$.

The remaining part of the Hamiltonian is 
\be H^{(1)}=\sum_{l=0}^5 b_l^\dagger b_l\sin k_l\ ,\ \  k_l=\frac{\left(2l+1\right)}{6}-eq \ee
Its matrix elements in the basis \eqref{eq:eigenstate L=6} can be calculated in the same fashion as for $L=2$. We obtain the non-vanishing matrix elements
\be \bra{p=n'e,000111;\theta}H^{(1)}\ket{p=ne,000111;\theta} = -\delta_{n',n+1} -\delta_{n',n-1} \ee
From this we may easily construct $H$ in the basis of parity eigenstates, keeping in mind that the energy levels of $H^{(0)}$ are invariant under $(n,\theta) \to (-n,-\theta)$ as before.

\subsection{Four spatial sites} \label{Section:L>6}

For $L=4$, the ground and first-excited states are constructed from three sets of states $\ket{p=ne,00001111;\theta}$, $\ket{p=ne,10010110;\theta}$, and $\ket{p=ne,01100110;\theta}$, with
\be \label{eq:eigenstate L=8} |n,\bm{x} ;\theta\rangle \propto \sum_{l=0}^7 e^{il\theta} U_1^l |p=ne\rangle \otimes \ket{\bm{x} } \ ,  \ \ \theta = 0, \pm\frac{\pi}{4}, \pm\frac{\pi}{2}, \pm\frac{3\pi}{4}, \pi\ee
The Hamiltonian in the strong-coupling limit $H^{(0)}$ is block-diagonal with blocks as large as $3\times 3$. Thus we see that the maximum block size increases with $L$. For $\theta = 0$, the strong-coupling vacuum $\ket{0; \theta =0}_{\text{SC}}$ lives in a $2\times2$ block corresponding to $n=0$. We obtain
\be \ket{0; \theta =0}_{\text{SC}}=\frac{1}{\sqrt2}\left(\ket{0,00001111;0}+\ket{0,10010110;0}\right) \ee
with energy $E^{(0)}=-4m$.  As might be expected from $L=2,3$ the two degenerate ground states for $\theta=\pi$ are
\be \frac{1}{\sqrt2}\left(\ket{p=\pm4e,00001111;\theta=\pi}+\ket{p=\pm4e,10010110;\theta=\pi}\right) \ee
with energy $E^{(0)}=e^2-4m$.

The non-vanishing matrix elements of $H^{(1)}$ in the basis \eqref{eq:eigenstate L=8} are 
\bea \bra{p=n'e,00001111;\theta}H^{(1)}\ket{p=ne,00001111;\theta}&=&-\sqrt{2}\cos\frac{\pi}{8}\left(\delta_{n',n+1}+\delta_{n',n-1}\right) \nonumber\\ \bra{p=n'e,10010110;\theta}H^{(1)}\ket{p=ne,10010110;\theta}&=&-\sqrt{2}\sin\frac{\pi}{8}\left(\delta_{n',n+1}+\delta_{n',n-1}\right) \eea

\subsection{Larger Lattices} \label{section:larger lattices}

As we increase $L$ beyond $4$, the number of relevant sets of gauge-invariant states increases quickly. While we only needed $L-1$ sets for $L\le 4$, we need $L+1=6$ sets of states for $L=5$ and $2L+2=14$ sets of states for $L=6$. 

What we can say is that for $L\le 5$ we have $\sim L$ sets of gauge-invariant states. As is shown in Appendix \ref{Section:Truncation}, for a wide range of parameters $e$ and $m$, we obtain fairly accurate results with a Hilbert space cutoff $n_{\text{max}} \sim 2L$. Therefore, the Hilbert space needed for the calculation of low-lying energy levels and the mass gap has dimension $\sim 2L^2$ for $L\le 5$. Thus, about $2\log_2 L+1$ qubits are needed for quantum calculations. For the case $L=4$ considered here for the quantum computation of the phase transition, only five qubits at most are needed. In Appendix \ref{Section:Truncation} we will show that in fact three qubits suffice.


\section{Role of the theta parameter}\label{app:B}

In this appendix we discuss the well-known connection between $\theta$ and the background electric field in the context of our formalism. We also study via several examples the dependence of the gap on the value of $\theta$. We demonstrate through these examples that a phase transition does not occur for $\theta\ne\pi$.

\subsection{Background field} \label{Section:theta equiv}
As mentioned in section \ref{Section:Continuum}, the parameter $\theta$, which determines the gauge sector one is working in, can also be thought of as a constant of integration when solving Gauss's Law,
\be\label{eq:B1} F=\frac{e\theta}{2\pi}, \ee
where $F$ acts as a background field to the system. Introducing $\theta$ as in \eqref{eq:eigenstate}, we can see its connection to a background electric field as follows. In terms of the gauge zero mode, we have
\be\label{eq:prefactor state} \braket{q}{p=ne,\bm{x};\theta} \propto e^{ineq} \sum_{l=0}^{2L-1}e^{i\left(\theta-\frac{\pi n}{L}\right)l} U_1^l\ket{\bm{x}} \ . \ee
This is an eigenstate of the zero mode of the electric field with eigenvalue $\frac{p}{2L} = \frac{ne}{2L}$.

Evidently, this state is related to the $\theta = 0$ state with the quantum number $n$ shifted by $\frac{L\theta}{\pi}$,
\be \braket{q}{p=ne,\bm{x};\theta} = e^{i\frac{L\theta}{\pi}eq} \braket{q}{p=ne - \frac{Le\theta}{\pi},\bm{x};\theta = 0} \ee
This shift corresponds to a shift in the zero mode of the electric field, $\frac{p}{2L} \to \frac{p}{2L} - \frac{e\theta}{2\pi}$, which coincides with the background field $F$ (Eq.\ \eqref{eq:B1}).


Coleman \cite{COLEMAN1976} includes the background field in the full electric field operator, and then describes the spectrum in terms of a Hamiltonian which depends on $F$, hence $\theta$. This may be viewed as an active transformation of the Hamiltonian. In our case, instead of modifying the operators, the Hilbert space of states is itself modified by the addition of a background field, as demonstrated by \eqref{eq:prefactor state}. In addition, it follows from \eqref{eq:prefactor state} that switching from the $\theta$-sector to $\theta=0$ is implemented with a unitary operator which is not gauge-invariant. In the continuum case \cite{bib1}, the generator of this transformation is related to the chiral charge $Q_5$, and the addition of a background field to the system is related to the phenomenon of spontaneous chiral symmetry breaking.

\subsection{Behavior of the gap for general theta} \label{Section:other theta}



Here, we study the behavior of the mass gap for $\theta \ne 0$ where no phase transition is expected.

\begin{figure}[ht]
    \centering
    \subfigure[\ $e=0.05$]{\includegraphics[scale=0.5]{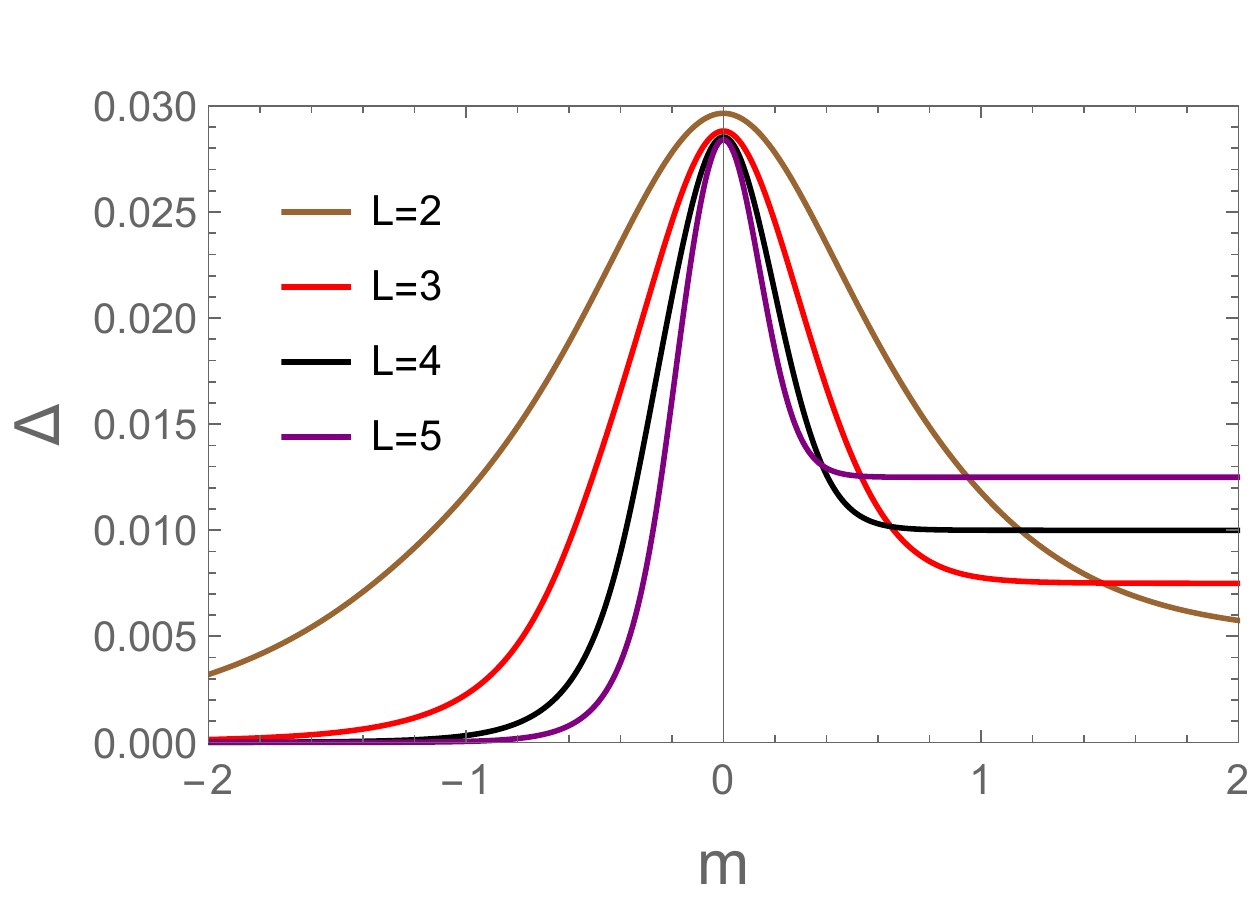}}
    \subfigure[\ $e=0.1$]{\includegraphics[scale=0.5]{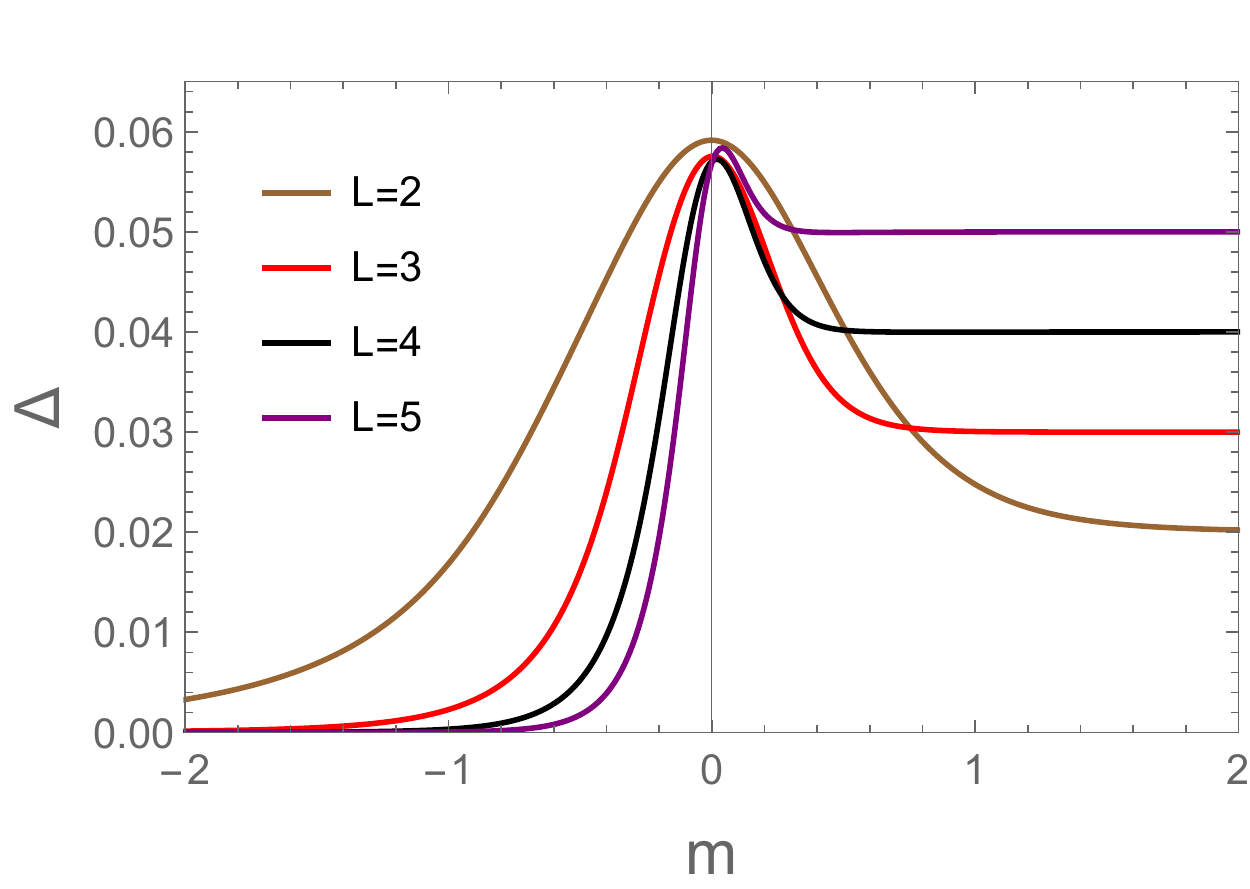}}
    \subfigure[\ $e=0.5$]{\includegraphics[scale=0.5]{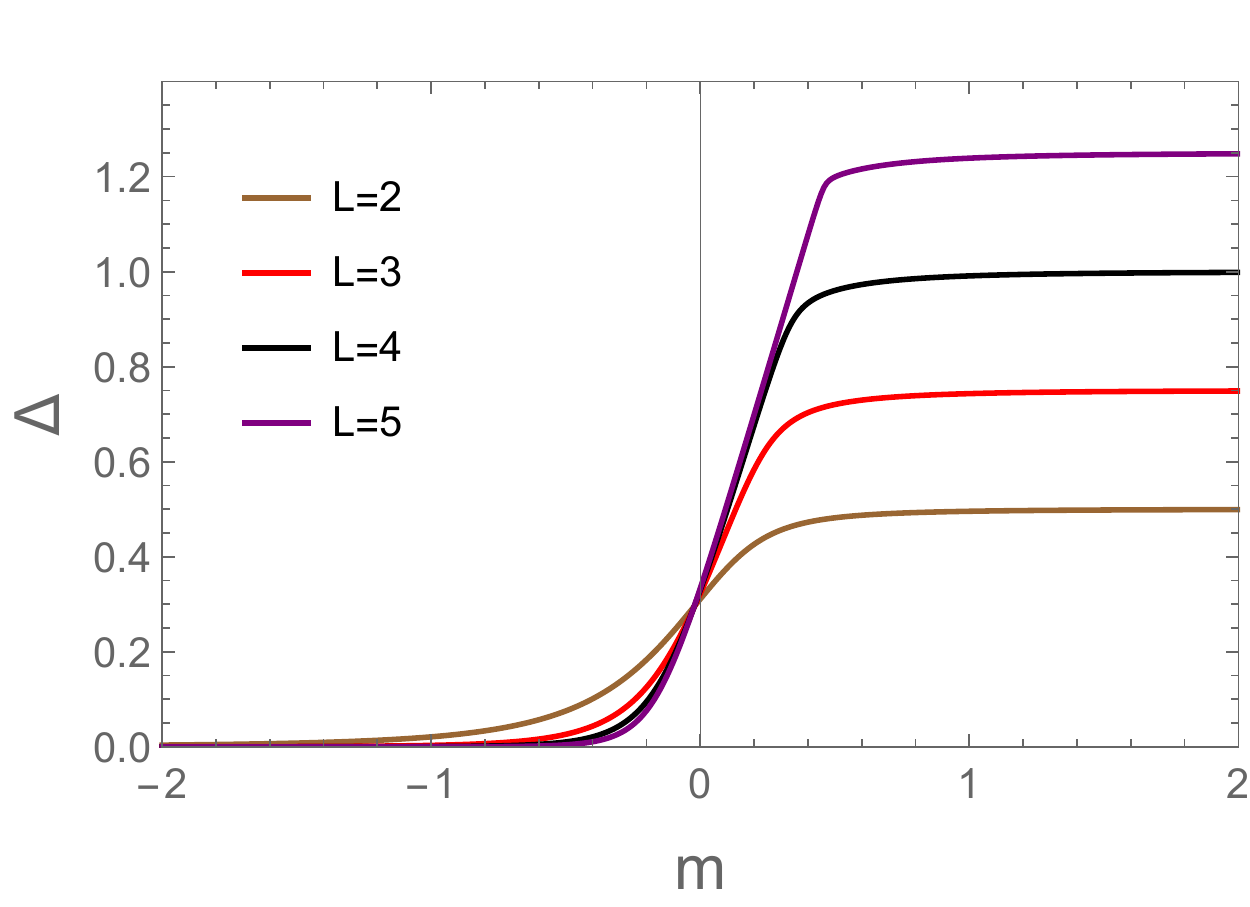}}
    \subfigure[\ $e=1$]{\includegraphics[scale=0.5]{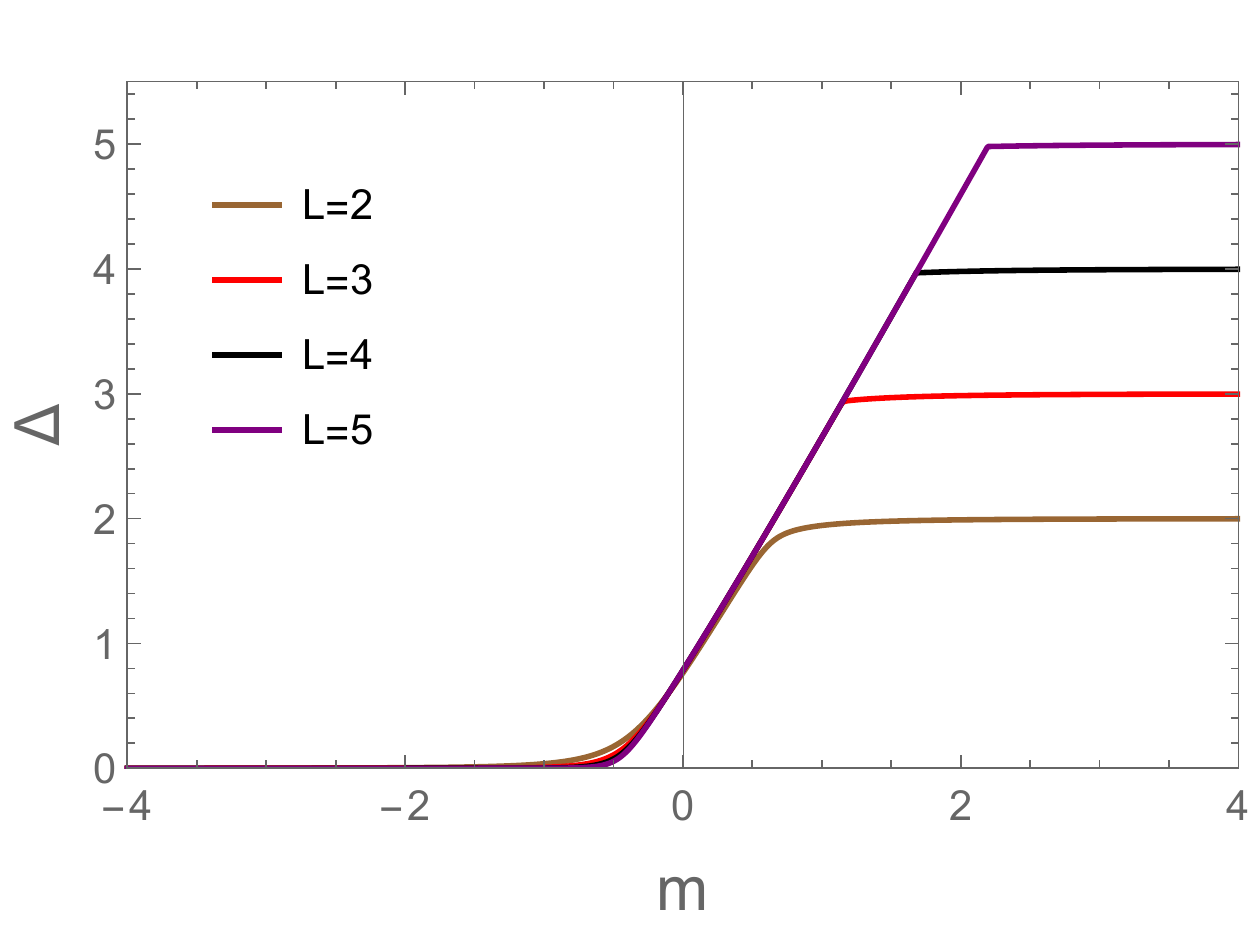}}
    \caption{Mass gap $\Delta$ vs.\ bare mass $m$ in the $\theta=0$ sector for $L=2,3,4,5$. }
    \label{fig:gap_vs_m_e=01_N=2,3,4,5,6_theta=0}
\end{figure}
In Figure \ref{fig:gap_vs_m_e=01_N=2,3,4,5,6_theta=0}, the mass gap is plotted for $\theta=0$. Evidently, it does not vanish for $m>0$, which is consistent with the expectation that no phase transition occurs. In the plots we included $m<0$ which is equivalent to the $\theta = \pi$ sector (the latter is obtained via a reflection about $m=-\frac{e^2}{8}$). The mass gap approaches zero for large negative $m$, which is consistent with the phase transition occurring for $\theta = \pi$. Notice that convergence to the continuum ($L\to\infty$) limit is quicker as the coupling $e$ increases for $m<0$. 

\begin{figure}[ht]
    \centering
    \subfigure[\ $e=0.1$]{\includegraphics[scale=0.45]{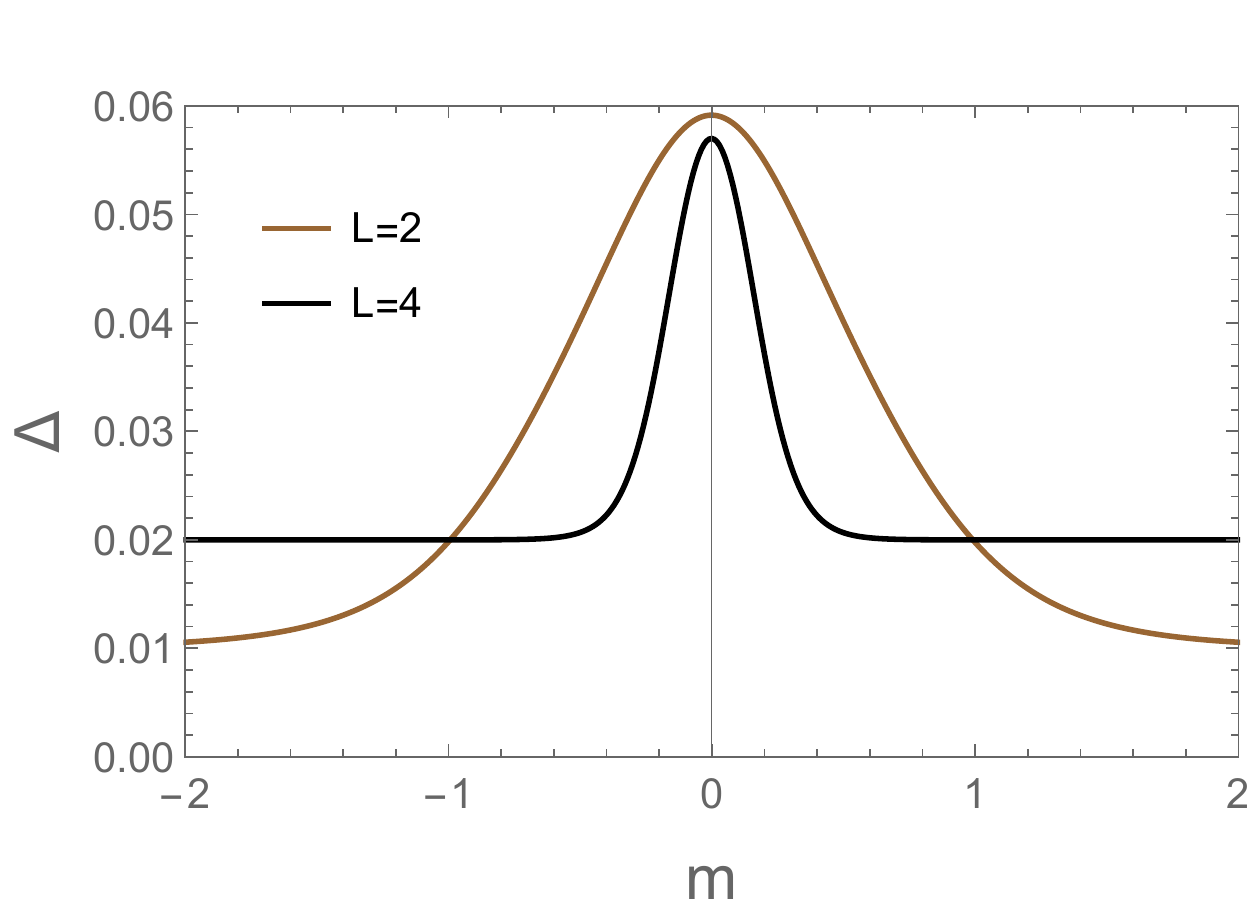}}
    \subfigure[\ $e=0.5$]{\includegraphics[scale=0.45]{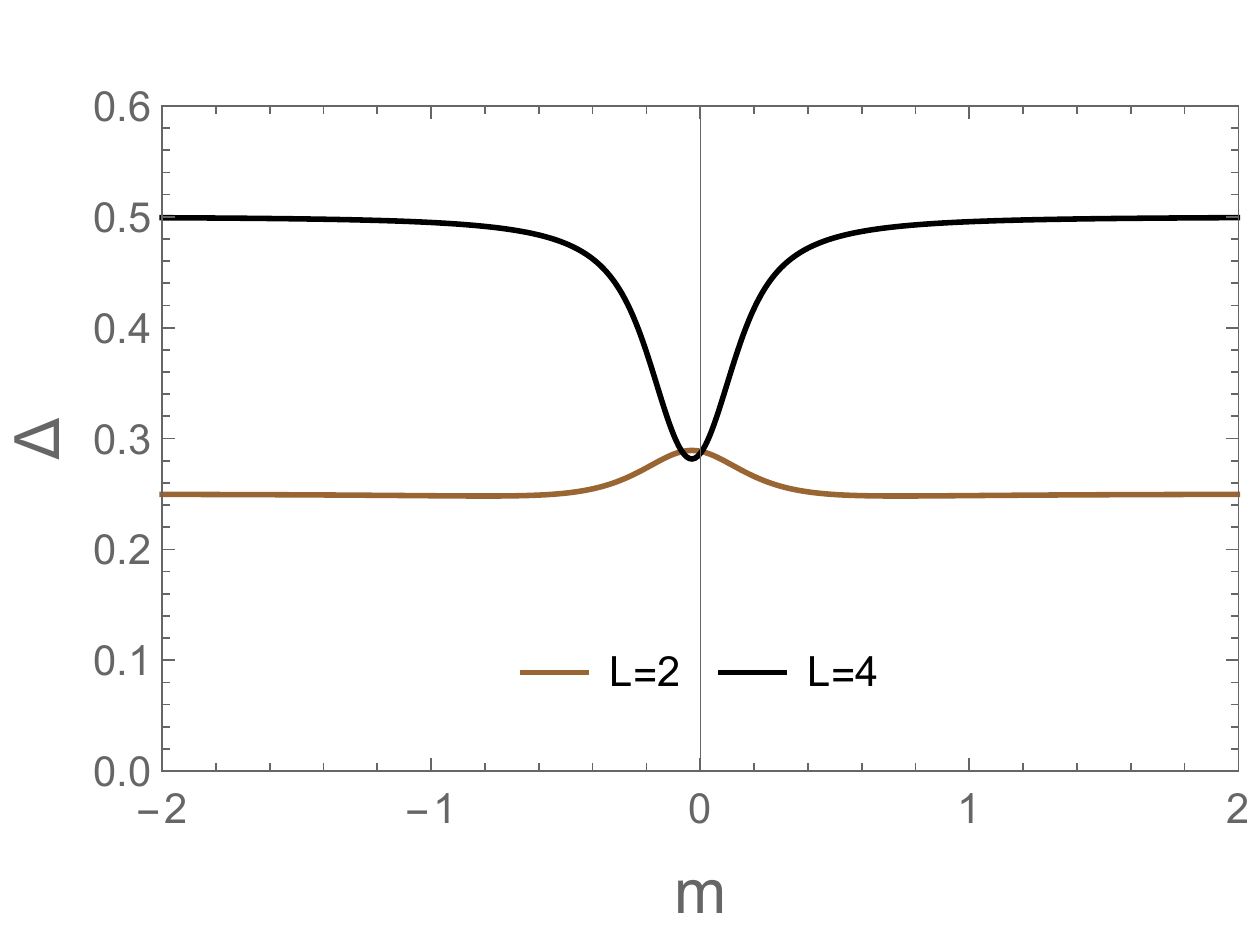}}
    \subfigure[\ $e=1$]{\includegraphics[scale=0.45]{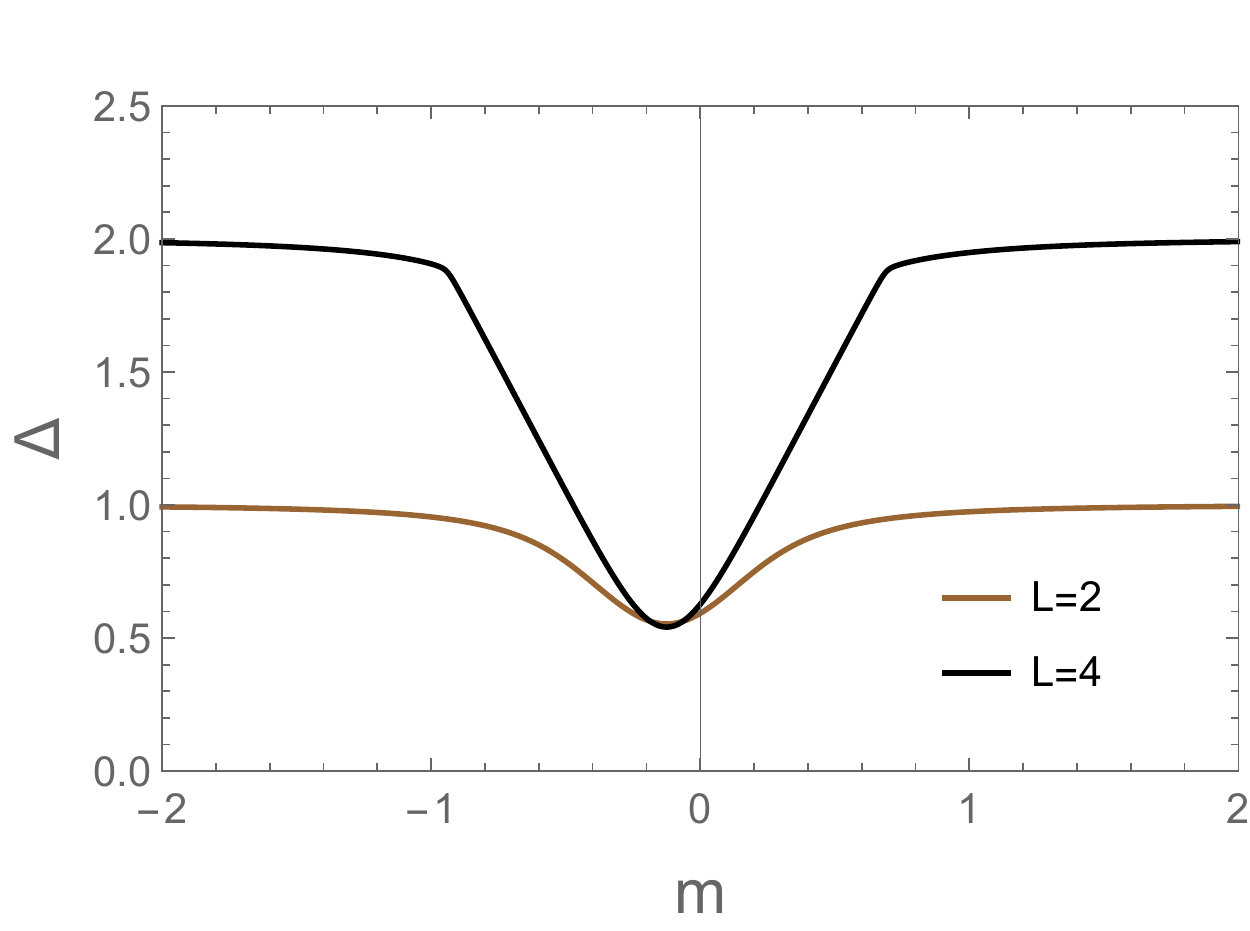}}
    \caption{Gap $\Delta$ vs.\ bare mass $m$ in the $\theta=\pm\frac{\pi}{2}$ sectors for $L=2,4$. }
    \label{fig:theta=piover2_e=01}
\end{figure}
\begin{figure}[ht]
    \centering
    \subfigure[\ $e=0.1$]{\includegraphics[scale=0.45]{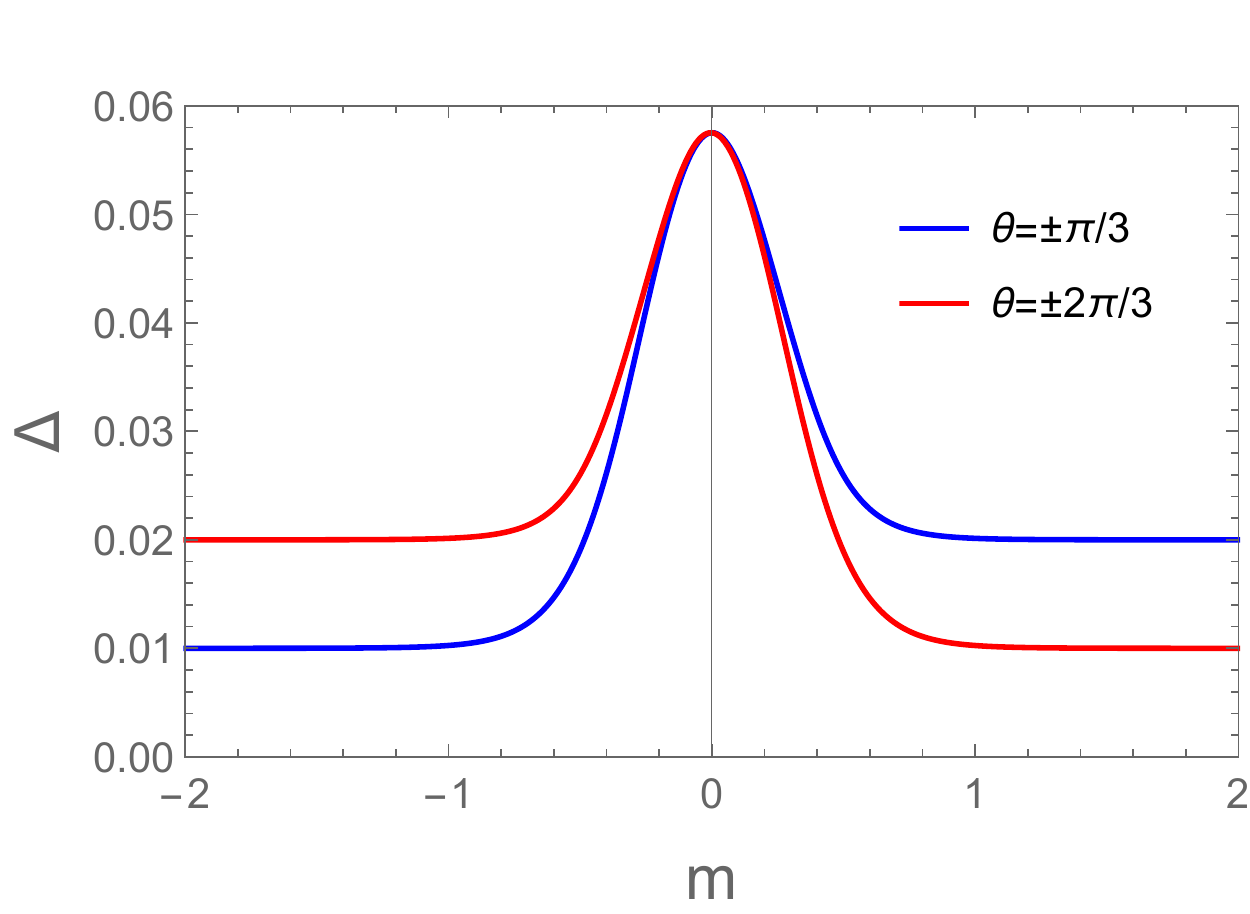}}
    \subfigure[\ $e=0.5$]{\includegraphics[scale=0.45]{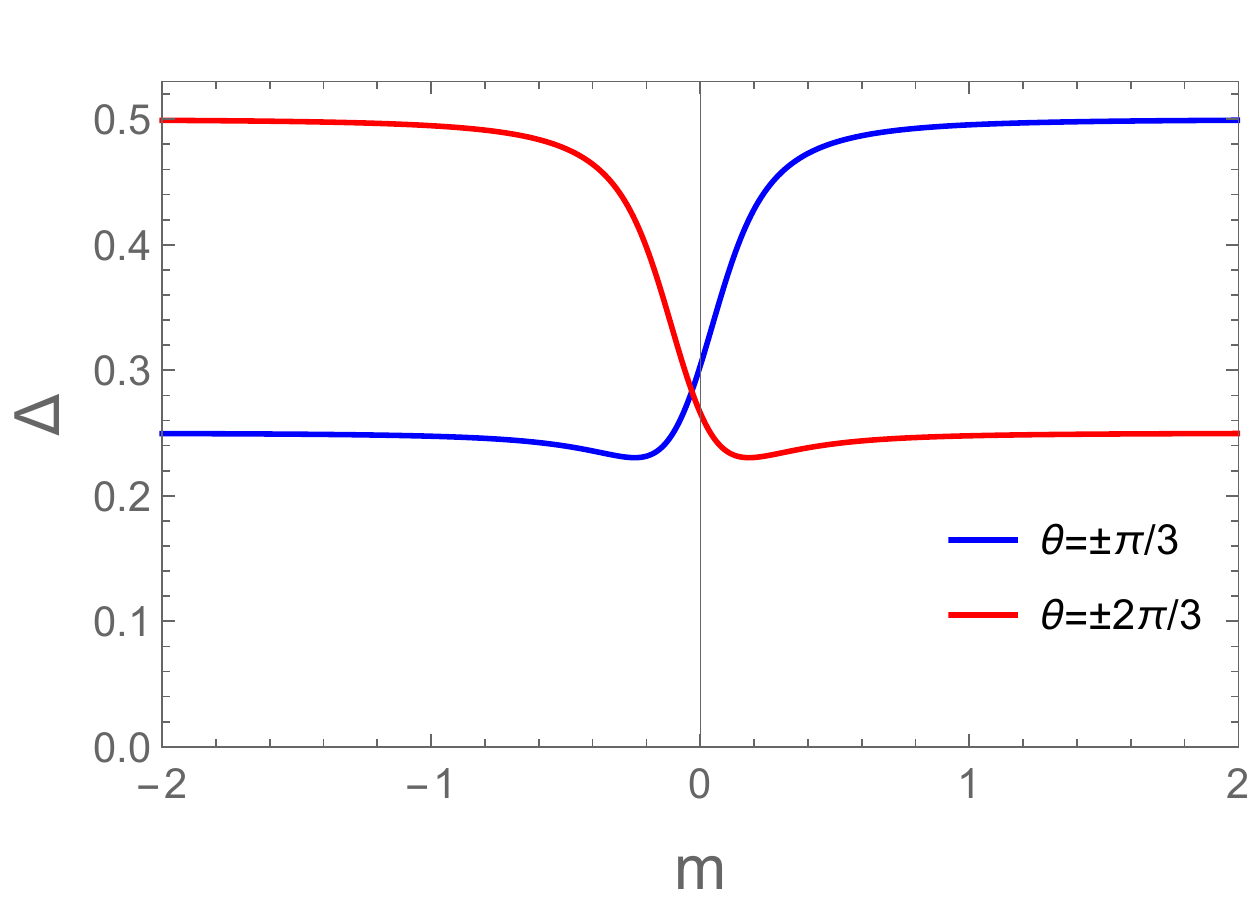}}
    \subfigure[\ $e=1$]{\includegraphics[scale=0.45]{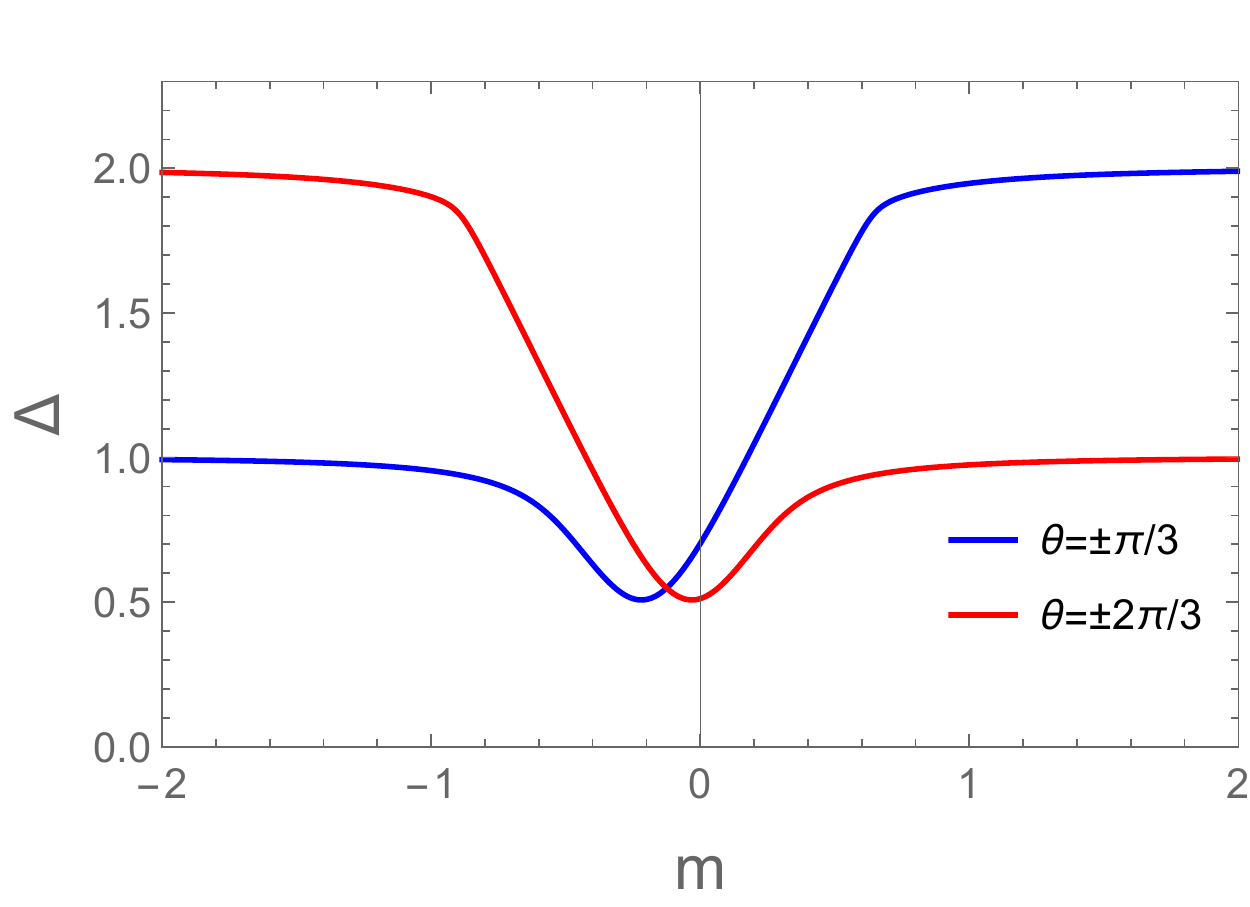}}
    \caption{Gap $\Delta$ vs.\ bare mass $m$ in the $\theta=\pm\frac{\pi}{3}$ and $\theta=\pm\frac{2\pi}{3}$ sectors for $L=3$.}
    \label{fig:N=3_theta=piover3_and_2piover3}
\end{figure}

In Figure \ref{fig:theta=piover2_e=01}, we plot the mass gap for $\theta = \pm \frac{\pi}{2}$. In these cases the gap does not approach zero, as no phase transition occurs. Similar behavior is observed for $\theta = \pm \frac{\pi}{3}, \pm \frac{2\pi}{3}$, as shown in Figure \ref{fig:N=3_theta=piover3_and_2piover3}.

\section{Truncation} \label{Section:Truncation}
\begin{figure}[ht]
    \centering
    \subfigure[\ $L=2$]{\includegraphics[scale=0.45]{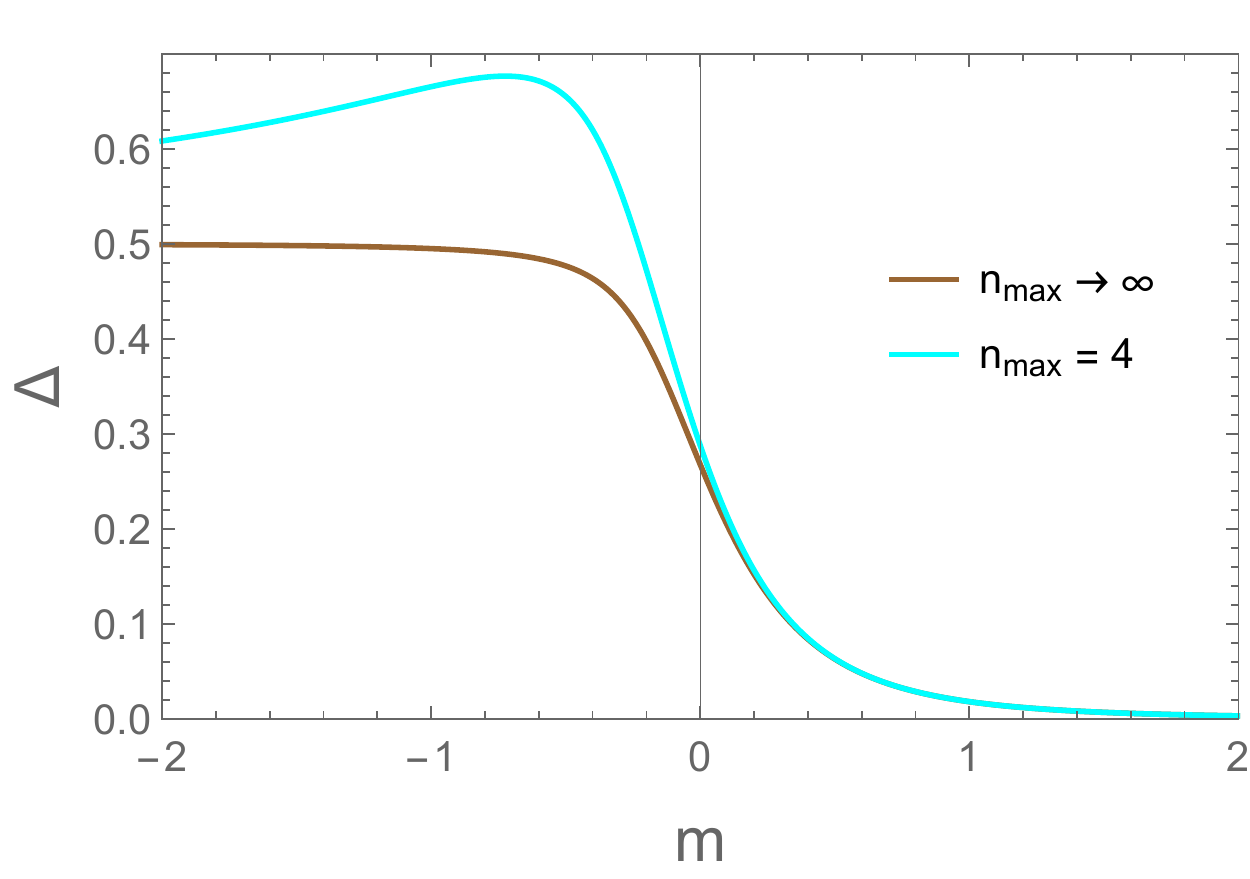}}
    \subfigure[\ $L=3$]{\includegraphics[scale=0.45]{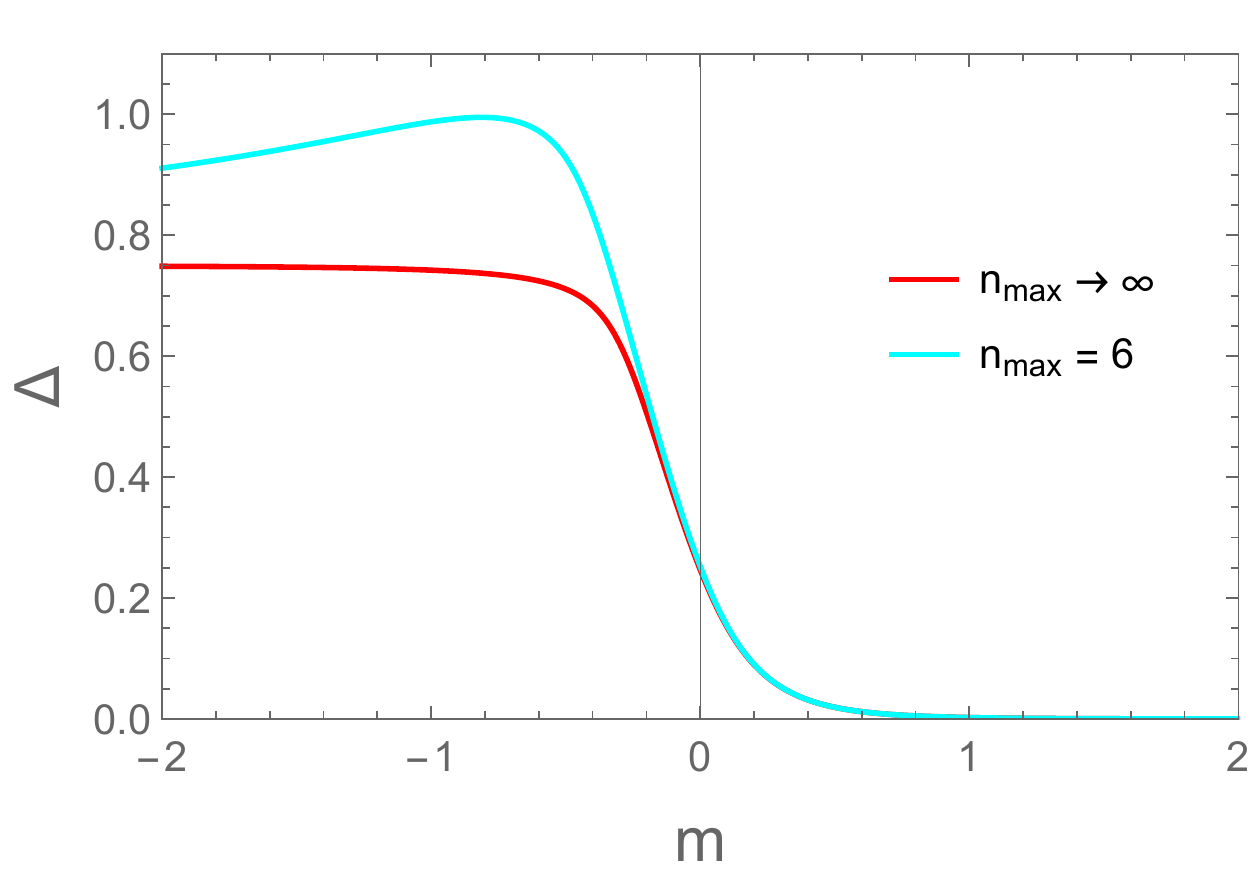}}
    \subfigure[\ $L=4$]{\includegraphics[scale=0.45]{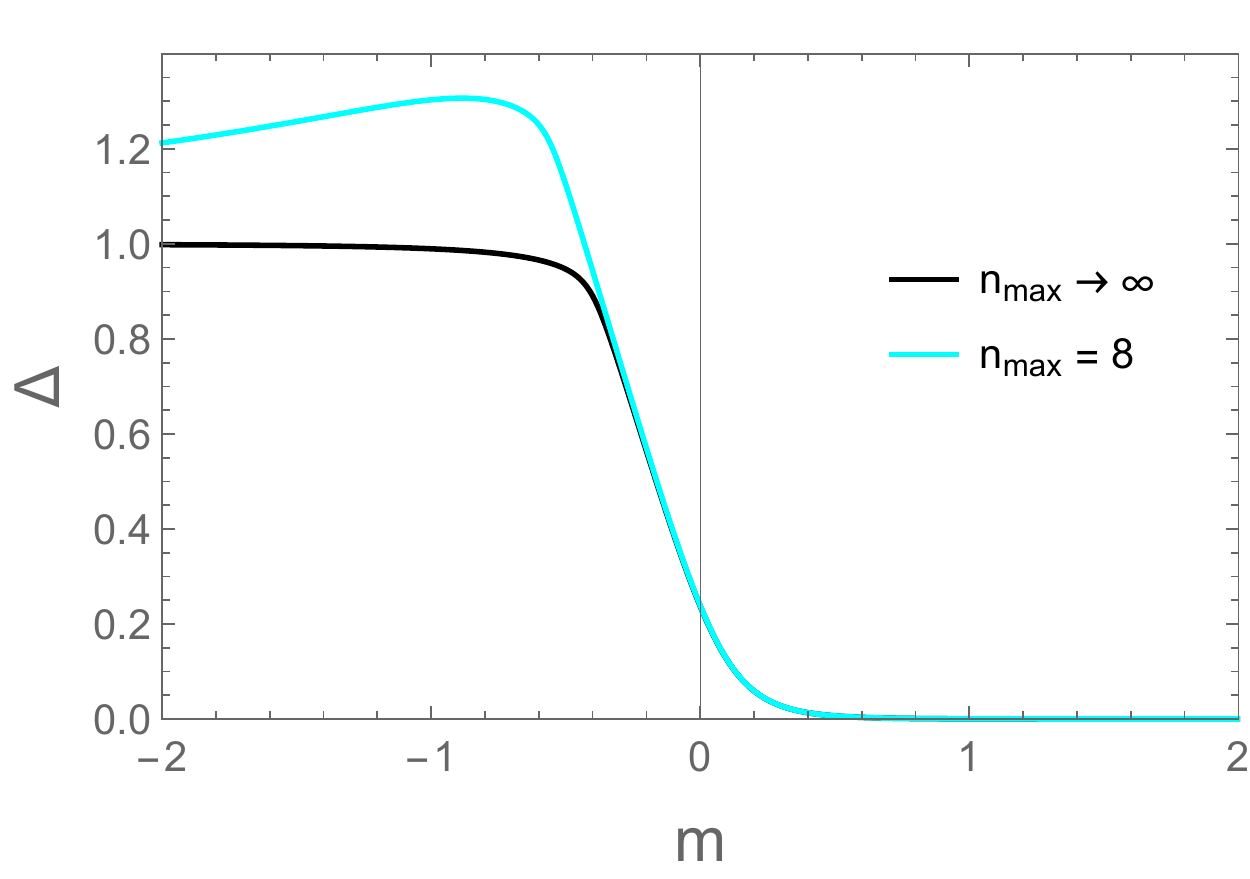}}
    \caption{Gap $\Delta$ vs.\ bare mass $m$ at coupling $e=0.5$ in the $\theta=\pi$ sector. The curve for the cutoff $n_{\text{max}} =2L$ is compared with the curve obtained when the cutoff is sent to infinity. }
    \label{fig:truncation_2N_vs_large_e=05}
\end{figure}

\begin{figure}[ht]
    \centering
    \subfigure[\ $L=2$]{\includegraphics[scale=0.45]{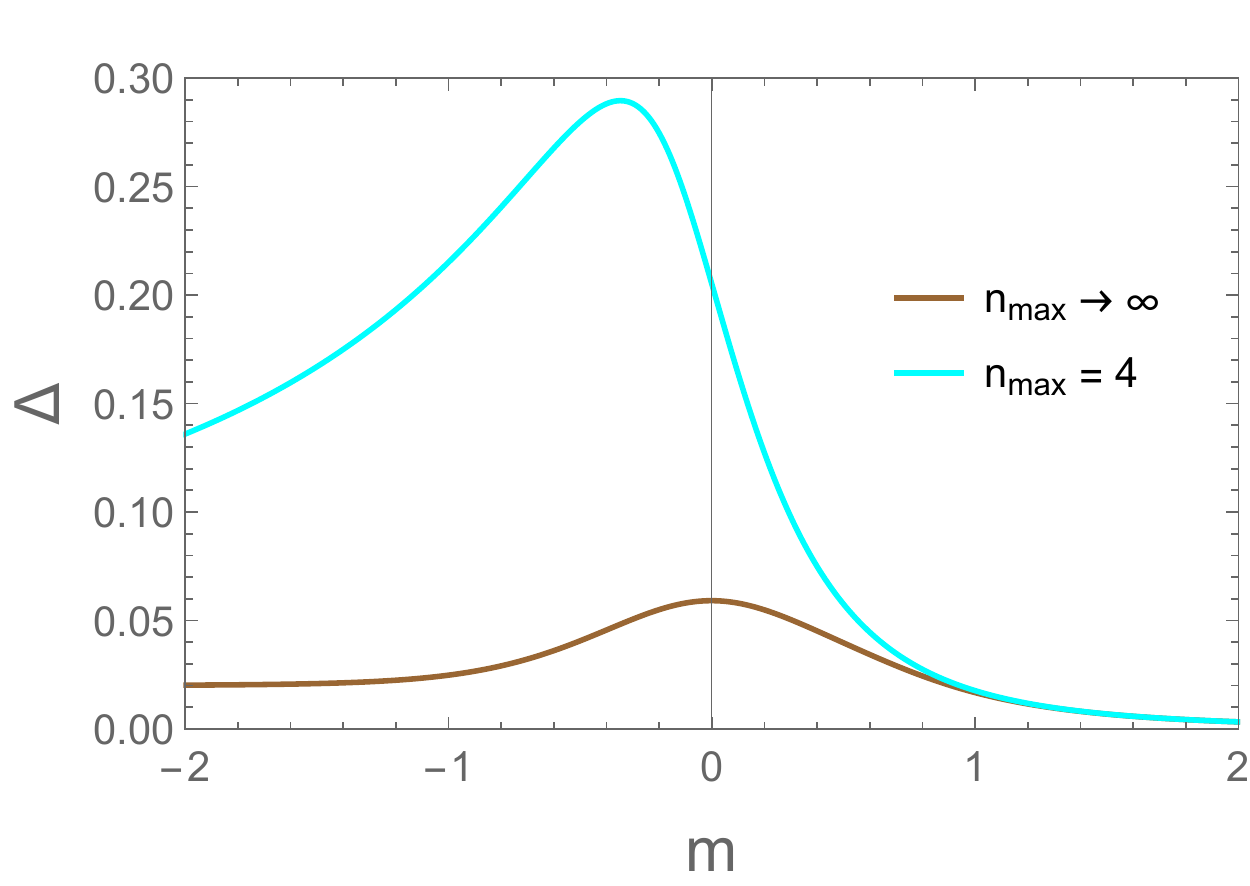}}
    \subfigure[\ $L=3$]{\includegraphics[scale=0.45]{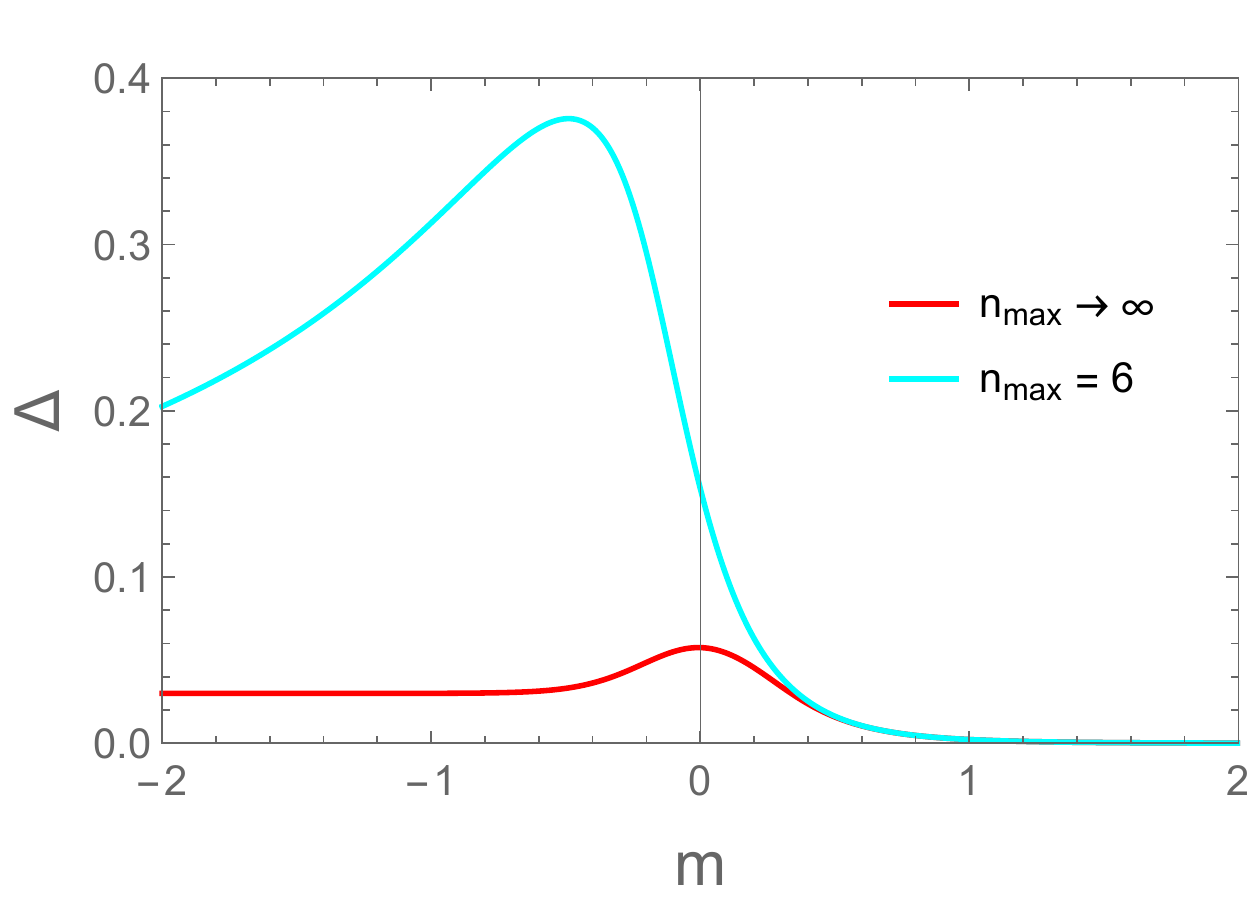}}
    \subfigure[\ $L=4$]{\includegraphics[scale=0.45]{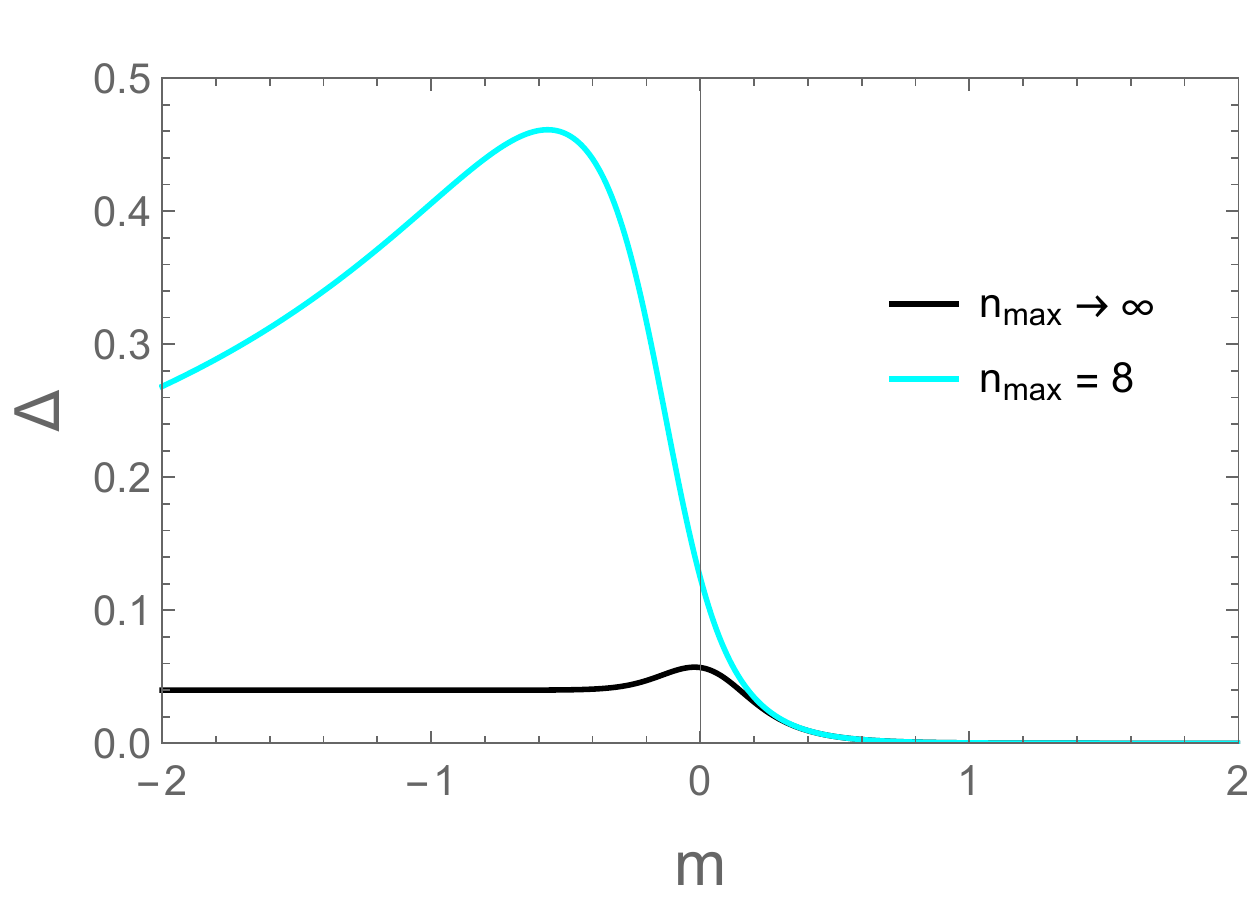}}
    \caption{Gap $\Delta$ vs.\ bare mass $m$ at coupling $e=0.1$ in the $\theta=\pi$ sector. The curve for the cutoff $n_{\text{max}} =2L$ is compared with the curve obtained when the cutoff is sent to infinity. }
    \label{fig:truncation_2N_vs_large_e=01}
\end{figure}

Interestingly a high cutoff for the gauge field is not required for many values of $m$ in the $\theta=\pi$ sector, namely for $m\gg 1$. This can be seen in Figures \ref{fig:truncation_2N_vs_large_e=05} and \ref{fig:truncation_2N_vs_large_e=01}.
Looking at these figures, we deduce that the cutoff $n_{\text{max}}\sim 2L$  suffices for the points in Figure \ref{fig:pseudo_vs_coupling} which have small error bars. Therefore any truncation error will correspond to points where there is already a good deal of uncertainty even with a cutoff $n_{\text{max}}\to\infty$. This further prompts us to focus on $e\ge 0.5$ for the quantum computation.
In Figures \ref{fig:truncation_2N_vs_large_e=05} and \ref{fig:truncation_2N_vs_large_e=01} agreement between the curves corresponding to a cutoff $n_{\text{max}} =2L$ and the exact curves holds for most $m>0$ (physical regime for $\theta=\pi$), but this is not the case for $m<0$ (physical regime for $\theta=0$). The range of agreement expands as $L$ increases. Note that the $\theta=0$ case can be obtained by reflecting across $m=-\frac{e^2}{8}$. 

\begin{figure}[ht]
    \centering
    \subfigure[\ $L=3$]{\includegraphics[scale=0.5]{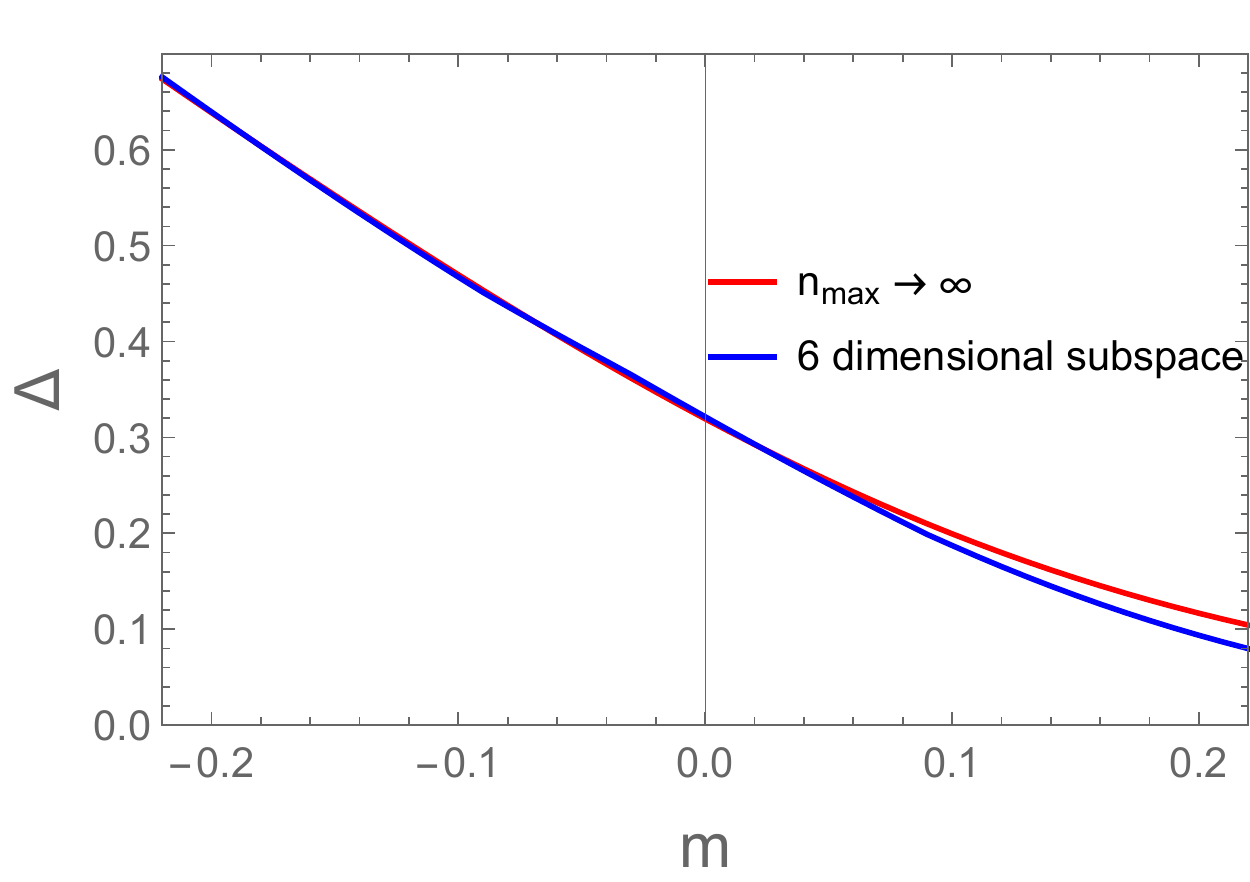}}
    \subfigure[\ $L=4$]{\includegraphics[scale=0.5]{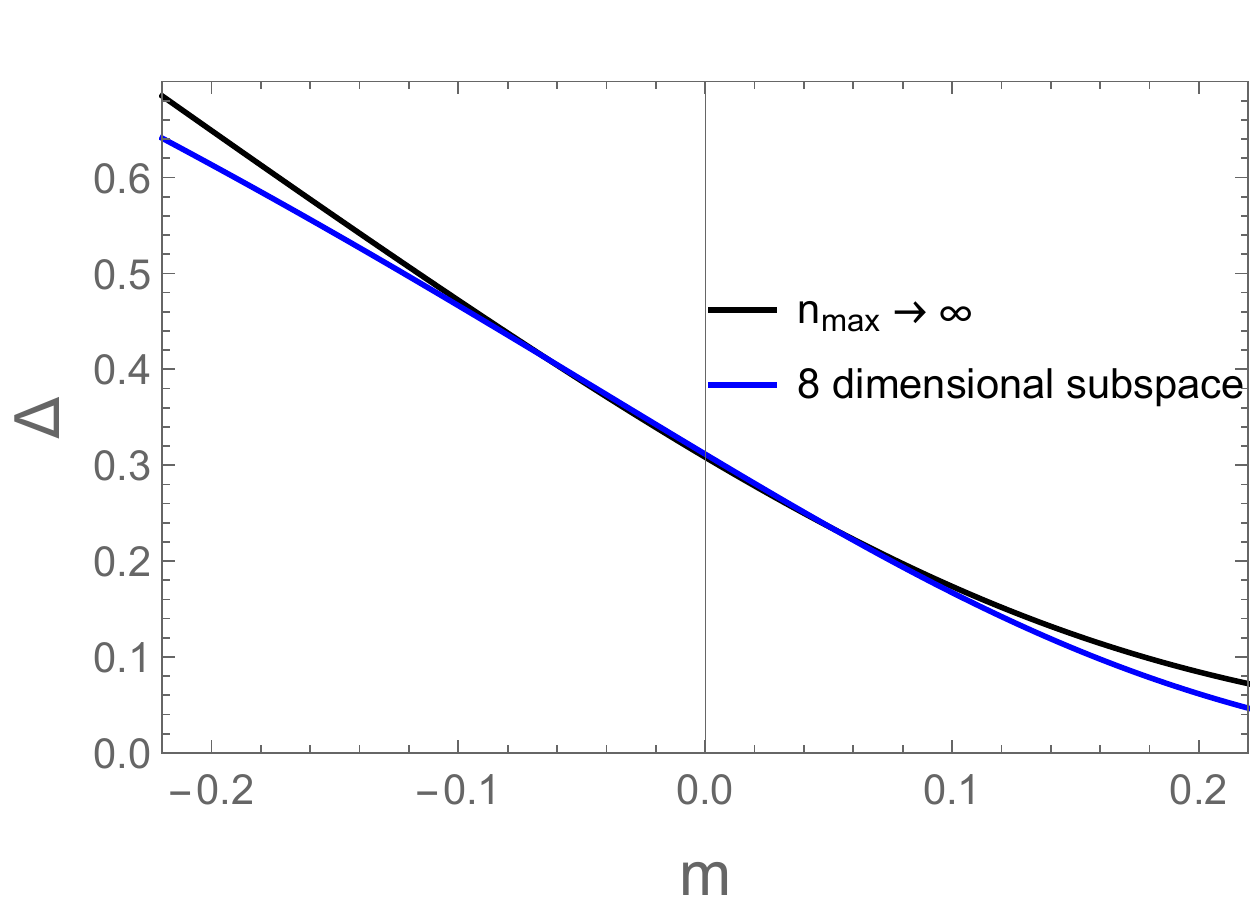}}
    \caption{Gap $\Delta$ vs.\ bare mass $m$ at coupling $e=0.75$ in the $\theta=\pi$ sector. The curve obtained from the eigenvalues of an ``optimal" six-dimensional Hamiltonian (for $L=3$) and eight-dimensional Hamiltonian (for $L=4$) is compared with the curve obtained when the cutoff is sent to infinity. }
    \label{fig:N=34_3qub}
\end{figure}

In our quantum calculation we work with very limited resources, and so even a cutoff of $n_{\text{max}}=2L$ is not practical for $L>3$. In our calculations we reduce $L=3$ and $L=4$ to a three-qubit problem. While immediately this may seem to be an extreme move, especially for $L=4$ which has a Hilbert space of dimension $20$ at truncation $n_{\text{max}}=8$, it turns out not to cause insurmountable problems. If we choose to populate our low-dimensional Hilbert space (six and eight-dimensional for $L=3,4$, respectively) with the most relevant basis states for the range of $e$ that is of interest to us, in this case $[0.5,1.0]$, then we get gap vs. mass curves such as those shown in Figure\ \ref{fig:N=34_3qub}.
In the figure, we zoom in so that we may analyze the region containing the pseudo-critical points: $m\in[0.12,0.22]$. We see a small degree of disagreement between the curves corresponding to the three-qubit truncation and the exact curves, in the region of interest, but there is improvement when $e$ is increased.

\end{document}